\documentclass[12pt, a4wide]{book}
\usepackage{setspace}
\usepackage{datetime}
\usepackage{ifpdf}
\usepackage{graphicx}
\usepackage{epsfig}
\usepackage{pgf}
\usepackage{slashbox}
\usepackage{enumerate}
\usepackage{amsmath}
\usepackage{amssymb}
\usepackage{braket}
\usepackage{tensor}
\usepackage{subfig}
\usepackage{wrapfig}
\usepackage{color}
\usepackage{units} 
\usepackage[pdftex,                
bookmarks=true,
bookmarksnumbered=true,
bookmarksopen=true, %
bookmarksopenlevel=2,%
hypertexnames=false,
breaklinks=true,
pdfpagemode=UseOutlines,
linkbordercolor={1 1 1}]{hyperref} 
\hypersetup{
pdfauthor   = {Victor Eugen \texorpdfstring{Ambru\c s}{Ambrus}},
pdftitle    = {Particle Production in a Roberston-Walker Space with a 
de Sitter Expansion Phase of Finite Extension},
pdfsubject  = {Master thesis},
pdfkeywords = {de Sitter, particle production, Bogoliubov coefficients},
}
\def\pprime{{\prime\prime}}
\def\Vec#1{\boldsymbol{#1}}

\newcommand{\partialdiv}[2]{\frac{\partial #1}{\partial #2}}

\newcommand{\abs}[1]{\left| #1 \right|}

\newcommand{\be}{ \begin{equation} }
\newcommand{\ee}{ \end{equation} }
\newcommand{\bea}{ \begin{eqnarray} }
\newcommand{\eea}{ \end{eqnarray} }
\newcommand{\ba}{ \begin{align} }
\newcommand{\ea}{ \end{align} }

\def\vx{\Vec{x}}
\def\vxp{\Vec{x}^\prime}

\newcommand{\vn}{\Vec{n}}

\def\vsigma{\boldsymbol\sigma}
\def\vp{{\Vec{p}}}
\def\vpp{{\Vec{p}^\prime}}

\def\vq{\Vec{q}}

\def\pout{p_f}
\def\vpin{\vp_i}
\def\vpout{\vp_f}
\def\vxin{\vx_i}
\def\vxout{\vx_f}
\def\dvx{\Vec{dx}}
\def\dvxin{\dvx_i}
\def\dvxout{\dvx_f}
\def\tin{t_i}
\def\tout{t_f}
\def\Min{\ensuremath{\text{M}_i}}
\def\Mout{\ensuremath{\text{M}_f}}
\def\fin{f^{\rm in}}
\def\fout{f^{\rm out}}
\def\fsin{f^{\textrm{in}*}}
\def\fsout{f^{\textrm{out}*}}

\def\foutp{\fout_{\vp}}

\def\Uin{U^{\rm in}}
\def\Uout{U^{\rm out}}
\def\Vin{V^{\rm in}}
\def\Vout{V^{\rm out}}

\def\ain{a_{\rm in}}
\def\aout{a_{\rm out}}
\def\adin{\ain^\dagger}
\def\adout{\aout^\dagger}
\def\bin{b_{\rm in}}
\def\bout{b_{\rm out}}
\def\bdin{\bin^\dagger}
\def\bdout{\bout^\dagger}
\def\din{d_{\rm in}}
\def\dout{d_{\rm out}}
\def\ddin{\din^\dagger}
\def\ddout{\dout^\dagger}

\def\halpha{{\hat{\alpha}}}
\def\hbeta{{\hat{\beta}}}
\def\hgamma{{\hat{\gamma}}}

\def\hrho{{\hat{\rho}}}

\def\ve1{\Vec{e}_1}
\def\ve2{\Vec{e}_2}
\def\ve3{\Vec{e}_3}

\def\snu{\sqrt{\mu}}

\newcommand{\DSol}[4]{#1_{#3,#4}^{\textrm{#2}}}
\def\lambdap{\lambda^{\prime}}

\def\llp{{\lambda\lambdap}}

\def\UdS{\DSol{U}{dS}{\vp}{\lambda}}
\def\VdS{\DSol{V}{dS}{\vp}{\lambda}}
\def\UdSp{\DSol{U}{dS}{\vpp}{\lambdap}}
\def\VdSp{\DSol{V}{dS}{\vpp}{\lambdap}}

\newcommand{\KGfSol}[2]{f_{#2}^{\textrm{#1}}}
\newcommand{\KGfsSol}[2]{f_{#2}^{\textrm{#1}\,*}}
\def\fdS{\KGfSol{dS}{\vp}}
\def\fsdS{\KGfsSol{dS}{\vp}}

\def\fdSout{\KGfSol{dS}{\vpout}}

\def\fsdSoutm{\KGfsSol{dS}{-\vpout}}

\def\spinor#1#2{\left(
\begin{matrix}
#1
\smallskip\\
#2
\end{matrix}
\right)}
\def\Hank#1#2{{H^{(#1)}_{#2}}}

\def\scprod#1#2{\mathinner{\langle{#1}, {#2}\rangle}}
\def\comm#1#2{\mathinner{\left[{#1}, {#2}\right]}}
\def\acomm#1#2{\mathinner{\left\{{#1}, {#2}\right\}}}
\def\ad#1{\ensuremath{\textrm{ad}_{#1}}}

\def\ztparant#1{\left(\frac{p}{\omega}e^{-\omega #1}\right)}
\def\pparant{\left(\frac{p}{\omega}\right)}
\def\qparant{\left(\frac{q}{\omega}\right)}

\def\qmpparant{\frac{q - p}{\omega}}
\def\mH{\mathcal{H}}
\def\bsq{\abs{\beta(p)}^2}

\def\cst#1{{\color{blue}{#1}}}
\def\cnd#1{{\color{red}{#1}}}
\def\crd#1{{\color{green}{#1}}}

\def\colthree#1#2#3{\cst{#1}, \cnd{#2}, \crd{#3}}

\title{Particle Production in a Roberston-Walker Space with a 
de Sitter Expansion Phase of Finite Extension}
\author{Victor Eugen Ambru\c s}
\date{}
\mmddyyyydate
\usepackage{chngcntr}
\counterwithin{equation}{section}

\usepackage{harvard} 

\def\imgpath{img}

\begin{document}

\begin{titlepage}
\begin{center}
\begin{flushleft}
{\bf \small West University of Timi\c soara\\
Department of Physics}
\end{flushleft}
\ \\[0.25\textheight]
\textsc{\Huge MASTER THESIS}\\[0.05cm]
\ \\[0.20\textheight]
\begin{minipage}{0.45\textwidth}
\begin{flushleft} \large
{\bf \small SUPERVISOR, \\
Lect.dr. NICOLAEVICI NISTOR}
\end{flushleft}
\end{minipage}
\begin{minipage}{0.45\textwidth}
\begin{flushright} \large
\textbf{\small AUTHOR,\\
AMBRU\c S VICTOR EUGEN}
\end{flushright}
\end{minipage}
\ \\[0.25\textheight]
{\bf \large Timi\c soara\\
2010}
\end{center}
\end{titlepage}
\thispagestyle{empty} 
\begin{titlepage}
\begin{center}
\begin{flushleft}
{\bf \small West University of Timi\c soara\\
Department of Physics}
\end{flushleft}
\ \\[0.15\textheight]

{\Huge\bf PARTICLE PRODUCTION IN \\A ROBERTSON-WALKER SPACE \\WITH A DE SITTER
PHASE\\OF FINITE EXTENSION\\}
\ \\[0.15\textheight]
\begin{minipage}{0.45\textwidth}
\begin{flushleft} \large
{\bf \small SUPERVISOR, \\
Lect.dr. NICOLAEVICI NISTOR}
\end{flushleft}
\end{minipage}
\begin{minipage}{0.45\textwidth}
\begin{flushright} \large
\textbf{\small AUTHOR,\\
AMBRU\c S VICTOR EUGEN}
\end{flushright}
\end{minipage}
\vfill
{\bf \large Timi\c soara\\
2010}
\end{center}
\end{titlepage}
\thispagestyle{empty} 
\chapter*{\centering \begin{normalsize}Abstract\end{normalsize}}
\begin{quotation}
\noindent
	We investigate the phenomenon of particle production in a
	Friedmann-Robertson-Walker universe which contains a phase of de Sitter 
	expansion for a finite interval, outside which it reduces to the flat
	Minkowski spacetime. We compute the particle number density for a massive 
	scalar and a spinorial field and point out differences between the two
	cases. We find that the resulting particle density approaches a constant 
	value at the scale of the Hubble time and that for a certain choice of 
	the parameters the spectrum is precisely thermal for the spinorial field,
	and almost thermal for the scalar field.
\end{quotation}
\clearpage
\tableofcontents
\cleardoublepage                
\pagenumbering{arabic}          
\chapter{Introduction}
The keywords describing this thesis are: quantum fields on curved spaces, 
de Sitter solutions, Bogoliubov transformation, particle production by the 
coupling with the gravitational field.

Quantum fields on curved spaces are the generalization of the Minkowski 
quantum theory of fields. The general approach is to consider the space 
curvature as a background field, described by the metric tensor, which 
obeys the classical (i.e. not quantum) Einstein field equations. This approach
suffers from a number of drawbacks, but it is nevertheless a bold step forward
towards the grand unification of all known interactions and the quantization
of the gravitational field. History tells us that a new theory is validated 
against results which are considered as classical. This pseudo-quantum 
treatment of quantum fields on a curved background provide a good way to 
produce some classical results.

Working in a background gravitational field is all but easy. With very few 
exceptions, there are no known analytical solutions to the resulting field
equations. One of these exceptions is the de Sitter spacetime, which describes
a Friedmann-Robertson-Walker Universe undergoing an exponential expansion
\cite{book:MTW,book:birrell_davies}.

Even though solutions to field equations might be derived in an external 
gravitational field, the quantum theory built on them is not as natural and 
intuitive as it is on Minkowski. For example, the Poincar\'e invariance of the 
Minkowski space and of the field equations automatically ensures the existence 
of ``positive'' and ``negative'' frequency states. However, a different choice 
of coordinate system (e.g. the spherical one), or the coordinates of a 
non-inertial observer (e.g. the Rindler coordinates) can also be used to define
particle states, which might not have the same physical meaning as the former.
The accelerated observer detects particles as if he would have been submerged
in a thermal bath of temperature related to his own acceleration 
\cite{book:birrell_davies}.

\begin{figure}[!ht]
   \centering
   \includegraphics[width=0.8\linewidth]{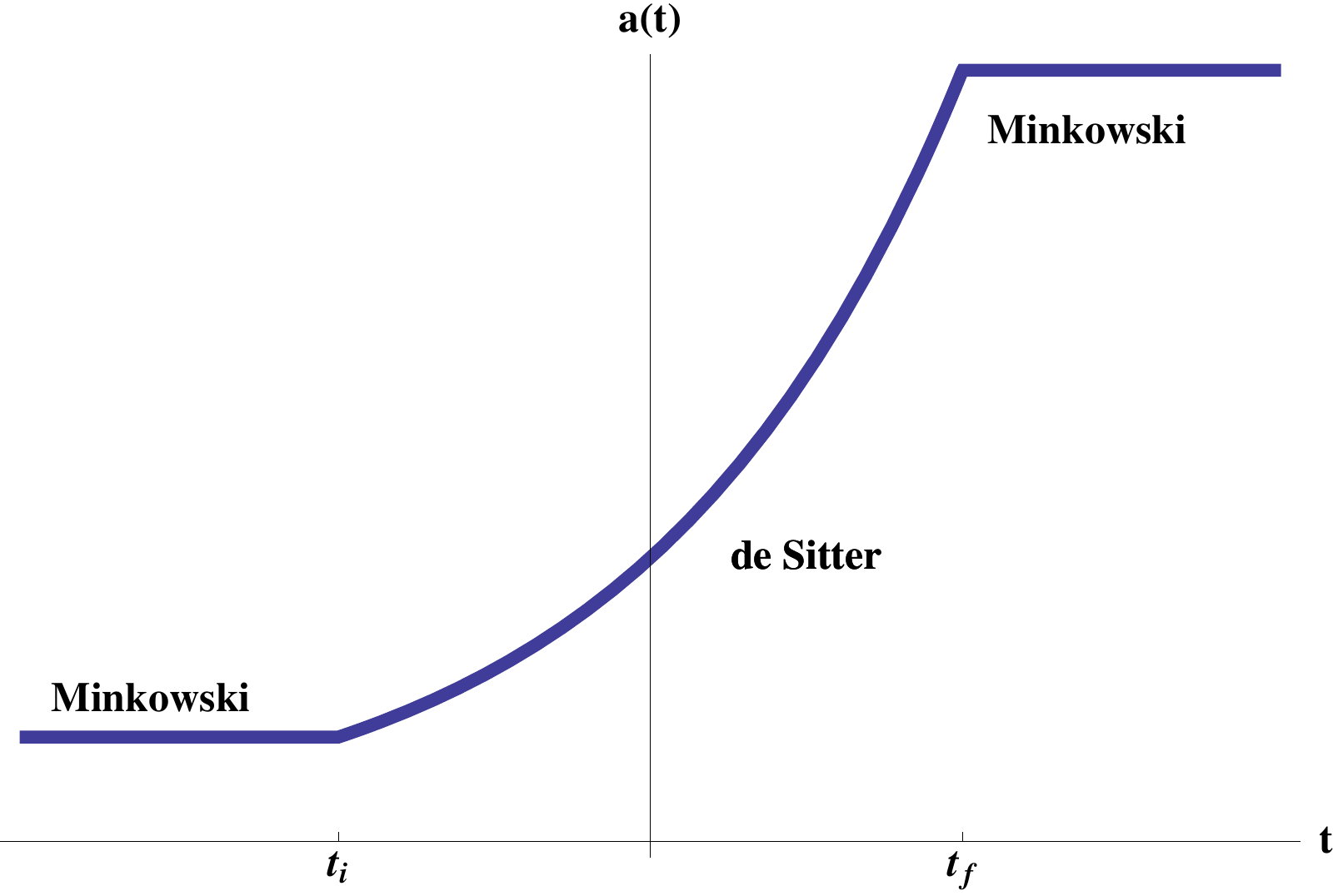}
   \caption{The scale factor for the FRW space under consideration}
   \label{fig:frw_space_minkdsmink}
\end{figure}

We shall circumvent the more philosophical issues regarding particle states
definition and interpretation, and instead consider the space to have only 
a finite region of a de Sitter expansion phase, outside which the space is 
flat (see \autoref{fig:frw_space_minkdsmink}).

Particle states are only defined on the flat regions of space. The method we 
employ is to let the {\em in} vacuum state evolve through the de Sitter phase,
and compute the expectation value of the particle number or the energy density 
operator in the {\em out} region. Poincar\'e invariance guarantees there is no
particle production on the Minkowski regions. This phenomenon takes place
only on the de Sitter phase. The particle and anti-particle modes used to 
construct the solution on de Sitter space are used merely as mathematical tools
that allow us to propagate the {\em in} modes, and do not receive any physical
interpretation as to their particle content. These key ingredients are 
summarized in \autoref{tab:minkdsmink_descr}.

The purpose of this thesis is to evaluate the density of created particles
with given momentum in a unit volume of the {\em out} space for a massive
scalar and a spinorial field. We find that there is a cutoff for particles
with momentum higher than the expansion factor. 
Using the spectral density we also evaluate
the particle number density (per unit volume), and the energy density.
We show that the particle number density approaches a constant value as
the expansion time approaches the Hubble time, and it increases with 
the expansion factor and with the mass of the created particles. 
The energy density is finite only for the case of a conformally coupled 
massive scalar field, in all other cases (including the spinorial field), 
it has a logarithmic divergence. It also increases with the expansion factor,
and exhibits a higher order increase with respect to the particle mass.

The thesis is structured as follows: \autoref{chap:qfcs} presents the 
spacetime under investigation, after which we recall the basics of the 
quantization procedure on curved background and the formalism for the 
description of the particle production phenomenon. 
A key ingredient in the calculation are the free field equation solutions
(the quantum modes), which we present in \autoref{chap:dSsol}.

Our main result is contained in \autoref{chap:kg} and \autoref{chap:dirac}, 
where we explicitly obtain the expression for the number density of the 
created particles. These chapters follow a common pattern for the
investigation of scalar and spinorial particle production respectively.
Each consists of two sections, the first gives the analytical
solution in terms of Hankel functions while the second applies approximation 
formulas to extract information on the particle production phenomenon. The 
results obtained are accompanied by a set of figures.

We summarize our results and present our conclusions in \autoref{chap:conc}, 
where we also point out possibilities for further development.

We have provided \autoref{app:hankel} as a small reference regarding Hankel
functions. \autoref{chap:pauli_spinors} gives insight on the underworks of the
Pauli spinors which occur in the polarized solutions of the Dirac equation.

\chapter{Quantum fields on curved spaces}\label{chap:qfcs}
This section summarizes the framework and main results needed for the
development of the thesis. We assume the reader to be familiar with the
quantum theory of free fields and general relativity (good books include
\cite{book:itzykson_zuber,book:bjorken_drell,book:MTW,book:wald}).
In \autoref{sec:frw} we present the space-time under consideration and list 
the results from general relativity which we shall use in subsequent 
chapters. In \autoref{sec:qfcs} we introduce the general
formalism for the construction of a quantum field theory on an arbitrary
space-time, following the method of \cite{book:birrell_davies}. Other
introductory texts include \cite{ebook:ford,ebook:jacobson}. For an
introduction to the vierbein formalism, used for the generalization of the
Dirac field to curved spaces, the reader can consult
\cite{art:cota_external_symm,art:cota_polarized_fermions}.
For a modern analysis on the construction of a meaningful symmetric
divergenceless stress-energy tensor, we refer to \cite{art:forger_romer}.
\begin{section}{Friedmann-Robertson-Walker spaces}\label{sec:frw}
The Friedmann-Robertson-Walker space is a spatially isotropic and homogeneous
spacetime described by the metric
\begin{equation}
	ds^2 = dt^2 - a^2(t) \Vec{dx}^2,\label{eq:ds_frw}
\end{equation}
where $a(t)$ is known as the scale factor.
Such a space undergous a dilation (or contraction) of distances by the factor
$a(t)$. Two special cases of the FRW space are the Minkowski space, which 
corresponds to 
\begin{equation}
	a(t) = \text{const}, \label{eq:frw_mink_def}
\end{equation}
and the de Sitter space, which corresponds to
\begin{equation}
	a(t) = e^{\omega t}. \label{eq:frw_dS_def}
\end{equation}
The expansion parameter $\omega$ is also known as the Hubble expansion rate 
$H$, and the hubble time is defined as $1/H$.

The space we will work with is a FRW space consisting of three regions, as 
illustrated in \autoref{fig:frw_space_minkdsmink}. The continuity of the metric
tensor requests $a(t)$ to be a continuous function of $t$, which leads to
a line element of the following form:
\begin{subequations}\label{eq:ds}
\begin{enumerate}[(1)]
	\item Minkowski {\em in} region, $t < \tin$:
\begin{equation}
	ds^2 = dt^2 - \dvxin \quad,\quad \vxin = e^{\omega \tin} \vx
	\label{eq:ds_mink_in}
\end{equation} 
	\item de Sitter expansion phase, $\tin < t < \tout$:
\begin{equation}
	ds^2 = dt^2 - \dvx^2
	\label{eq:ds_dS}
\end{equation} 
	\item Minkowski {\em in} region, $t > \tout$:
\begin{equation}
	ds^2 = dt^2 - \dvxout \quad,\quad \vxout = e^{\omega \tout} \vx
	\label{eq:ds_mink_out}
\end{equation}
\end{enumerate}
\end{subequations}
We shall refer to $\tin$ as the ``initial time'', to $\tout$ as the
``final time'' and to $\Delta t = \tout - \tin$ as the ``expansion time''.
Note that the effect of the expansion of space is to increase the physical 
distances, which in turn produces a redshift in particle wavelengths. We shall
use ``physical quantities'' (e.g. physical length, physical momentum) to refer
to the results of measurements performed by an observer in a Minkowski region.
For example, the momentum operator for the {\em out} region is 
\begin{equation}
	\Vec{P}^{\text{out}} = -i \nabla_f = -i e^{-\omega \tout} \nabla.
	\label{eq:Pout_def}
\end{equation}
This is the natural definition of the momentum operator associated with the
Killing vector of unit length which generates space translations. The length
of the Killing vector is evaluated using the Minkowski metric 
\begin{equation}
	\eta_{\mu\nu} = \text{diag}(1,-1,-1,-1), \label{eq:etamunu}
\end{equation}
used for the construction of the line element \eqref{eq:ds_mink_out}. 
Similarly, the momentum operator for the {\em in} region is naturally defined
as
\begin{equation*}
	\Vec{P}^{\text{in}} = -i e^{-\omega \tin} \nabla.
\end{equation*}
This establishes the relation between the momentum operators measuring 
physical momenta in the {\em in} and {\em out} regions:
\begin{equation}
	\Vec{P}^{\text{out}} = e^{-\omega(\tout - \tin)} \Vec{P}^{\text{in}}.
	\label{eq:Pout_Pin}
\end{equation}
Particles which have a measured momentum of $p$ in the {\em out} region had
a measured momentum of 
\begin{equation}
	\vq = \vp e^{\omega \Delta t}
	\label{eq:q_def}
\end{equation}
in the {\em in} region. In terms of the de Sitter momentum $\vp_{\text{dS}}$, 
given by the Killing vector $-i \partial_i$, we have
\begin{equation}
	\vq = \vp_{\text{dS}} e^{-\omega \tin} ,\qquad 
	\vp = \vp_{\text{dS}} e^{-\omega \tout}.
	\label{eq:qp_pdS}
\end{equation}
\begin{table}\renewcommand{\arraystretch}{1.5}\addtolength{\tabcolsep}{1pt}
\centering
\begin{tabular}{r|c|c|c|}
	& {\em in} & de Sitter phase & {\em out}\\\hline
	time span & $t < \tin$ & $\tin < t < \tout$ & $\tout < t$\\
	$ds^2$ & $dt^2 - e^{2\omega \tin} \Vec{dx}^2$ & 
	$dt^2 - e^{2\omega t} \Vec{dx}^2$ & $dt^2 - e^{2\omega \tout} \Vec{dx}^2$\\	
	physical coordinates & $\vxin = \vx e^{\omega \tin}$ & $\vx e^{\omega t}$ &
	$\vxout = \vx e^{\omega \tout}$\\
	physical momenta & $q = p_{\text{dS}} e^{-\omega \tin}$ & 
	$p_{\text{dS}} e^{-\omega t}$ & 
	$p = p_{\text{dS}} e^{-\omega \tout}$\\\hline
	%
	%
\end{tabular}
\caption{summary of the studied space-time}
\label{tab:minkdsmink_descr}
\end{table}
An important feature of FRW spaces is that the metric is conformal with
the Minkowski metric, allowing to be cast in the form
\begin{equation}
	ds^2 = a^2(\eta)(d\eta^2 - \Vec{dx}^2) ,\qquad
	\eta = \int \frac{dt}{a(t)},
	\label{eq:ds_conformal}
\end{equation}
where $\eta$ is called the conformal time. We recall that a conformal 
transformation may be described by
\begin{equation}
	g_{\mu\nu} \mapsto \overline{g}_{\mu\nu} = \Omega^2(x) g_{\mu\nu}.
	\label{eq:conf_transf}
\end{equation}
As the metric undergoes this transformation, all metric-dependent quantities
(the connection coefficients, Riemann tensor, curvature) transform in a 
non-trivial way. In particular, the transformation law for the 
four-dimensional d'Alembert operator is:
\begin{equation}
	\left( \Box + \frac{1}{6} R \right) \phi \mapsto 
	\left( \overline{\Box} + \frac{1}{6} \overline{R} \right) \overline{\phi}
	\quad,\quad
	\overline{\phi}(x) = \frac{1}{\Omega} \phi(x).
	\label{eq:kg_conformal_transf}
\end{equation}
All barred quantities are evaluated using the conformally transformed
metric \eqref{eq:conf_transf}. The generally covariant d'Alembert 
operator on a curved space is 
\begin{equation}
	\Box \phi = g^{\mu\nu} \nabla_{\mu} \nabla_{\nu} \phi = \frac{1}{\sqrt{-g}} 
	\partial_\mu\left(\sqrt{-g} g^{\mu\nu} \partial_\nu \phi\right),\qquad
	g = \det(g_{\mu\nu}).
	\label{eq:dalembert}
\end{equation}
This result is of special interest for the theory of scalar fields on curved 
spaces, described by the lagrangian \eqref{eq:kg_lagrangian}. If
the coupling parameter is set to $\xi = 1/6$ (conformal coupling), the 
massless scalar field obeys a conformal field equation, and thus the solutions 
are proportional to the Minkowski ones.

The connection coefficients in a coordinate basis $\partial_\mu$, also known
as Christofell symbols, are defined as 
\begin{equation}
	\Gamma\indices{^\lambda_\mu_\nu} = \frac{1}{2} g^{\lambda\kappa}
	\left(g_{\kappa\mu,\nu} + g_{\kappa\nu,\mu} - g_{\mu\nu,\kappa}\right).
	\label{eq:christofell}
\end{equation}
We shall use these in the construction of the covariant derivative appearing
in the d'Alembert operator \eqref{eq:dalembert}. In a FRW space of line 
element \eqref{eq:ds_frw} the Christofell symbols are
\begin{equation}
	\Gamma\indices{^t_i_j} = a(t) a^\prime(t) \delta_{ij} ,\qquad
	\Gamma\indices{^i_t_j} = \frac{a^\prime(t)}{a(t)} \delta\indices{^i_j}.
	\label{eq:gamma_frw}
\end{equation}
The prime denotes differentiation with respect to the argument. In 
the conformal chart $(\eta, \vx)$, which we shall use in \autoref{chap:dSsol}, 
these coefficients read
\begin{equation}
	\Gamma\indices{^\eta_\eta_\eta} = \frac{a^\prime(\eta)}{a(\eta)},\qquad 
	\Gamma\indices{^\eta_i_j} = \frac{a^\prime(\eta)}{a(\eta)} \delta_{ij},\qquad
	\Gamma\indices{^i_\eta_j} = \frac{a^\prime(\eta)}{a(\eta)} \delta\indices{^i_j}.
	\label{eq:gamma_conformal}
\end{equation}
The Ricci scalar (the curvature) is given by
\begin{equation}
	R = R\indices{^\nu_\nu} = R\indices{^\mu^\nu_\mu_\nu} = 
	- 6\frac{a^\pprime}{a} - 6 \left(\frac{a^\prime}{a}\right)^2.
	\label{eq:ricci_frw}
\end{equation}
%
%
The Minkowski metric $\eta_{\mu\nu}$ \eqref{eq:etamunu} is point-independent
($a(t) = \text{const}$), thus the Christofell symbols 
\eqref{eq:gamma_frw} and the Ricci curvature \eqref{eq:ricci_frw}
vanish. In particular, the d'Alembert operator \eqref{eq:dalembert}
is the wave operator
\begin{equation}
	\Box = \partial_t^2 - \Delta ,\qquad \Delta = \sum_{i=1}^3 \partial^2_i,
	\label{eq:dalembert_mink}
\end{equation}
with $\Delta$ being the Laplace operator, acting on all space components.
Care must be taken when reading the space coordinates of the 
Minkowski {\em in} and {\em out} regions, among which we shall distinguish
by appending a subscript ${}_i$ or ${}_f$.

Let us now focus on our case of interest, i.e. that of the de Sitter space 
\eqref{eq:ds_dS} with the scale factor $a(t) = e^{\omega t}$. 
The conformal time \eqref{eq:ds_conformal} is easily integrated to yield
\begin{subequations}
\begin{equation}
	\eta = -\frac{1}{\omega} e^{-\omega t},
	\label{eq:conformal_time_dS}
\end{equation}
therefore $a(\eta) = -1/\omega\eta$. Substituting in \eqref{eq:ds_conformal}, 
the conformal line element reads
\begin{equation}
	ds^2 = \frac{1}{\omega^2 \eta^2} (d\eta^2 - \Vec{dx}^2).
\end{equation}
\end{subequations}
The connection coefficients in this chart are
\begin{equation}
	\Gamma\indices{^\eta_\eta_\eta} = -\frac{1}{\eta} ,\qquad 
	\Gamma\indices{^\eta_i_j} = -\frac{1}{\eta} \delta_{ij} ,\qquad
	\Gamma\indices{^i_\eta_j} = -\frac{1}{\eta} \delta\indices{^i_j},
	\label{eq:gamma_conformal_dS}
\end{equation}
and the d'Alembert operator is
\begin{equation}
	\Box \phi = (\omega^2 \eta^2) 
	\left(\partial_\eta^2 - 2/\eta \partial_\eta - \Delta\right) \phi.
	\label{eq:dalembert_dS}
\end{equation}
The Ricci scalar reads
\begin{equation}
	R = 12 \omega^2 \label{eq:ricci_dS}.
\end{equation}
\end{section}
\begin{section}{Quantization procedure}\label{sec:qfcs}
One of the main ingredients in the construction of the quantum field theory
in Minkowski spoace is the requirement of Poincar\'e invariance. 
In particular, the invariance to time translations allow for the construction
of positive and negative frequency modes, which naturally define particle and
anti-particle states. This invariance is not guaranteed in general 
relativity, where the choice of particle states is echivocal. Poincar\'e
inertial observers all agree on the definitions of particles. On the other
hand, only a special class of freely falling observers will register the same
particle content in a given quantum states.

Following the development in \cite{book:birrell_davies}, we shall restrict 
ourselves to $C^{\infty}$ 4-dimensional, globally hyperbolic, pseudo-Riemannian
manifolds. The differentiability conditions ensure the existence of 
differential equations and the global hyperbolicity ensures the existence of
Cauchy hypersurfaces.

The formalism of quantum field theory is generalized to curved spacetime in a 
straightforward way, but the physical interpretation is not. Except 
for some cases such as static spacetimes, the physical interpretation of 
particle states is ambiguous.

In this section we shall outline the general framework of a quantum theory of
fields in a background gravitational field, and apply it to general FRW spaces,
discussed in \autoref{sec:frw}.
\begin{subsection}{The Klein-Gordon field}\label{sec:qfcs_kg}
	The generally covariant action for a charged scalar field is
	\begin{subequations}
	\begin{equation}
		\mathcal{S}[\phi] = \int d^4x \sqrt{-g} \left\{ g^{\mu\nu}
		\nabla_\mu \phi^{\dagger}(x) \nabla_\nu \phi(x) - 
		(m^2 + \xi R(x)) \phi^{\dagger}(x) \phi(x) \right\},
		\label{eq:chkg_action}
	\end{equation}	
	 and the corresponding field equation is
	\begin{equation}		
		(\Box + m^2 + \xi R(x)) \phi(x) = 0 \quad,\quad
		\Box \phi(x) = g^{\mu\nu}\nabla_\mu \nabla_\nu \phi(x)
		\label{eq:kg_fieldeq}
	\end{equation}
	\end{subequations}
	The most frequently considered values for the coupling parameter are 
	$\xi = 0$ (minimal coupling) and $\xi=\nicefrac{1}{6}$ (conformal coupling).
	Although the natural coupling to the gravitational field is the minimal one,
	we shall keep $\xi$ arbitrary when developing the theory. As we shall see,
	the conformal coupling will emerge as a case of special interest, not only 
	because for this choice the particle production ceasses when the scalar field
	is massless (see \autoref{sec:kg_massless}), but also because the energy
	density of the created particles is at a minimum 
	(see, for example, \autoref{sec:kg_asympt_large}).
	
	The $\mathcal{U}(1)$ invariance of the action assures the conserved current 
	density
	\begin{equation}
		\mathfrak{j}^\mu = i \sqrt{-g}\, g^{\mu\nu} (\phi^{\dagger}(x) 
		\overleftrightarrow{\partial_{\nu}} \phi(x)) \quad,\quad
		\partial_\mu \mathfrak{j}^{\mu} = 0,
		\label{eq:kg_current_density}
	\end{equation}
	and the corresponding current vector
	\begin{equation}
		j^\mu = i g^{\mu\nu} (\phi^{\dagger}(x) 
		\overleftrightarrow{\partial_{\nu}} \phi(x)) \quad,\quad
		\nabla_\mu j^{\mu} = 0,
	\end{equation}
	where the bilateral derivative is
	\begin{equation}
		f\overleftrightarrow{\partial_\mu} g = f \partial_\mu g - g \partial_\mu f.
		\label{eq:bilaterald}
	\end{equation} 
	The conserved charge follows from \eqref{eq:kg_current_density}
	\begin{equation}
		Q = \int j^\mu d\Sigma_\mu ,
		\label{eq:kg_charge_general}
	\end{equation}
	where the integral extends over a Cauchy surface. The result is independent 
	of the choice of surface. If we pick $x^0$ to be the temporal coordinate, 
	the integral \eqref{eq:kg_charge_general} reads
	\begin{equation}
		Q = i \int d^3x \sqrt{-g}\, g^{\mu 0} (\phi^{\dagger}(x) 
		\overleftrightarrow{\partial_{\mu}} \phi(x)),\qquad
		\partial_0 Q = 0.
	\end{equation}
	This suggests to introduce the scalar product 
	\begin{equation}
		\scprod{f_1}{f_2} = i \int d^3x \sqrt{-g} g^{\mu 0} (f_1^{\dagger}(x) 
		\overleftrightarrow{\partial_{\mu}} f_2(x)) ,\qquad
		\partial_0 \scprod{f_1}{f_2} = 0,
		\label{eq:kg_scprod}
	\end{equation}
	which can be easily shown to be independent of time if and only if 
	$f_1(x), f_2(x)$ are solutions to the Klein-Gordon equation 
	\cite{ebook:ford}.
	In the discussion above we referred to a charged scalar field only for
	establishing the form of the scalar product \eqref{eq:kg_scprod}. In our 
	investigation we shall restrict to the case of an uncharged field 
	(the difference is unessential) for which the Lagrangian density reads
	\begin{equation}
		\mathcal{L}(\phi, \partial_\mu \phi) = \sqrt{-g} \left\{g^{\mu\nu}
		\frac{1}{2} \nabla_\mu \phi(x) \nabla_\nu \phi(x) - 
		\frac{1}{2} (m^2 + \xi R(x)) \phi^2(x)\right\}.
		\label{eq:kg_lagrangian}
	\end{equation}
	The resulting field equation is identical to \eqref{eq:kg_fieldeq}, for 
	which the same scalar product \eqref{eq:kg_scprod} can be introduced.
	
	The conjugate momentum is defined as the derivative of the Lagrangian density
	\eqref{eq:kg_lagrangian} with respect to the time derivative of the field
	\begin{equation}
		\pi(x) = \partialdiv{\mathcal{L}}{\partial_0 \phi(x)} = 
		\sqrt{-g} g^{\mu 0} \nabla_\mu \phi(x).
		\label{eq:kg_pidef}
	\end{equation}
	The quantization of the field is achieved by imposing the canonical 
	commutation rules:
	\begin{subequations}\label{eq:kg_canonicalcomm}
	\begin{gather}
		\comm{\phi(t, \vx)}{\phi(t, \vxp)} = \comm{\pi(t, \vx)}{\pi(t, \vxp)} = 0\\
		\comm{\phi(t, \vx)}{\pi(t, \vxp)} = i \delta^3(\vx - \vxp)
	\end{gather}
	\end{subequations}
	The field equation is linear, and therefore admits a complete set of 
	solutions, with respect to which the field $\phi(x)$ can be expanded:
	\begin{equation}
		\phi(x) = \sum_{i} \left(a_i f_{i}(x) + a_i^\dagger f^*_i(x)\right)
		\label{eq:kg_phi_modes}	
	\end{equation}
	Since we have a well defined scalar product, we require these modes 
	to satisfy the orthonormalization condition
	\begin{equation}
		\scprod{f_i}{f_j} = \delta_{ij}, \qquad \scprod{f_i}{f_j^*} = 0
		,\qquad \scprod{f_i^*}{f_j^*} = -\delta_{ij}.
		\label{eq:kg_orthonormij}
	\end{equation}
	Using the canonical commutation rules \eqref{eq:kg_canonicalcomm} and 
	the completeness relation
	\begin{equation}
		i \sum_i \sqrt{-g} g^{\mu 0} \left( f_i^*(t, \vxp) \partial_\mu f_i(t, \vx)
		- f_i(x)\partial_\mu f^*_i(t, \vxp)\right) = \delta^3(\vx - \vxp)
		\label{eq:kg_completeij}
	\end{equation}
	we arrive at
	\begin{equation}
		\comm{a^{\phantom{\dagger}}_i}{a^{\phantom{\dagger}}_j} = 0,\qquad
		\comm{a^\dagger_i}{a^\dagger_j} = 0, \qquad
		\comm{a_i}{a^\dagger_j} = \delta_{ij}.
		\label{eq:kg_commij}
	\end{equation}
	These operators can be obtained by taking
	the scalar product between the corresponding mode and the field operator
	\begin{equation}
		a_i = \scprod{f_i}{\phi} ,\qquad 
		a^\dagger_j = -\scprod{f_j^*}{\phi}.
		\label{eq:kg_aoperator_scprod}
	\end{equation}  
	The Fock space can be constructed by defining a vacuum (or no-particle) 
	state such that
	\begin{equation}
		a_i \ket{0} = 0 ,\qquad	\braket{0|0} = 1.
		\label{eq:kg_vacuum_def}
	\end{equation}
	With respect to this state, we call $a_i$ annihilation operators and
	$a^\dagger_j$ creation operators. The former operators annihilate quanta
	in mode $f_i$, while the latter creates quanta in mode $f_j$. From the
	vacuum state, succesive application of the creation operators $a^\dagger_j$
	generates the particle states. The operator that counts the number
	of particles in a given state is the particle number operator given by
	\begin{equation}
		N_i = a^\dagger_i a_i.
		\label{eq:kg_pnumberi}
	\end{equation}
	Finally, the symmetric, divergence-less stress-energy tensor is obtained 
	with the general prescription
	\begin{equation}
		T_{\mu\nu} = \frac{2}{\sqrt{-g}} 
		\frac{\delta \mathcal{L}}{\delta g^{\mu\nu}}.
		\label{eq:kg_Tmunu_def}
	\end{equation}
	The Lagrangian density \eqref{eq:kg_lagrangian} depends on the metric through 
	$\sqrt{-g}$ and through the derivative term. The variation of the latter 
	with respect to the metric is $\sim \nabla_\mu \phi \nabla_\nu \phi$,
	while the former's variation is given by
	\begin{equation}
		\frac{\delta\sqrt{-g}}{\delta g^{\mu\nu}} = 
		-g_{\mu\nu} \frac{\sqrt{-g}}{2},
		\label{eq:detg_variation}
	\end{equation}
	and thus we arrive at
	\begin{gather}
		T_{\mu\nu} = \nabla_\mu \phi \nabla_\nu \phi - 
		g_{\mu\nu} \frac{\mathcal{L}}{\sqrt{-g}},
		\label{eq:kg_Tmunu}\\
		\frac{\mathcal{L}}{\sqrt{-g}} = \frac{1}{2} \left(g^{\mu\nu}
		\nabla_\mu \phi(x) \nabla_\nu \phi(x) - 
		(m^2 + \xi R(x)) \phi^2(x)\right).\nonumber
	\end{gather}
	There is a problem with this tensor: the energy density of the vacuum state is 
	infinite. However, we are only interested in energy differences, therefore
	we shall substract the vacuum expectation value and define the new tensor 
	as the normally ordered (Wick ordered) stress-energy tensor:
	\begin{equation}
		:T_{\mu\nu}: = T_{\mu\nu} - \braket{0|T_{\mu\nu}|0}.
		\label{eq:kg_Tmunu_normal}
	\end{equation}
	In terms of creation and annihilation operators we have
	\begin{multline}
		:T_{\mu\nu}: = \sum_{i, j} \left\{ 
		a_i a_j (\partial_\mu f_i \partial_\nu f_j - g_{\mu\nu} \mathcal{L}(f_i, f_j))
		+ a^\dagger_i a^\dagger_j (\partial_\mu f^*_i \partial_\nu f^*_j - 
		g_{\mu\nu} \mathcal{L}(f^*_i, f^*_j)) \right.\\
		\left. + a^\dagger_j a_i (\partial_\mu f_i \partial_\nu f^*_j
		+\partial_\nu f_i \partial_\mu f^*_j  - g_{\mu\nu} \mathcal{L}(f_i, f^*_j)) \right\},
		\label{eq:kg_Tmunu_aij}
	\end{multline}
	where $\mathcal{L}(f_i, f_j)$ is the bilinear form
	\begin{equation}
		\mathcal{L}(f_i, f_j) = \frac{1}{2}\left(g^{\mu\nu} \partial_\mu f_i 
		\partial_\nu f_j - (m^2 + \xi R(x)) f_i f_j\right).
		\label{eq:kg_Lij}
	\end{equation}
\end{subsection}
\begin{subsection}{The Dirac field}\label{sec:qfcs_dirac}
	Fields with non-zero spin require multi-component wavefunctions
	to describe the extra degrees of freedom. In Minkowski space-time these
	components correspond naturally to the cartesian coordinate system. In
	curved spaces, things tend to become ambiguous because of the requirement of
	invariance with respect to general coordinate transformations. 
	Further, because the metric is not
	necessarily homogeneous, there must be a mechanism that decouples the 
	field and the equation it obeys from the specific choice of coordinates.
	This is achieved by working in natural frames, described by the frame 
	vectors $\{e_{\halpha}\}$ and the coframe 1-forms
	$\{\omega^{\hbeta}\}$. The hatted indices refer to components with
	respect to this basis, while unhatted ones refer to components in an 
	holonomic reference frame, e.g. $e_{\halpha}=e_{\halpha}^{\mu}\partial_\mu$,
	$\omega^{\halpha} = \omega^{\halpha}_\mu dx^{\mu}$.
	The coframe 1-forms are defined such that
	\begin{subequations}
	\begin{equation}
		g_{\mu\nu} = \eta_{\halpha\hbeta}\omega^{\halpha}_{\mu} \omega^{\hbeta}_{\nu} 
		\quad,\quad
		g = \eta_{\halpha\hbeta} \omega^{\halpha} \otimes \omega^{\hbeta},
		\label{eq:coframe_def}
	\end{equation}
	where $\eta_{\halpha\hbeta} = {\rm diag}(1,-1,-1,-1)$ is the Minkowski 
	metric. The corresponding frame vectors are chosen such that 
	\begin{equation}
		\omega^{\halpha}_{\mu} e^{\mu}_{\hbeta} = \delta\indices{^{\halpha}_{\hbeta}}		
		,\qquad
		e^{\mu}_{\hbeta} \omega^{\hbeta}_{\nu} = \delta\indices{^\mu_\nu}.
		\label{eq:frame_coframe_delta}
	\end{equation}
	The frame plays the same r\^ole with respect to the inverse of the metric
	tensor $g^{-1}$ as the coframe plays with respect $g$:
	\begin{equation}
		g^{\mu\nu} = \eta^{\halpha\hbeta}e^{\mu}_{\halpha}e^{\nu}_{\hbeta}.
		\label{eq:frame_def}
	\end{equation}
	Any vector $x = x^\mu \partial_\mu$ can be written in terms of the 
	tetrad frame vectors as $x = x^{\halpha}e_{\halpha}$, with the components
	given by:
	\begin{equation}
		x^{\halpha} = x^\mu \omega^{\halpha}_{\mu}.\label{eq:vec_frame_comp}
	\end{equation}
	\end{subequations}
	The tetrad description has the advantage that the hatted components of
	vectors do not change on a change of coordinates. The Lagrangian density 
	for the Dirac field is
	\begin{subequations}
	\begin{equation}
		\mathcal{L}(x) = \sqrt{-g} \left\{
		\frac{i}{2} \left(\overline{\psi} \gamma^{\halpha} D_{\halpha} \psi - 
		\overline{D_{\halpha} \psi} \gamma^{\halpha} \psi \right)- 
		m \overline{\psi} \psi\right\},\label{eq:dirac_lagrangian}
	\end{equation}
	and the corresponding field equation is
  \begin{equation}
		(i \gamma^{\halpha} D_\halpha - m) \psi(x) = 0 .
		\label{eq:dirac_fieldeq}		
  \end{equation}
	\end{subequations}
	The vierbein formalism comes into play when defining the $\gamma$ matrices.
	Similar to the Minkowski theory, these are matrices with the anticommuting
	property
	\begin{equation}
		\acomm{\gamma^{\halpha}}{\gamma^{\hbeta}} = 2\eta^{\halpha\hbeta}.
		\label{eq:dirac_gamma_acomm}
	\end{equation}
	The coordinate dependence is contained in the covariant derivative 
	\begin{equation}
		D_{\halpha} = e^{\mu}_\halpha \partial_\mu + \frac{1}{2}
		\Gamma_{\hbeta\hgamma\halpha} D(\Sigma^{\hbeta\hgamma}).
		\label{eq:dirac_Dalpha}
	\end{equation}
	The connection coefficients with hatted indices are the equivalent of the 
	Christofell symbol \eqref{eq:christofell}, considered in a natural frame,
	and are defined as
	\begin{equation}
		\nabla_{\hbeta} e_{\halpha} = \Gamma\indices{^\hgamma_\halpha_\hbeta} e_{\hgamma} 
		,\qquad
		\Gamma_{\halpha\hbeta\hgamma} = \frac{1}{2}\left(c_{\halpha\hbeta\hgamma} + 
		c_{\halpha\hgamma\hbeta} - c_{\hbeta\hgamma\halpha}\right).
		\label{eq:Gamma_natural_def}
	\end{equation}
	These coefficients can be computed using the Cartan coefficients 
	$c_{\halpha\hbeta\hgamma}$ defined as	
	\begin{equation}
		\comm{e_\halpha}{e_\hbeta} = c\indices{_\halpha_\hbeta^\hgamma} e_{\hgamma}
		,\qquad
		\comm{e_\halpha}{e_\hbeta}^\mu = e^\nu_\halpha\partial_\nu e^\mu_\hbeta -
		e^\nu_\hbeta\partial_\nu e^\mu_\halpha.
		\label{eq:cartan_def}
	\end{equation}
	The antiadjoint generators for the spinorial representation of the 
	Lorentz transformation are
	\begin{subequations}	
	\begin{equation}
		D(\Sigma_{\halpha\hbeta}) = \frac{1}{4} \comm{\gamma_{\halpha}}{\gamma_{\hbeta}},
		\label{eq:dirac_Sigma_def}
	\end{equation}
	corresponding to the anti-hermitian generator of the definition 
	representation of the Lorentz group
	\begin{equation}
		\left(\Sigma_{\halpha\hbeta}\right)\indices{^\hgamma_\hrho} = 
		\delta\indices{^\hgamma_\halpha} \eta_{\hbeta\hrho} - 
		\delta\indices{^\hgamma_\hbeta} \eta_{\halpha\hrho},
		\label{eq:lorentz_Sigma_comp}		
	\end{equation}
	which obey the commutation rules
	\begin{equation}
		\comm{D(\Sigma_{\halpha\hbeta})}{D(\Sigma_{\hgamma\hrho})} = 
		D(\Sigma_{\halpha\hgamma})\eta_{\hbeta\hrho} - 
		D(\Sigma_{\halpha\hrho})\eta_{\hbeta\hgamma} -
		D(\Sigma_{\hbeta\hgamma})\eta_{\halpha\hrho} + 
		D(\Sigma_{\hbeta\hrho})\eta_{\halpha\hgamma}.
		\label{eq:lorentz_Sigma_commutation}
	\end{equation}	
	\end{subequations}	
	The covariant derivative \eqref{eq:dirac_Dalpha} ensures the covariance of
	the lagrangian density and of the field equation on an arbitrary 
	change of coordinates and of the tetrad vectors. Similar to the scalar case,
	there is a $\mathcal{U}(1)$ symmetry of the Lagrangian 
	\eqref{eq:dirac_lagrangian} which assures the conserved current vector
	\begin{equation}
		j^{\mu}(x) = e^{\mu}_{\halpha} \overline{\psi}(x) \gamma^{\halpha} \psi(x) 
		,\qquad \nabla_{\mu} j^{\mu} = 0.
		\label{eq:dirac_conserved_current}
	\end{equation}	
	The time-independent charge associated with this current is
	\begin{equation}
		Q = \int d^3x \sqrt{-g} e^{0}_{\halpha} \overline{\psi}(x) \gamma^{\halpha} \psi(x)
		,\qquad \partial_0 Q = 0.
		\label{eq:dirac_conserved_charge}
	\end{equation}
	The scalar product
	\begin{equation}
		\scprod{\psi}{\chi} = \int d^3x \sqrt{-g} e^{0}_{\halpha} \overline{\psi}(x) \gamma^{\halpha} \chi(x)
		\quad,\quad \partial_0 \scprod{\psi}{\chi} = 0
		\label{eq:dirac_scprod}
	\end{equation}
	is well defined since it is time-independent. The choice of conjugate 
	momenta corresponding to the field components $\psi_a$ (latin indices 
	from the beginning of the alphabet will label spinorial indices) is not 
	straightforward because the lagrangian density is 0 when the fields obey
	the field equations. The traditional choice for the momenta corresponding to
	the field $\psi$ (written in spinorial form) is
	\begin{equation}
		\pi(x) = \sqrt{-g} e^0_{\halpha} \overline{\psi} \gamma^{\halpha}
		\label{eq:dirac_pidef}
	\end{equation}
	The quantization for half-integer spin fields (fermions) is performed by 
	imposing the anti-commutation rules
	\begin{subequations}
	\begin{gather}
		\acomm{\psi_a(t,\vx)}{\psi_b(t,\vxp)} = \acomm{\pi_a(t,\vx)}{\pi_b(t,\vxp)}
		= 0 \\
		\acomm{\psi_a(t,\vx)}{\pi_b(t, \vxp)} = \delta_{ab} \delta^3(\vx - \vxp)
		\label{eq:dirac_cannonicalacomm}
	\end{gather}
	\end{subequations}
	The Dirac field equation \eqref{eq:dirac_fieldeq} is linear, therefore we 
	can expand the field operator $\psi(x)$ in terms of a complete set of
	solutions (modes):
	\begin{equation}
		\psi(x) = \sum_i \left(b_i u_i + d_i^\dagger v_i\right)
	\end{equation}
	The set of modes must be orthonormal, e.g.
	\begin{equation}
		\scprod{u_i}{u_j} = \delta_{ij} \quad,\quad \scprod{v_i}{v_j} = \delta_{ij}
		,\qquad \scprod{u_i}{v_j} = 0,
		\label{eq:dirac_orthonormij}
	\end{equation}
	and complete,
	\begin{equation}
		\sum_i \sqrt{-g} e^0_{\halpha} 
		\left(u_{ia}(t,\vx) (\overline{u}_i(t, \vxp) \gamma^{\halpha})_b +
		v_{ia}(t,\vx) (\overline{v}_i(t, \vxp) \gamma^{\halpha})_b \right)
		= \delta_{ab} \delta^3(\vx - \vxp).
		\label{eq:dirac_completeij}
	\end{equation}
	These properties entail the anticommutation relations for the operators 
	$b_i$ and $d_i^\dagger$
	\begin{equation}
		\acomm{b_i}{b^\dagger_j} = \delta_{ij} ,\qquad
		\acomm{d_i}{d^\dagger_j} = \delta_{ij}.
		\label{eq:dirac_acommij}
	\end{equation}
	The operators $b_i^\dagger$ ($b_i$) create (annihilate) particles 
	corresponding to the modes $u_i$, while $d_j^\dagger$ ($d_j$) are the 
	corresponding anti-particle 
	operators, which create (destroy) anti-particles corresponding to the
	modes $v_j$. These operators can be expressed as scalar products between
	the corresponding modes and the field:
	\begin{equation}
		b_i = \scprod{u_i}{\psi} ,\qquad 
		d^\dagger_j = \scprod{v_j}{\psi}.
		\label{eq:dirac_bdoperator_scprod}
	\end{equation}
	The construction of the Fock space is identical to the scalar case 
	\eqref{eq:kg_vacuum_def}. The particle number operator, whose expectation 
	value gives the number of particles in a given state, is defined as the sum
	of both particles and anti-particles:
	\begin{equation}
		N_i = b^\dagger_i b_i + d^\dagger_i d_i
		\label{eq:dirac_pnumberi}
	\end{equation}	
	The stress-energy tensor, defined by \eqref{eq:kg_Tmunu_def}, is more 
	easily computed by noting that
	\begin{equation*}
		\frac{\delta \mathcal{L}}{\delta e^{\lambda}_\halpha} =
		\frac{\delta \mathcal{L}}{\delta g^{\mu\nu}} 
		\frac{\delta g^{\mu\nu}}{\delta e^{\lambda}_\halpha}
	\end{equation*}
	The last term can be computed using \eqref{eq:frame_def} to yield
	\begin{equation*}
		\frac{\delta g^{\mu\nu}}{\delta e^{\lambda}_\halpha} = \eta^{\halpha\hbeta}
		(e^\mu_\hbeta \delta\indices{^\nu_\lambda} + e^\nu_\hbeta \delta\indices{^\mu_\lambda})
	\end{equation*}
	This expression is symmetric in the indices $\mu, \nu$, and can easily be 
	inverted, thus
	\begin{equation}
		T_{\mu\nu} = \frac{1}{\sqrt{-g}} \frac{\delta \mathcal{L}}{\delta e^{\mu}_\halpha}
		\eta_{\halpha\hbeta} \omega^{\hbeta}_\nu
		\label{eq:dirac_Tmunu_def}
	\end{equation}
	Applying this result to the Dirac lagrangian density 
	\eqref{eq:dirac_lagrangian}, we get the stress-energy tensor for the 
	Dirac field:
	\begin{equation}
		T_{\mu\nu} = \frac{i}{4} \left(\overline{\psi} \gamma_{\nu} \overleftrightarrow{D}_\mu \psi + 
		\overline{\psi} \gamma_{\mu} \overleftrightarrow{D}_\nu \psi\right)
		\label{eq:dirac_Tmunu}
	\end{equation}
	Only the fields $\psi$ and $\overline{\psi}$ are differentiated, and when
	differentiating the $\overline{\psi}$ field, the bar also runs over the
	derivative \eqref{eq:dirac_Dalpha}. The term $g_{\mu\nu} \mathcal{L}$ was
	omitted since the Lagrangian density vanishes when $\psi$ is a solution
	to the Dirac equation.
\end{subsection}
\end{section}
\begin{section}{Particle production}\label{sec:particle_prod}
The coupling of a quantum field with a time-dependent background gravitational
field gives rise to an interaction, similar to the effect of coupling
to a classical time-dependent electric field. This interaction can feed energy
into the field, which in turn can manifest as the phenomenon of spontaneous 
particle creation from an initial vacuum. 
The formalism employed for the analysis of the particle
production process is that of the Bogoliubov transformations. Let us assume
that the field is in the vacuum state before the initial time $\tin$ (up to 
where the space is identical to the Minkowski space). We let this
state evolve subject to the interaction with the gravitational field.
At a future time $\tout$ we perform a measurement of particle numbers in
this state. If the result is non-zero, then the interaction has
produced particles. There is one problem with this approach: due to the
non-unique definition of a particle, one might fool himself by detecting
particles because of a, say, specific reference frame change. This is indeed
the case in general relativity, where the lack of a general symmetry, like the
Poincar\'e invariance
in Minkowski, prevents us to define a vacuum state on which all
freely falling observers would agree.

To obtain a more objective probe of the state of a field one must construct
locally-defined quantities, such as expectation values of tensors 
(e.g., $T_{\mu\nu}$), which assume a particular value at the point $x$ of
spacetime. The stress-tensor is objective in the sense that, for a fixed state
$\ket{\psi}$, the results of different measuring devices can be related in the 
familiar fashion by the usual tensor transformation. For example, if
$\Braket{\psi|T_{\mu\nu}(x)|\psi} = 0$ for one observer, it will vanish for all
observers. This is in contrast to the particle concept, where one observer may
detect no particles while another may disagree \cite{book:birrell_davies}.

However, many problems arise when one tries to define a stress-energy tensor 
that is not infinite, but we shall sweep these issues under the rug by 
computing energy differences only, and noticing wether they are null or not.
The Bogoliubov formalism is slightly different for the scalar field and
the spinorial one, because of the different definition of scalar products,
and therefore must be discussed separately.
\begin{subsection}{The Klein-Gordon field}\label{sec:particle_prod_kg}
	The field $\phi(x)$ can be expressed in terms of any complete set of
	solutions. The Klein-Gordon equation \eqref{eq:kg_fieldeq} is linear, thus
	any solution of the equation can be expressed in terms of a complete set.
	Let $\{\fin_i, \fsin_i\}$ be the complete orthonormal set of
	modes describing the {\em in} particle states:
	\begin{subequations}\label{eq:kg_bogo_in}
	\begin{gather}
		\phi(x) = \sum_i \left(\fin_i \ain(i) + \fsin_i \adin(i)\right),\\
		\scprod{\fin_i}{\fin_j} = \delta_{ij} ,\qquad
		\scprod{\fsin_i}{\fsin_j} = -\delta_{ij} ,\qquad
		\scprod{\fin_i}{\fsin_j} = 0.
	\end{gather}
	\end{subequations}
	Simliarly, let $\{\fout_i, \fsout_i\}$ describe {\em out} particle states:
	\begin{subequations}\label{eq:kg_bogo_out}
	\begin{gather}
		\phi(x) = \sum_j \left(\fout_j \aout(j) + \fsout_j \adout(j)\right),\\
		\scprod{\fout_i}{\fout_j} = \delta_{ij} ,\qquad
		\scprod{\fsout_i}{\fsout_j} = -\delta_{ij} ,\qquad
		\scprod{\fout_i}{\fsout_j} = 0.
	\end{gather}
	\end{subequations}
	The {\em out} modes can be expressed in terms of the {\em in} modes using
	Bogoliubov coefficients:
	\begin{subequations}
	\begin{gather}\label{eq:kg_bogo_inout}
		\fout_i = \sum_j\left(\alpha_{ij} \fin_j + \beta_{ij} \fsin_j\right),\\
  	\alpha_{ij} = \scprod{\fin_j}{\fout_i} ,\qquad
  	\beta_{ij} = -\scprod{\fsin_j}{\fout_i}.
	\end{gather}
	\end{subequations}
	The orthonormalization of the {\em in} and {\em out} modes imply the
	following orthonormalization condition for the Bogoliubov coefficients:
	\begin{equation}
		\sum_k \left(\alpha^*_{ik} \alpha_{jk} - \beta^*_{ik}\beta_{jk}\right) = \delta_{ij}
		,\qquad
		\sum_k \left(-\alpha_{ik} \beta_{jk} + \beta_{ik}\alpha_{jk}\right) = 0.
		\label{eq:kg_bogo_orthonormij}
	\end{equation}
	Further manipulations of the scalar products defined by \eqref{eq:kg_scprod}
	give the relations
	\begin{equation*}
  	\alpha^*_{ij} = \scprod{\fout_i}{\fin_j} ,\qquad
  	\beta_{ij} = \scprod{\fsout_i}{\fin_j},
	\end{equation*}
	which are useful for the inverse of the transformation 
	\eqref{eq:kg_bogo_inout}:
	\begin{equation}
		\fin_j = \sum_{i} \left(\alpha^*_{ij} \fout_i -
	  \beta^*_{ij} \fsout_i\right). \label{eq:kg_bogo_outin}
	\end{equation}
	By expressing the {\em in} modes in terms of the {\em out} modes using
	\eqref{eq:kg_bogo_inout} in the field expansion \eqref{eq:kg_bogo_in} and
	equating with \eqref{eq:kg_bogo_out}, we can express the operators
	$\aout, \adout$ in terms of the {\em in} operators:
	\begin{subequations}\label{eq:kg_bogo_aad}
	\begin{align}
		\ain(j) &= \sum_i \left(\alpha_{ij} \aout(i) +
		\beta_{ij}^* \adout(i) \right), &
		\aout(i) &= \sum_j \left(\alpha^*_{ij} \ain(j) -
		\beta^*_{ij} \adin(j) \right),
		\label{eq:kg_bogo_a}\\
		\adin(j) &= \sum_i \left(\beta_{ij} \aout(i) +
		\alpha_{ij}^* \adout(i) \right), &
		\adout(i) &= \sum_j \left(-\beta_{ij} \ain(j) +
		\alpha_{ij} \adin(j) \right).
		\label{eq:kg_bogo_ad}
	\end{align}
	\end{subequations}
	If the mean value of the {\em out} number operator \eqref{eq:kg_pnumberi}
	in the {\em in} vacuum state, given by
	\begin{gather}
		n_i = \braket{0_{\rm in} | N^{\rm out}_i | 0_{\rm in}} =
		\sum_k \abs{\beta_{ik}}^2,
		\label{eq:kg_bogo_pnumberi}
	\end{gather}
	is nonzero, then particle creation occured between the {\em in} 
	and {\em out} states. The mean value for other products of creation and
	annihilation operators are
	\begin{subequations}
	\begin{gather}
		\braket{0_{\rm in} | \aout(i) \aout(j) | 0_{\rm in}} =
		-\sum_k \alpha^*_{ik} \beta^*_{jk},\\
		\braket{0_{\rm in} | \adout(i) \adout(j) | 0_{\rm in}} =
		-\sum_k \beta_{ik} \alpha_{jk},\\
		\braket{0_{\rm in} | \adout(i) \aout(j) | 0_{\rm in}} =
		\sum_k \beta_{ik} \beta^*_{jk}.
	\end{gather}
	\end{subequations}
	These are useful when evaluating the expectation value of the {\em out}
	stress-energy tensor \eqref{eq:kg_Tmunu_aij} in the {\em in} vacuum state:
		\begin{multline}
		\braket{0_{\rm in} | :T^{\rm out}_{\mu\nu}(x): | 0_{\rm in}} =\\ \sum_{i, j, k} \left\{
		-\alpha^*_{ik} \beta^*_{jk} (\partial_\mu f_i \partial_\nu f_j - g_{\mu\nu} \mathcal{L}(f_i, f_j))
		- \beta_{ik} \alpha_{jk} (\partial_\mu f^*_i \partial_\nu f^*_j -
		g_{\mu\nu} \mathcal{L}(f^*_i, f^*_j)) \right.\\
		\left. + \beta^*_{ik}\beta_{jk} (\partial_\mu f_i \partial_\nu f^*_j
		+\partial_\nu f_i \partial_\mu f^*_j  - g_{\mu\nu} \mathcal{L}(f_i, f^*_j)) \right\}.
		\label{eq:kg_bogo_Tmunu_aij}
	\end{multline}
\end{subsection}
\begin{subsection}{The Dirac field}\label{sec:particle_prod_dirac}
	Similarly to the case of the scalar field described in
	\autoref{sec:particle_prod_kg}, the field $\psi$ is expanded in
	terms of {\em in} and {\em out} modes, presumed to be orthonormal.
	The Bogoliubov coefficients are slightly different, as can be seen
	by writing the {\em out} modes in terms of the {\em in} ones:
	\begin{subequations}\label{eq:dirac_bogo_outin}
	\begin{gather}
		\Uout_i = \sum_j\left(\alpha_{ij} \Uin_j + \beta_{ij} \Vin_j\right)
		,\qquad	
		\Vout_i = \sum_j\left(\beta^*_{ij} \Uin_j + \alpha^*_{ij} \Vin_j\right),\\
  	\alpha_{ij} = \scprod{\Uin_j}{\Uout_i} ,\qquad
  	\beta_{ij} = \scprod{\Vin_j}{\Uout_i}.
	\end{gather}
	\end{subequations}
	There is a sign difference for the $\beta$ coefficient compared to
	\eqref{eq:kg_bogo_outin}, which becomes manifest in the orthonormalization
	condition:
	\begin{equation}
		\sum_k \left(\alpha^*_{ik} \alpha_{jk} + \beta^*_{ik}\beta_{jk}\right) = \delta_{ij}
		,\qquad
		\sum_k \left(\alpha_{ik} \beta_{jk} + \beta_{ik}\alpha_{jk}\right) = 0
		\label{eq:dirac_bogo_orthonormij}
	\end{equation}
	The inverse transformation of \eqref{eq:dirac_bogo_outin} is
	\begin{subequations}
	\begin{equation}
		\Uin_j = \sum_i\left(\alpha^*_{ij} \Uout_i + \beta_{ij} \Vout_i\right)
		,\qquad
		\Vin_j = \sum_i\left(\beta^*_{ij} \Uout_i + \alpha_{ij} \Vout_i\right)
	\end{equation}\label{eq:dirac_bogo_inout}
	\end{subequations}
	The same coefficients link the {\em in} creation and annihilation operators
	to the {\em out} ones:
	\begin{subequations}\label{eq:dirac_bogo_bd}
	\begin{align}
		\bin(j) &= \sum_i \left(\alpha_{ij} \bout(i) +
		\beta_{ij}^* \ddout(i) \right), &
		\bdin(j) &= \sum_i \left(\alpha^*_{ij} \bdout(i) +
		\beta_{ij} \dout(i) \right),
		\label{eq:dirac_bogo_binout}\\
		\ddin(j) &= \sum_i \left(\beta_{ij} \bout(i) +
		\alpha^*_{ij} \ddout(i) \right), &
		\din(j) &= \sum_i \left(\beta^*_{ij} \bdout(i) +
		\alpha_{ij} \dout(i) \right).
		\label{eq:dirac_bogo_dinout}
	\end{align}
	The inverse equations follow:
	\begin{align}
		\bout(i) &= \sum_j \left(\alpha^*_{ij} \bin(j) +
		\beta^*_{ij} \ddin(j) \right), &
		\bdout(i) &= \sum_j \left(\alpha_{ij} \bdin(j) +
		\beta_{ij} \din(j) \right),
		\label{eq:dirac_bogo_boutin}\\
		\ddout(i) &= \sum_j \left(\beta_{ij} \bin(j) +
		\alpha_{ij} \ddin(j) \right), &
		\dout(i) &= \sum_j \left(\beta^*_{ij} \bdin(j) +
		\alpha^*_{ij} \din(j) \right).
		\label{eq:dirac_bogo_doutin}
	\end{align}
	\end{subequations}
	The expectation value for the particle number \eqref{eq:dirac_pnumberi} 
	in the {\em in} vacuum state is twice the one for the 
	(uncharged) scalar field \eqref{eq:kg_bogo_pnumberi}:
	\begin{equation}
		n_i = \braket{0_{\rm in} | N^{\rm out}_i | 0_{\rm in}} =
		2 \sum_k \abs{\beta_{ik}}^2
		\label{eq:dirac_bogo_pnumberi}
	\end{equation}
	The mean value for other products of creation and annihilation operators are
	\begin{subequations}
	\begin{align}
		\braket{0_{\rm in} | \dout(i) \bout(j) | 0_{\rm in}} &=
		\sum_k \alpha^*_{ik} \beta_{jk},\\
		\braket{0_{\rm in} | \bdout(i) \ddout(j) | 0_{\rm in}} &=
		\sum_k \beta^*_{ik} \alpha_{jk},\\
		\braket{0_{\rm in} | \bdout(i) \bout(j) | 0_{\rm in}} &=
		\sum_k \beta^*_{ik} \beta_{jk},\\
		\braket{0_{\rm in} | \ddout(i) _{\rm out}(j) | 0_{\rm in}} &=
		\sum_k \beta^*_{ik} \beta_{jk}.
	\end{align}
	\end{subequations}
\end{subsection}
\end{section}

\chapter{The free field equation on de Sitter space-time}\label{chap:dSsol}
The equations for the free Klein-Gordon \eqref{eq:kg_fieldeq} and 
Dirac \eqref{eq:dirac_fieldeq} fields are linear in the field, thus the solutions
form a linear vector space. The construction of a basis in a vector space can be 
done by choosing a set of commuting conserved operators, and solve the 
corresponding eigenvalue problems. This produces a set of labels, which we have 
collectively denoted by the subscripts
${}_i, {}_j$ in the preceding section. After applying these labels to the
modes, the field equation simplifies, ideally to an algebraic relation 
between the labels (as in the Minkowski case), or if the set of operators
was not complete, to a simpler differential equation.

On the de Sitter space with line element \eqref{eq:ds_frw}, we recognize the
rotational and translational symmetry familiar from the Minkowski theory. 
Since the field equations are invariant on such transformations, we conclude
the momentum operators $P_i = -i \partial_i$ and angular mometum operators
$J_{ij} = x^i P_j - x^j P_i + S_{ij}$, with $S_{ij}$ being the spin generators
from the Minkowski theory, are both conserved and useful for the construction 
of modes. A thorough discution on symmetries on the de Sitter space are given
in \cite{art:cota_external_symm,art:cota_polarized_fermions,art:cota_kg}. 

In the construction of the solution to the Klein-Gordon field we shall follow
the work of \cite{art:cota_kg}, but other good references are 
\cite{book:birrell_davies,art:haro_elizade}. The construction of polarized
fermions solutions to the Dirac equation follows 
\cite{art:cota_polarized_fermions}, but the reader could also refer to 
\cite{art:haro_elizade}.
\begin{section}{The Klein-Gordon field}\label{sec:dSsol_kg}
For the most part of the derivation we will work on the conformal chart
\eqref{eq:ds_conformal}. The scalar product in this chart is given by
\begin{subequations} \label{eq:kg_scprod_dS}
\begin{equation}
	\scprod{f}{g} = i \int d^3x (-\omega \eta)^{-2} 
	(f^*(\eta,\vx) \overleftrightarrow{\partial_\eta} g(\eta,\vx)),
	\label{eq:kg_scprod_dS_conf}
\end{equation}	
while in the FRW chart it reads
\begin{equation}
	\scprod{f}{g} = i \int d^3x e^{3\omega t} 
	(f^*(t,\vx) \overleftrightarrow{\partial_t} g(t,\vx)).
	\label{eq:kg_scprod_dS_frw}
\end{equation}
\end{subequations}
The equation \eqref{eq:kg_fieldeq} in conformal coordinates translates to
\begin{equation}
	\omega^2\left(\eta^2 \partial_\eta^2 - 2 \eta \partial_\eta - \eta^2\Delta +
	\frac{m^2}{\omega^2} + 12 \xi\right) \phi(x) = 0,
	\label{eq:kg_fieldeq_dS}
\end{equation}
where we have used \eqref{eq:dalembert_dS} for the d'Alembert operator and 
\eqref{eq:ricci_dS} for the Ricci scalar. 
If $m=0$ and $\xi = 1/6$, the equation is conformal to the Minkowski case for 
$\phi/(-\omega \eta)$ (see \eqref{eq:kg_conformal_transf}).
We note that for a change of function
\begin{subequations}\label{eq:function_change_poly}
\begin{equation}
	F(\eta) = \eta^\alpha f(\eta),
\end{equation}
the derivative terms change as
\begin{align}
	\eta \frac{d}{d\eta} F(\eta) &= \eta^\alpha (\alpha f(\eta) + \eta f^\prime(\eta)),\\
	\eta^2 \frac{d^2}{d\eta^2} F(\eta) &= \eta^\alpha 
	(\alpha(\alpha - 1) f(\eta) + 2 \alpha \eta f^\prime(\eta) + \eta^2 f^\pprime(\eta)),
\end{align}
\end{subequations}
Primes denote differentiation with respect to the argument. 
If we let $\alpha = 1$, equation \eqref{eq:kg_fieldeq_dS} writes
\begin{equation}
	\left(\eta^2 \partial_\eta^2 - \eta^2 \Delta + 
	\frac{m^2}{\omega^2} + 12 (\xi - 1/6)\right) \frac{\phi(x)}{-\omega\eta} = 0
	\label{eq:kg_conf_fieldeq_dS}
\end{equation}
For $m = 0$ and $\xi = 1/6$, this equation reduces to the equation 
for a massless scalar field in flat spacetime.
Thanks to the space translation symmetry, we can construct the solutions 
as eigenfunctions of the momentum operator $P_i = -i \partial_i$
such that
\begin{equation}
	P_i f_{\vp}(x) = p^i f_{\vp}(x),
	\label{eq:kg_eigen_momentum}
\end{equation}
i.e. the $\vx$ dependence is in a plane wave factor $\sim e^{i\vp\vx}$. Thus
we introduce
\begin{equation}
	f_{\vp}(x) = \frac{1}{(2 \pi)^{\frac{3}{2}}} (-\omega \eta) 
	\varphi_p(\eta) e^{i \vp \vx}.
	\label{eq:kg_varphi_def}
\end{equation}
We note that the orthonormalization condition \eqref{eq:kg_orthonormij} 
translates to a Wronskian condition on $\varphi_p$:
\begin{equation}
	\scprod{f_{\vpp}}{f_{\vp}} = i W(\varphi^*_p(\eta), \varphi_p(\eta)) \delta^3(\vp - \vpp),
	\qquad W(\varphi^*_p(\eta), \varphi_p(\eta)) = -i.
	\label{eq:kg_varphi_norm}
\end{equation}
The Wronskian of two functions is defined as
\begin{equation}
	W\left(f(z), g(z)\right) = f(z) g^{\prime}(z) - f^{\prime}(z) g(z).
	\label{eq:wronskian_def}
\end{equation}
The equation \eqref{eq:kg_conf_fieldeq_dS} reads
\begin{equation}
	\left(\eta^2 \frac{d^2}{d\eta^2} + \eta^2 p^2 + 
	\frac{m^2}{\omega^2} + 12 (\xi - 1/6)\right) \varphi_{\vp}(\eta) = 0,
	\label{eq:varphi_eq}
\end{equation}
where we have introduced 
\begin{equation}
	p = \abs{\vp} = \sqrt{\vp^2}\label{eq:p_def}
\end{equation} 
for the length of the vector $\vp$.
This is the equation of a harmonic oscillator of variable frequency. The 
solution of this equation can be expressed in terms of Hankel functions 
(see \autoref{app:hankel}). Note that the argument of the Hankel function is
$z = -\eta p$, with $z > 0$. After a function change 
\eqref{eq:function_change_poly} with $\alpha = 1/2$ we arrive at the 
Bessel equation:
\begin{equation*}
	\left(z^2 \frac{d^2}{dz^2} + z \frac{d}{dz} + z^2 + 
	\frac{m^2}{\omega^2} + 12 (\xi - 1/6) - 1/4\right) 
	\frac{\varphi_{\vp}(z)}{\sqrt{z}} = 0	
\end{equation*}
We define
\begin{equation}
	M^2 = \frac{m^2}{\omega^2} + 12 (\xi - \frac{1}{6}) ,\qquad
	\mu = \frac{1}{4} - M^2,
	\label{eq:kg_def_M}
\end{equation}
and write the solution as
\begin{equation}
	\varphi_{\vp}(\eta) = \mathcal{N} \sqrt{-p\eta} Z_{\nu}(-p \eta) ,\qquad
	\nu = \sqrt{\abs{\mu}},
	\label{eq:kg_nudef}
\end{equation}
where $Z_\nu$ is given by
\begin{equation}
	Z_\nu(z) = \begin{cases}
		\phantom{e^{-\pi\nu/2}}\Hank{1}{\nu}(z) & \mu > 0\\
		e^{-\pi\nu/2} \Hank{1}{i \nu}(z) & \mu < 0
	\end{cases}
	,\qquad
	Z_\nu^*(z) = \begin{cases}
		\phantom{e^{\pi\nu/2}}\Hank{2}{\nu}(z) & \mu > 0\\
		e^{\pi\nu/2} \Hank{2}{i \nu}(z) & \mu < 0
	\end{cases}.
	\label{eq:kg_Z_def}
\end{equation}
More insight on this particular choice of solution is provided in 
\autoref{sec:hank_kg}. Using the wronskian relation \eqref{eq:kg_wronski}, we 
can determine the normalization constant $\mathcal{N}$ from the normalization 
condition \eqref{eq:kg_varphi_norm}:
\begin{equation*}
	\mathcal{N} = \sqrt{\frac{\pi}{4p}}
\end{equation*}
The result, written in the FRW chart, is
\begin{equation}
	\varphi_p(t) = \sqrt{\frac{\pi}{4 \omega}} e^{- \frac{1}{2} \omega t} 
	Z_\nu\ztparant{t}
	\label{eq:kg_varphi_sol_frw}
\end{equation}
Thus the plane wave solution \eqref{eq:kg_varphi_def} reads:
\begin{equation}
	f^{\rm dS}_{\vp}(t, \vx) = \frac{1}{(2\pi)^{3/2}} 
	\sqrt{\frac{\pi}{4 \omega}} e^{- \frac{3}{2} \omega t} 
	Z_\nu\ztparant{t} e^{i \vp \vx}
	\label{eq:kg_sol_frw}
\end{equation}
Together with their complex conjugates, these solutions form a complete set 
with respect to which the field can be expanded in the usual way:
\begin{equation}
	\phi(x) = \int d^3 x \left(\fdS(t, \vx) a_{\rm dS}(t, \vx) + 
	\fsdS(t, \vx) a^\dagger_{\rm dS}(t, \vx)\right),
	\label{eq:kg_phi_dS}
\end{equation}
where $a_{\rm dS}(t, \vx), a^\dagger_{\rm dS}(t, \vx)$ are the destruction
and creation operators corresponding to the modes $\fdS(t, \vx)$. The
superscript ${\rm dS}$ is used to distinguish between the solutions in this 
standard set and other combinations we shall use to define other particle 
states. 
\end{section}
\begin{section}{The Dirac field}\label{sec:dSsol_dirac}
As with the previous case, we will work in the conformal chart 
\eqref{eq:ds_conformal}. The scalar product in this chart is given by
\begin{subequations} \label{eq:dirac_scprod_dS}
\begin{equation}
	\scprod{\psi}{\chi} = \int d^3x e^{3 \omega t}\,
	\overline{\psi}(t, \vx)\gamma^0 \chi(t, \vx),
	\label{eq:dirac_scprod_dS_frw}
\end{equation}
while in the FRW chart it reads
\begin{equation}
	\scprod{\psi}{\chi} = i \int d^3x (-\omega \eta)^{-3}\,
	\overline{\psi}(\eta, \vx)\gamma^0 \chi(\eta, \vx)
	\label{eq:dirac_scprod_dS_conf}
\end{equation}
\end{subequations}
The field equation is expressed in terms of the tetrad frame vectors 
\eqref{eq:frame_def} and of the connection coefficient \eqref{eq:dirac_Dalpha}.
We choose the following tetrad vectors:
\begin{equation}
	e_{\hat{\eta}} = (-\omega\eta) \partial_\eta ,\qquad
	e_{\hat{i}} = (-\omega\eta) \partial_i.
	\label{eq:frame_dS_conformal}
\end{equation}
The corresponding Cartan coefficients \eqref{eq:cartan_def} evaluate to:
\begin{equation}
	c_{\hat{\eta}\hat{i}\hat{j}} = - \omega \eta_{\hat{i}\hat{j}}.
	\label{eq:cartan_dS_conformal}
\end{equation}
The corresponding connection coefficients \eqref{eq:Gamma_natural_def} follow:
\begin{equation}
	\Gamma_{\hat{\eta}\hat{i}\hat{j}} = - \omega \eta_{\hat{i}\hat{j}}.
	\label{eq:Gamma_natural_dS_conformal}
\end{equation}
Note the $\eta$ followed by hatted indices refers to the Minkowski metric
\eqref{eq:etamunu}.
All unlisted coefficients vanish. Substituting this result in 
\eqref{eq:dirac_Dalpha}, we evalute the covariant derivative 
\eqref{eq:dirac_Dalpha}:
\begin{equation}
	D_{\hat{\eta}} = (-\omega\eta)\partial_\eta ,\qquad
	D_{\hat{i}} = (-\omega\eta)\partial_i - \omega \eta_{\hat{i}\hat{j}}
	D(\Sigma^{\hat{0}\hat{j}}).
\end{equation}
The Dirac equation \eqref{eq:dirac_fieldeq} reads
\begin{equation}
	\left((-\omega \eta)
	(i \gamma^{\hat{0}} \partial_\eta + i \gamma^{\hat{i}} \partial_i) + 
	\frac{3 i \omega}{2} \gamma^{\hat{0}} - m\right) \psi(x) = 0.
\end{equation}
Using the general prescription \eqref{eq:function_change_poly}, we can 
eliminate the free $\gamma^{\hat{0}}$ term by using $\alpha = 3/2$:
\begin{equation}
	\left((-\omega \eta)(i \gamma^{\hat{0}} \partial_\eta - \gamma^{\hat{i}} P_i)
	 - m\right) \frac{\psi(x)}{(-\omega\eta)^{\frac{3}{2}}} = 0.
	\label{eq:dirac_fieldeq_simplified} 
\end{equation}
As with the scalar case, the equation is invariant to space translations and
rotations. Moreover, because of the spin degree of freedom, we can also use the
helicity operator $W = D(\Vec{S})\Vec{P}$
(the time component of the Pauli-Lubanski vector operator), 
along with the momentum operator $P_i = -i \partial_i$
to label the resulting modes:
\begin{subequations}\label{eq:dirac_UVlabel}
\begin{align}
	P_i U_{\vp, \lambda}(\eta, \vx) &= \phantom{-}p^i U_{\vp,\lambda}(\eta, \vx)&
	W U_{\vp,\lambda}(\eta, \vx) &= p \lambda U_{\vp, \lambda}(\eta, \vx)
	\label{eq:dirac_Ulabel}\\
	P_i V_{\vp, \lambda}(\eta, \vx) &= -p^i V_{\vp, \lambda}(\eta, \vx) &
	W V_{\vp, \lambda}(\eta, \vx) &= p \lambda V_{\vp, \lambda}(\eta, \vx)
	\label{eq:dirac_Vlabel}
\end{align}
\end{subequations}		
In analogy with the scalar case \eqref{eq:kg_varphi_def}, we choose
\begin{subequations}\label{eq:dirac_UVdef}
\begin{align}
	U_{\vp, \lambda}(\eta, \vx) &= \frac{1}{(2\pi)^{3/2}} (-\omega \eta)^{3/2} 
	u(\eta, \vp, \lambda) e^{i \vp\vx},
	\label{eq:dirac_Udef}\\
	V_{\vp, \lambda}(\eta, \vx) &= \frac{1}{(2\pi)^{3/2}} (-\omega \eta)^{3/2}
	v(\eta, \vp, \lambda) e^{-i \vp\vx}.
	\label{eq:dirac_Vdef}
\end{align}
\end{subequations}
The $V$ spinors are connected to the $U$ spinors through the familiar charge
conjugation operation
\begin{equation}
	V_{\vp, \lambda}(\eta, \vx) = C \overline{U}^T_{\vp, \lambda}(\eta, \vx),
	\qquad C = i \gamma^2 \gamma^0.
	\label{eq:dirac_C_def}
\end{equation}
The explicit form of a solution depends on our choice of $\gamma$ matrices. 
Throughout this paper we shall work in the Dirac representation:
\begin{gather}
	\gamma^{\hat{0}} = \begin{pmatrix}
	1 & 0\\
	0 & -1\end{pmatrix} ,\qquad
	\gamma^{\hat{i}} = \begin{pmatrix}
	0 & \sigma_i\\
	-\sigma_i & 0
	\end{pmatrix}.
	\label{eq:dirac_gamma_diracrep}
\end{gather}
The elements indicated are $2 \times 2$ matrices, and $\sigma_i$ are 
the Pauli matrices \eqref{eq:pauli_matrices}. In this representation,
the charge conjugation operator \eqref{eq:dirac_C_def} relating 
particle wavefunctions to anti-particle wavefunctions is 
\begin{subequations}
\begin{equation}
	C = \begin{pmatrix}
	0 & -i \sigma_2\\
	-i \sigma_2 & 0
	\end{pmatrix}.\label{eq:dirac_C_diracrep}
\end{equation}
The spin generators $D(S_i) = i \varepsilon_{ijk} D(\Sigma^{jk})$ are diagonal:
\begin{equation}
	D(S_i) = \frac{1}{2}\begin{pmatrix}
		\sigma_i & 0\\
		0 & \sigma_i
	\end{pmatrix},
	\label{eq:dirac_Si_diracrep}
\end{equation}
and so the helicity operator $W$ is
\begin{equation}
	W = \frac{1}{2} \begin{pmatrix}
		\vsigma \cdot \Vec{P} & 0\\
		0 & \vsigma \cdot \Vec{P}
	\end{pmatrix}.
	\label{eq:dirac_W_diracrep}
\end{equation}
\end{subequations}
This operator is block-diagonal, therefore we can write the four-component 
Dirac spinor part of the solutions \eqref{eq:dirac_UVdef} as
\begin{subequations}
\begin{align}
	u(t, \vp, \lambda) = \begin{pmatrix}
		f(t, \vp, \lambda) \xi_{\lambda}(\vp)\\
		g(t, \vp, \lambda) \xi_{\lambda}(\vp)
	\end{pmatrix}\quad,\quad
		v(t, \vp, \lambda) = \begin{pmatrix}
		g^*(t, \vp, \lambda) \eta_{\lambda}(\vp)\\
		-f^*(t, \vp, \lambda) \eta_{\lambda}(\vp)
	\end{pmatrix},
	\label{eq:dirac_uvdef}
\end{align}
where $f$ and $g$ are scalar functions and $\xi$ and $\eta$ are 
two-component Pauli spinors satisfying the eigenvalue equations
\begin{equation}
	\vp \vsigma \xi_{\lambda}(\vp) = 2p \lambda \xi_{\lambda}(\vp) ,\qquad
	\vp \vsigma \eta_{\lambda}(\vp) = -2p \lambda \eta_{\lambda}(\vp).
	\label{eq:dirac_xietadef}
\end{equation}
These spinors are related through the charge conjugation operation 
\eqref{eq:dirac_C_def} 
\begin{equation}
	\eta_{\lambda}(\vp) = i \sigma_2 \xi^*_{\lambda}(\vp).
	\label{eq:dirac_etaxicc}
\end{equation}
\end{subequations}
The explicit construction and some properties of these spinors are derived
in \autoref{chap:pauli_spinors}. For the purpose of this section 
we will only use the orthogonality
relations \eqref{eq:xieta_ortho_vn} and \eqref{eq:eta_mvn_xi_vn_prod}, with 
which we can evaluate the scalar product \eqref{eq:dirac_scprod_dS_conf} as
a normalization condition for the functions $f$ and $g$:
\begin{equation}
	\abs{f(t,\vp, \lambda)}^2 + \abs{g(t, \vp, \lambda)}^2 = 1.
	\label{eq:dirac_fg_norm}
\end{equation}
To determine these functions we write the spinorial equation 
\eqref{eq:dirac_fieldeq_simplified} for the $U$ spinor
\eqref{eq:dirac_Udef}:
\begin{equation*}
	\left(i \gamma^{\hat{0}} \eta \frac{d}{d\eta} - \eta \vp \Vec{\hat{\gamma}}
	 + k\right)
	\begin{pmatrix}
		f\xi_{\lambda}(\vp)\\ g\xi_{\lambda}(\vp) 
	\end{pmatrix} = 0 ,\qquad
	k = \frac{m}{\omega}.
\end{equation*} 
This is a coupled system of differential equations:
\begin{align}
\begin{split}
	2\lambda p\eta g &= \left(i \eta \frac{d}{d\eta} + k\right) f,\\
	2\lambda p\eta f &= \left(i \eta \frac{d}{d\eta} - k\right) g.
\end{split}
\label{eq:dirac_eqfg}
\end{align}
Combining the two equations, we arrive at a second order differential 
equation for $f$:
\begin{equation*}
	\left(\eta^2 \frac{d^2}{d\eta^2} + k^2 + ik + p^2 \eta^2\right) f = 0.
\end{equation*}
By making a function change \eqref{eq:function_change_poly} with 
$\alpha = 1/2$, and a change of variable $z = -p\eta$, 
the above equation becomes
\begin{equation}
	\left(z^2 \frac{d^2}{dz^2} + z \frac{d}{dz} + z^2 - (1/2 - ik)^2\right)
	\frac{f}{\sqrt{z}} = 0,
\end{equation}
which is the equation for a Bessel function of order $\nu_- = \nicefrac{1}{2} - ik$
\eqref{eq:bessel_equation}, thus the solution is
\begin{equation}
	f(\eta, p, \lambda) = i \mathcal{N} \sqrt{-p\eta} \Hank{1}{\nu_-}(-p\eta).
	\label{eq:dirac_solf}
\end{equation}
$f$ is chosen such that the solution of the massless case will reduce to the
Minkowski case, since in this case the field equation is conformally Minkowski.
We refer the reader to \autoref{sec:hank_dirac} for further insight. The 
function $g$ follows from \eqref{eq:dirac_eqfg}:
\begin{equation*}
	g = \frac{2i\lambda \mathcal{N}}{p \eta} \sqrt{-p \eta} \left(
	i \nu_- \Hank{1}{\nu_-}(-p\eta) - ip\eta\Hank{1}{\nu_-}^\prime(-p\eta)
	\right).
\end{equation*}
Replacing the derivative of the Hankel function using \eqref{eq:hank1_nupm_prime},
we obtain 
\begin{equation}
	g(\eta, p, \lambda) = 2i\lambda \mathcal{N} \sqrt{-p \eta} 
	e^{-\pi k} \Hank{1}{\nu_+}(-p\eta).
	\label{eq:dirac_solg}
\end{equation}
The normalization constant follows from the normalization condition 
\eqref{eq:dirac_fg_norm}:
\begin{equation*}
	\mathcal{N} = \sqrt{\frac{\pi}{4}} e^{\pi k/2},
\end{equation*}
derived using \eqref{eq:hank_nupms} for the complex conjugate of the
Hankel functions and the identity \eqref{eq:hank_idcota}. We list the
solutions in the FRW chart, in the form \eqref{eq:dirac_UVdef}:
\begin{subequations}\label{eq:dirac_uvsol_frw}
\begin{gather}
	u(t, \vp, \lambda) = i \sqrt{\frac{\pi p}{4\omega}} e^{-\frac{1}{2} \omega t}
	\spinor{\phantom{2\lambda}e^{\pi k/2} \Hank{1}{\nu_-}\ztparant{t} \xi_{\lambda}(\vp)}
	{2\lambda e^{-\pi k/2} \Hank{1}{\nu_+}\ztparant{t} \xi_{\lambda}(\vp)},
	\label{eq:dirac_usol}\\ 
	v(t, \vp, \lambda) = i \sqrt{\frac{\pi p}{4\omega}} e^{-\frac{1}{2} \omega t}
	\spinor{-2\lambda e^{-\pi k/2} \Hank{2}{\nu_-}\ztparant{t} \eta_{\lambda}(\vp)}
	{\phantom{-2\lambda}e^{\pi k/2} \Hank{2}{\nu_+}\ztparant{t} \eta_{\lambda}(\vp)},
	\label{eq:dirac_vsol}\\
	\overline{u}(t, \vp, \lambda)\gamma^{\hat{0}} u(t, \vp, \lambdap) = \delta_{\lambda\lambdap}
	,\qquad
	\overline{v}(t, \vp, \lambda)\gamma^{\hat{0}} v(t, \vp, \lambdap) = \delta_{\lambda\lambdap},
	\label{eq:dirac_uvnorm}
\end{gather}
\end{subequations}
and thus
\begin{subequations}\label{eq:dirac_UVsol_frw}
\begin{gather}
	\UdS(t, \vx) = \frac{i}{(2\pi)^{3/2}} \sqrt{\frac{\pi p}{4\omega}} e^{-2 \omega t}
	\spinor{\phantom{2\lambda}e^{\pi k/2} \Hank{1}{\nu_-}\ztparant{t} \xi_{\lambda}(\vp)}
	{2\lambda e^{-\pi k/2} \Hank{1}{\nu_+}\ztparant{t} \xi_{\lambda}(\vp)}
	e^{i \vp \vx},
	\label{eq:dirac_Usol_frw}\\ 
	\VdS(t, \vx) = \frac{i}{(2\pi)^{3/2}} \sqrt{\frac{\pi p}{4\omega}} e^{-2 \omega t}
	\spinor{-2\lambda e^{-\pi k/2} \Hank{2}{\nu_-}\ztparant{t} \eta_{\lambda}(\vp)}
	{\phantom{-2\lambda}e^{\pi k/2} \Hank{2}{\nu_+}\ztparant{t} \eta_{\lambda}(\vp)}
	e^{-i \vp \vx},
	\label{eq:dirac_Vsol_frw}\\ 	
	\scprod{\UdS}{\UdSp} = \delta_{\lambda\lambdap}\delta^3(\vx - \vxp)
	,\qquad
	\scprod{\VdS}{\VdSp} = \delta_{\lambda\lambdap}\delta^3(\vx - \vxp).
	\label{eq:dirac_UVnorm}
\end{gather}
\end{subequations}
\end{section}
\begin{section}{The Minkowski solutions}
As immediate from the line element \eqref{eq:ds}, there is an essential 
difference between the initial ({\em in}) and the final ({\em out}) Minkowski
spaces: the {\em out} distances are dilated by a factor 
$e^{\omega(\tout - \tin)}$. The usual plane wave solutions are constructed 
with respect to these coordinates, and the mode labels will refer to the 
physical momentum. The killing vectors of unit norm associated to the 
translational symmetry which define the hamiltonian and the momentum are:
\begin{align}
	H^{\rm in} &= i \partialdiv{}{t}, &
	P^{\rm in}_j &= -i \partialdiv{}{x^j_i}, &
	\vxin &= e^{\omega \tin} \vx, 
	\\
	H^{\rm out} &= i \partialdiv{}{t}, &
	P^{\rm out}_j &= -i \partialdiv{}{x^j_f}, &
	\vxout &= e^{\omega \tout} \vx.
\end{align}
For the scalar field, the {\em out} modes are
\begin{gather}
	\KGfSol{\Mout}{\vp}(t, \vxout) = \frac{1}{\sqrt{2E (2 \pi)^3 }} 
	e^{- i E t + i \vp\vxout} ,\qquad
	E = \sqrt{m^2 + \vp^2},
\end{gather}
and obey the equation
\begin{gather}
	\left( \frac{d^2}{dt^2} - \vp^2 + m^2\right) \KGfSol{\Mout}{\vp}(t, \vxout) = 0
	\label{eq:fout_mink_eq}
\end{gather}
These modes are orthonormal with respect to the scalar product
\begin{gather}
	\scprod{f}{g} = i \int d^3x_f f^*(t,\vxout) \overleftrightarrow{\partial_t} g(t,\vxout)
	\label{eq:kg_scprod_mink}
\end{gather}

These functions have a phase factor $e^{-i\vp\vxout}$, while de Sitter modes
have phase factors $e^{\pm i \vp\vx}$. We shall use $\vp$ for the physical 
momentum as measured in the {\em out} state, and $\vq$ for the corresponding
physical momentum measured in the {\em in} state. The corresponding de Sitter 
momenta are 
\begin{equation}
	p_i = q e^{\omega \tin}, \qquad 
	p_f = p e^{\omega \tout}
	\label{eq:pipf_def}
\end{equation}
As the particle propagates through the expansion phase, its momentum, defined
with respect to the de Sitter momentum operator $-i \partial_j$, is conserved, 
and thus $p_i = p_f$, which implies that
\begin{equation*}
	q = p e^{\omega(\tout - \tin)},
\end{equation*}
as discussed in \autoref{sec:frw}. When we write the Minkowski modes as 
functions of the de Sitter coordinate $\vx$ (rather than $\vxout$), 
the dilation factor shifts to the momentum:
\begin{gather}
	\KGfSol{\Mout}{\vp}(t, \vx) = \frac{1}{\sqrt{2E (2 \pi)^3 }} 
	e^{- i E t + i \vpout\vx} \quad,\quad
	\vpout = \vp e^{\omega \tout}
	\label{eq:kg_out_sol}
\end{gather}	
Simliarly, the Dirac Minkowski solutions in the {\em out} region are
\begin{align}
\begin{split}
	\DSol{U}{\Mout}{\vp}{\lambda}(t, \vx) &= 
	\frac{1}{(2\pi)^{3/2}}
	\sqrt{\frac{m}{E}}
	  \spinor{\frac{p}{\sqrt{2m(E - m)}} \xi_{\lambda}(\vp)}
	   {2\lambda \sqrt{\frac{E - m}{2m}} \xi_{\lambda}(\vp)}
	e^{-iE t + i \vpout \vx}
	\\
	\DSol{V}{\Mout}{\vp}{\lambda}(t,\vx) &= 
	\frac{1}{(2\pi)^{3/2}}
	\sqrt{\frac{m}{E}}
	  \spinor{2\lambda \sqrt{\frac{E - m}{2m}} \eta_{\lambda}(\vp)}
	   {-\frac{p}{\sqrt{2m(E - m)}} \eta_{\lambda}(\vp)}
	e^{iE t - i \vpout \vx}
\end{split}
\label{eq:dirac_UVsol_out}
\end{align}
and satisfy the equation
\begin{gather}
	\left( i \gamma^{0} \partial_t - \vp \Vec{\gamma} - m\right) 
	\DSol{U}{\Mout}{\vp}{\lambda}(t, \vx) = 0
	\label{eq:dirac_fieldeq_mink}
\end{gather}	
These modes obey orthonormalization conditions with respect to the scalar product
\begin{gather}
	\scprod{\psi}{\chi} = \int d^3x \overline{\psi}(t, \vx)\gamma^0 \chi(t, \vx)
	\label{eq:dirac_scprod_mink}
\end{gather}
Similar relations can be written for the {\em in} modes.
\end{section}

\chapter{Creation of massive scalar particles}\label{chap:kg}
Having established our notations and the necessary formalism, we shall delve
into the analysis of the production of scalar particles.
Important results shall be followed by graphical illustrations.

The preferred type of plots is the 
{\em lin-log} plot ($\bsq$ as a function of $\ln p$), with a few 
{\em log-log} plots necessary to capture the hyperbolic character of the 
low-mass Klein-Gordon field ($\mu > 0$) (this distinction shall become clear
in the development of this chapter).

All graphical representations follow the following conventions: exact solutions
are ploted in blue, red and green colors while asymptotic forms are plotted 
in black. Most images contain multiple plots, which are distinguished
through a number representing the parameter ($\omega, m$ or $\tout$) that
differs for each curve. Dashed lines indicate ``transition'' points between
the three relevant regions, i.e. $q = \omega$ and $p = \omega$. The parameter
$\tin$ is fixed at $\tin = 0$ because $\bsq$ depends only on the difference
$\Delta t = \tout - \tin$. We shall refer to this difference both by $\Delta t$
and by $\tout$. The images were obtained using Mathematica 7.0.

In the first section we derive the analytical formula for the $\beta$ 
Bogoliubov coefficient, expressed in terms of Hankel functions. We use this 
formula to show that there is no particle production when the field is 
conformal (conformal coupling and no mass). This section ends with some
figures depicting $\bsq$, which we shall use for orientation in 
the asymptotic analysis. This will be the subject of 
\autoref{sec:kg_asympt}, where we investigate the form of $\bsq$ in
regions where asymptotic analysis is valid. This enables us to 
define an approximation of $\bsq$ which we shall plot at the end of the
section for comparison with the exact form $\bsq$. The asymptotic forms are 
used to evaluate the particle number density $n(\vx)$ (density of created 
particles per unit volume), and the results are compare with numerical 
integration in \autoref{sec:kg_asympt_fig}.
\begin{section}{Bogoliubov coefficients}\label{sec:kg_bogo}
Particle states with a well defined energy exist only in the flat regions of
space (see \autoref{fig:frw_space_minkdsmink}). During the expansion phase, 
such states undergo a non-trivial evolution dictated by the corresponding 
field equation on de Sitter space time, which has the effect of mixing 
particle and anti-particle states. 

We shall use the de Sitter and Minkowski solutions derived in 
\autoref{chap:dSsol} for the continuation of particle states through the
expansion phase, discussed in \autoref{sec:kg_junction}. 
To these modes we shall apply the Bogoliubov transformation
formalism of \autoref{sec:particle_prod}, and we shall compute the 
Bogoliubov coefficients in \autoref{sec:kg_bogo}.
\begin{subsection}{de Sitter \texorpdfstring{{\em in} and {\em out}}{in and out} modes}
\label{sec:kg_junction}
We have constructed the momentum base solutions to 
the Klein-Gordon and Dirac equations, both on de Sitter and on 
Minkowski spaces (see \autoref{chap:dSsol}). In this section we will describe
a method of continuing 
modes from the Minkowski flat regions of space into the de Sitter expanding
phase. This is done by constructing a linear combination of de Sitter
modes \eqref{eq:kg_sol_frw}, \eqref{eq:dirac_UVsol_frw} such that the Minkowski mode
entering the de Sitter expansion phase is continuous, with 
its first derivative continuous, at the junction:
\begin{gather}
	\foutp(t, \vx) = \left\{\begin{aligned}
		&\KGfSol{\Mout}{\vp}(t, \vx) & \tout < &t,\\
		&A(p, \tout) \KGfSol{dS}{\vpout}(t, \vx) + 
		B(p, \tout) \KGfsSol{dS}{-\vpout}(t, \vx) & \tin < &t < \tout,\\
		&e^{\frac{3}{2}(\tout - \tin)} \left(\alpha(p) \KGfSol{\Min}{\vq}(t, \vx) + 
		\beta(p) \KGfsSol{\Min}{-\vq}(t, \vx)\right) & & t < \tin,
	\end{aligned}\right.
	\label{eq:junction}\\
	\vpout = \vp e^{\omega \tout} ,\qquad	\vq = \vp e^{\omega (\tout - \tin)}.
	\nonumber
\end{gather}
The Minkowski modes of momentum $\vp$ are matched by de Sitter modes of
momentum $\vpout = \vp e^{\omega \tout}$, since de Sitter wavelengths 
are increased as the space expands. This can be checked by applying the
de Sitter momentum operator $-i \partial_i$ on both the Minkowski mode and
on the de Sitter combination and equating the two eigenvalues.

The continuity at the two junction points unambiguously define the matching 
coefficients $A(p, \tout), B(p, \tout)$
and the Bogoliubov coefficients $\alpha(p, \tin, \tout), \beta(p, \tin, \tout)$
These coefficients describe the mode mixing which 
occures during the expansion phase, as described in \autoref{sec:particle_prod}.
Of the two coefficients, $\beta$
will be extensively analysed in the following sections and chapters.

Using the same notations, a similar definition can be written for the {\em in} 
modes:
\begin{equation}
	\KGfSol{in}{\vq}(t, \vx) = \left\{\begin{aligned}
		&\KGfSol{\Min}{\vq}(t, \vx) & & t < \tin,\\
		&A(q, \tin) \KGfSol{dS}{\vpout}(t, \vx) + 
		B(q, \tin) \KGfsSol{dS}{-\vpout}(t, \vx) & \tin < &t < \tout,\\
		&e^{\frac{3}{2}(\tout - \tin)} \left(
		\alpha(p) \KGfSol{\Mout}{\vp}(t, \vx) + 
		\beta(p) \KGfsSol{\Mout}{-\vp}(t, \vx)\right) & \tout < &t.
	\end{aligned}\right.
	\label{eq:junction_in}
\end{equation}
The Klein-Gordon equation is a second order differential equation, and 
therefore it requires initial values for both the solution and its derivative.
Before applying the continuity conditions described in \eqref{eq:junction}, 
we first note that the de Sitter Klein-Gordon equation for 
$\varphi(t) \sim \phi(t) e^{\omega t}$ \eqref{eq:varphi_eq} reduces to the 
Minkowski Klein-Gordon equation \eqref{eq:kg_out_sol} if the expansion factor
is constant (i.e. $-\omega \eta = e^{-\omega t} = \text{const}$). Therefore, the 
continuity conditions shall be applyied for the $e^{\omega t} \phi(t,\vx)$ 
part of the modes rather than for the mode itself. An incorrect junction 
conditions leads to unphysical results, such as infinite density of created 
particles. Thus we require
\begin{subequations}\label{eq:kg_outdef}
\begin{align}
	\KGfSol{f}{\vp}(\tout, \vx) &= \KGfSol{out}{\vp}(\tout, \vx)
	\label{eq:kg_outjunction},\\
	\left(\frac{\partial_t e^{\omega t} \KGfSol{f}{\vp}(\tout, \vx)}{e^{\omega t}}\right)_{t = \tout}
	\mspace{-20.0mu} &= 
	\left(\partialdiv{\KGfSol{out}{\vp}(\tout, \vx)}{t}\right)_{t = \tout}.
	\label{eq:kg_outderivjunction}
\end{align}
\end{subequations}
Substituting the coefficients from \eqref{eq:junction} we arrive at the system
of equations:
\begin{subequations}
\begin{align}
 	A(p, \tout) \fdSout(\tout, \vx) + B(p, \tout) \fsdSoutm(\tout, \vx) &= \KGfSol{\Mout}{\vp}(\tout, \vx),\\
	A(p, \tout) \left(\frac{\partial_t e^{\omega t} \fdSout(t, \vx)}{e^{\omega t}}\right)_{t = \tout}\mspace{-20.0mu} +
	B(p, \tout) \left(\frac{\partial_t e^{\omega t} \fsdSoutm(t, \vx)}{e^{\omega t}}\right)_{t = \tout}\mspace{-20.0mu}
	 &= \left(\partialdiv{\KGfSol{\Mout}{\vp}}{t}\right)_{t = \tout}.
	\label{eq:kg_junction_eq}
\end{align}	
\end{subequations}
Substituting \eqref{eq:kg_out_sol} for $f^{\Mout}$ and \eqref{eq:kg_sol_frw} for
$\fdSout$, and using $\pout = p \exp{\omega \tout}$, we arrive at the 
equivalent matrix equation
\begin{multline}
	\begin{pmatrix}
		Z_\nu\pparant & 
		Z^*_\nu\pparant\smallskip\\
		\frac{1}{2} Z_\nu\pparant + \frac{p}{\omega} Z^{\, \prime}_\nu\pparant&
		\frac{1}{2} Z^*_\nu\pparant + \frac{p}{\omega} Z^{*\, \prime}_\nu\pparant
	\end{pmatrix}
	\begin{pmatrix}
		A(p, \tout) \smallskip\\
		B(p, \tout)
	\end{pmatrix} =\\
	  \sqrt{\frac{2\omega}{\pi E_p}} e^{-i E_p \tout + \frac{3}{2}\omega \tout}
	\begin{pmatrix}
		1 \smallskip\\
		i	
	\end{pmatrix}.\label{eq:kg_junction_matrixeq}
\end{multline}
The determinant of the matrix in the LHS can be computed using the 
Wronskian of the $Z$ functions \eqref{eq:kg_wronski}, and we 
find
\begin{subequations}\label{eq:kg_junction_solAB}
\begin{align}
  A(p, \tout) &= \phantom{-}\sqrt{\frac{\pi E_p}{8\omega}} e^{-i E_p \tout + \frac{3}{2} \omega \tout}
  \left\{\left(1 + \frac{i \omega}{2 E_p}\right) Z_\nu^*\pparant + 
  \frac{i p}{E_p} Z_\nu^{\prime\, *}\pparant\right\}
  \label{eq:kg_junction_solA},\\
  B(p, \tout) &= -\sqrt{\frac{\pi E_p}{8\omega}} e^{-i E_p \tout + \frac{3}{2} \omega \tout} 
  \left\{\left(1 + \frac{i \omega}{2 E_p}\right) Z_\nu\pparant +
  \frac{i p}{E_p} Z_\nu^{\prime}\pparant\right\}.
  \label{eq:kg_junction_solB}
\end{align}
\end{subequations}
Using the same wronskian relation, we arrive at the normalization relation
\begin{equation}
	\abs{A(p, \tout)}^2 - \abs{B(p, \tout)}^2 = e^{3 \omega \tout},
	\label{eq:kg_AB_norm}
\end{equation}
and thus the modes are orthonormal throughout all space, both in the
expansion phase, with respect to the de Sitter scalar product 
\eqref{eq:kg_scprod_dS_frw} and on the Minkowski region with respect to the 
scalar product \eqref{eq:kg_scprod_mink}:
\begin{equation*}
	\scprod{\fout_{\vpp}}{\fout_{\vp}} = \delta^3(\vp - \vpp),
\end{equation*}
since 
\begin{equation*}
	\delta^3(\vpout - \vpout^\prime) = e^{-3\omega \tout} \delta^3(\vp - \vpp).
\end{equation*}
It is convenient to define a new set of coefficients normalized to unity:
\begin{equation}
	\tilde{A}(p, \tout) = e^{-\frac{3}{2} \omega t} A(p, \tout),\qquad
	\tilde{B}(p, \tout) = e^{-\frac{3}{2} \omega t} B(p, \tout).
\label{eq:kg_junction_solAB_tilde}
\end{equation}
One might argue about the junction continuity condition for the derivative 
\eqref{eq:kg_outderivjunction}. We note that if we had chosen the continuity
of $a^{1-\alpha}(t) \phi(t)$ instead of $a(t) \phi(t)$, the resulting 
$A$ and $B$ coefficients would have had a similar form, with the following
replacement:
\begin{equation}
	\frac{i \omega}{2 E_p} \rightarrow (1 + 2\alpha)\frac{i \omega}{2 E_p}.
	\label{eq:kg_junctionalphaAB}
\end{equation}
As will be shown in \autoref{sec:kg_asympt_large}, the supplimentary term 
produces a leading term of order $1/p^2$ in the ultraviolet region, which
makes the volumic density of produced particles $n(x)$ an infinite number, 
since the particle number spectral density $n_p$ approaches a constant value.
\end{subsection}
\begin{subsection}{Mode mixing and density of created particles}
\label{sec:kg_bogo_mix}
In this section we apply the general theory of Bogoliubov transformation 
outlined in \autoref{sec:particle_prod} to the case in which mode mixing
occurs only for a certain label, selected through delta functions.

We have previously determined coefficients $A$ and $B$ such that
\begin{subequations}
\begin{equation}
	\foutp(t, \vx) = A(p, \tout) \fdSout(t, \vx) + B(p, \tout) \fsdSoutm(t, \vx).
\end{equation}
The same procedure applies in defining {\em in} modes:
\begin{align}
	\KGfSol{in}{\vp}(t, \vx) &= A(p, \tin) \KGfSol{dS}{\vpin}(t, \vx) + 
	B(p, \tin) \KGfsSol{dS}{-\vpin}(t, \vx)	,\\
	\KGfsSol{in}{-\vp}(t, \vx) &= B^*(p, \tin) \KGfSol{dS}{\vpin}(t, \vx) + 
	A^*(p, \tin) \KGfsSol{dS}{-\vpin}(t, \vx).
\end{align}
\end{subequations}
Now we must use the Bogoliubov coefficients \eqref{eq:kg_bogo_inout} to link
the two sets:
\begin{equation*}
	\foutp(t, \vx) = \int d^3p^\prime \left\{
	\alpha(\vp, \vpp) \KGfSol{in}{\vpp}(t, \vx) + 
	\beta(\vp, -\vpp) \KGfsSol{in}{-\vpp}(t, \vx)\right\}.
\end{equation*}
The Bogoliubov coefficients are readily evaluated as scalar products between
the {\em in} particle and anti-particle modes and the {\em out} particle 
modes:	
\begin{subequations}
\begin{align}
	\alpha(\vp, \vpp) &= e^{-3\omega \tin} \delta^3(\vpp - \vq) 
	\left(A^*(q, \tin) A(p, \tout) - B^*(q, \tin) B(p, \tout)\right),\\
	\beta(\vp, -\vpp) &= e^{-3\omega \tin} \delta^3(\vpp - \vq) 
	\left(A(q, \tin) B(p, \tout) - B(q, \tin) A(p, \tout)\right).
\end{align}
\end{subequations}
The {\em in} momentum $\vq = \vp e^{\omega(\tout - \tin)}$ corresponds to 
the {\em out} momentum $\vp$, in accord with the dilation of 
wavelengths occuring in the expansion phase.
The minus sign of the coefficient $B$ appeared because the scalar product 
of anti-particle modes is negative. It is convenient to define two reduced 
Bogoliubov coefficients:
\begin{subequations}
\begin{align}
	\alpha(\vp, \vpp) &= e^{\frac{3}{2}\omega (\tout - \tin)} 
	\delta^3(\vpp - \vq) \alpha(p),
	\label{eq:kg_alphadef}\\
	\beta(\vp, -\vpp) &= e^{\frac{3}{2}\omega (\tout - \tin)} 
	\delta^3(\vpp - \vq) \beta(p).
	\label{eq:kg_betadef}
\end{align}
explicitly given by
\begin{align}
	\alpha(p) &= \tilde{A}^*(q, \tin) \tilde{A}(p, \tout) - 
	\tilde{B}^*(q, \tin) \tilde{B}(p, \tout),
	\label{eq:kg_alphaAB}\\
	\beta(p) &= \tilde{A}(q, \tin) \tilde{B}(p, \tout) - 
	\tilde{B}(q, \tin) \tilde{A}(p, \tout).
	\label{eq:kg_betaAB}
\end{align}
such that the following normalization condition is obeyed:
\begin{equation}
	\abs{\alpha(p)}^2 - \bsq = 1.
	\label{eq:kg_orthonorm}
\end{equation}
\end{subequations}	
This is just the normalization condition \eqref{eq:kg_orthonormij}, while 
the orthogonality relation is automatically fulfilled. The {\em out} 
one-particle operators are expressed as
\begin{subequations}
\begin{align}
	\aout(\vp) &= e^{\frac{3}{2}\omega(\tout - \tin)} \left(
	\alpha^*(p) \ain(\vq) - \beta^*(p) \adin(-\vq)\right),\\
	\adout(\vp) &= e^{\frac{3}{2}\omega(\tout - \tin)} \left(
	-\beta(p) \ain(-\vq) + \alpha(p) \adin(\vq)\right),
\end{align}
\end{subequations}	
and the expectation value of these operators in the {\em in} vacuum state is
\begin{subequations}
\begin{align}
	\braket{0_{\rm in} | \aout(\vp) \aout(\vpp) | 0_{\rm in}} &=
	-\alpha^*(p) \beta^*(p) \delta^3(\vp + \vpp),\\
	\braket{0_{\rm in} | \adout(\vp) \adout(\vpp) | 0_{\rm in}} &=
	-\beta(p) \alpha(p) \delta^3(\vp + \vpp),\\
	\braket{0_{\rm in} | \aout(\vp) \adout(\vpp) | 0_{\rm in}} &=
	\abs{\alpha(p)}^2\delta^3(\vp - \vpp),\\
	\braket{0_{\rm in} | \adout(\vp) \aout(\vpp) | 0_{\rm in}} &=
	\abs{\beta(p)}^2 \delta^3(\vp - \vpp).
\end{align}
\end{subequations}
With these expectation values we can evaluate the particle number density
\eqref{eq:kg_bogo_pnumberi}
\begin{equation}
	n_{\vp} = \abs{\beta(p)}^2 \delta^3(\vp - \vp).
	\label{eq:kg_bogo_pnumberp_inf}
\end{equation}
The delta function in the RHS can be regarded as the volume of the infinite 
space
\begin{equation}
	V = \int d^3x e^{i \vx(\vp -\vp)} = (2\pi)^3 \delta^3(\vp - \vp),
\end{equation}
therefore we can consider the volumic particle density
\begin{equation}
	n_{\vp}(\vx) = \frac{1}{(2\pi)^3} \abs{\beta(p)}^2.
\end{equation}
In order to evaluate the number of particles with the magnitude of the 
momentum $p$, we integrate away the spherical coordinates and arrive at
\begin{equation}
	n_p(\vx) = \frac{p^2}{2\pi^2} \abs{\beta(p)}^2.
	\label{eq:kg_bogo_np}
\end{equation}
This is in agreement with the expectation value of the energy component 
of the (Minkowski) stress-energy tensor, normally ordered with respect to
the {\em out} vacuum:
\begin{equation}
	:T^{\rm out}_{\mu\nu}(x): = 
	T_{\mu\nu}(x) - \braket{0_{\rm out} | T_{\mu\nu}(x) | 0_{\rm out}},
	\label{eq:Tmunu_out}
\end{equation}
which evaluates to
\begin{multline}
	\braket{0_{\rm in} | :T^{\rm out}_{\mu\nu}(x): | 0_{\rm in}} = 
	\int \frac{d^3p}{(2\pi)^3} \\
	\left\{
	\left(\frac{p_\mu \tilde{p}_{\nu}}{E } - \eta_{\mu\nu} E\right) \frac{1}{2}
	\left(\alpha^*(p) \beta^*(p) e^{-2i E t} + \alpha(p) \beta(p) e^{2i E t}\right)
	+ \abs{\beta(p)}^2 \frac{p_\mu p_\nu}{E_p} \right\},
	\label{eq:kg_bogo_Tmunu_beta}
\end{multline}
with $\widetilde{p}^\mu = (E_p, -\vp)$, from which we read the expectation 
value for the energy 
\begin{equation}
	\braket{0_{\rm in} | :T^{\rm out}_{00}(x): | 0_{\rm in}} = \mathcal{E}(\vx) =
	\int \frac{d^3p}{(2\pi)^3} E_p \bsq.
	\label{eq:kg_T00_beta}
\end{equation}
The energy spectral density follows:
\begin{equation}
	\mathcal{E}_p = \frac{p^2}{2\pi^2} E_p \bsq.
	\label{eq:kg_Epdef}
\end{equation}
The pressure can also be read from the stress-energy tensor:
\begin{multline}
	\braket{0_{\rm in} | :T^{\rm out}_{ij}(x): | 0_{\rm in}} = - \delta_{ij}
	\int \frac{d^3p}{(2\pi)^3} \frac{\vp^2}{3 E_p} \times\\
	\left\{
	\left(1 - \frac{3E^2}{\vp^2}\right)\frac{1}{2} 
	\left(\alpha^*(p) \beta^*(p) e^{-2i E t} + \alpha(p) \beta(p) e^{2i E t}\right) - 
	\abs{\beta(p)}^2\right\}.
	\label{eq:kg_Tij_beta}
\end{multline}
By substituting \eqref{eq:kg_junction_solAB} for $A$ and $B$ in the formula
for $\beta(p)$ \eqref{eq:kg_betaAB}, we arrive at the following expression:
\begin{multline}
	\beta(p) = -\frac{\pi}{8\omega} \sqrt{E_q E_p} e^{-i(E_q \tin + E_p \tout)} \times\\
	\left\{\left[\left(\frac{\omega}{2 E_q} + \frac{\omega}{2 E_p}\right) Z_1(p, q) + 
	\frac{q}{E_q} Z_2(p, q) + 
	\frac{p}{E_p} Z_3(p, q)\right]+\right.\\
	\left.i \left[ \left(1 - \frac{\omega^2}{4E_qE_p}\right) (-Z_1(p, q)) + 
	\frac{\omega}{2E_q E_p} \left( 
	q Z_2(p, q) + p Z_3(p, q)\right)
	+ \frac{qp}{E_qE_p} Z_4(p, q) \right] \right\}.
	\label{eq:kg_beta_explicit}
\end{multline}
We have introduced the notation $E_p$ for the Minkowski energy of a particle 
of mass $m$ and momentum $p$
\begin{equation}
	E_p = E(p) = \sqrt{m^2 + \vp^2}, \label{eq:mink_e}
\end{equation}
and the functions $Z_i$ are given by:
\begin{subequations}\label{eq:kg_beta_Z}
\begin{gather}
	Z_1(q, p) = i\left(Z_\nu^*\qparant Z_\nu\pparant - 
	Z_\nu\qparant Z_\nu^*\pparant\right)
	\label{eq:kg_beta_Z1}\\
	Z_2(q, p) = i\left(Z_\nu^{*\prime}\qparant Z_\nu\pparant - 
	Z^\prime_\nu\qparant Z_\nu^*\pparant\right)
	\label{eq:kg_beta_Z2}\\
	Z_3(q, p) = i\left(Z_\nu^*\qparant Z_\nu^\prime\pparant - 
	Z_\nu\qparant Z_\nu^{*\prime}\pparant\right)
	\label{eq:kg_beta_Z3}\\
	Z_4(q, p) = i\left(Z_\nu^{*\prime}\qparant Z^\prime_\nu\pparant - 
	Z^\prime_\nu\qparant Z_\nu^{*\prime}\pparant\right)
	\label{eq:kg_beta_Z4}
\end{gather}
\end{subequations}
All $Z_i(q, p)$ are real, and as a consequence of the odd behaviour of 
$\beta(p)$ under $p \leftrightarrow q$, we have
\begin{gather*}
	Z_1(p, q) = -Z_1(q, p),\\
	Z_4(p, q) = -Z_4(q, p),\\
	Z_2(p, q) = -Z_3(q, p).
\end{gather*}
We note that the first group of terms is real, while the second is
imaginary, for all (positive or negative) values of $M^2$, since
the normalization constant $e^{\pm\pi\nu/2}$ appearing in \eqref{eq:kg_Z_def}
vanishes in products of the form $Z_\nu(z_1) Z^*_\nu(z_2)$.	
We find that, except for a wronskian produced by the term $Z_1 Z_4 - Z_2 Z_3$,
squaring this coefficient brings little analytical improvement.

Apart from the leading phase $e^{-iE_q \tin - i E_p \tout}$, $\beta$ only 
depends on the expansion time $\Delta t = \tout - \tin$, through 
$q = p e^{\omega \Delta t}$. From this we conlcude that the particle 
production phenomenon is invariant to translations in time and depends only
on the relative inflation of space rather than on independent 
{\em in} and {\em out} states.
\end{subsection}
\begin{subsection}{Particle production of conformal massless scalar particles}
\label{sec:kg_massless}
The $\beta$ coefficient determined in the previous section describes the 
phenomenon of particle production through the coupling between scalar or 
spinorial fields and the gravitational field. If the field equations are 
conformal to the Minkowski equations (albeit expressed in conformal time),
there should be no mode mixing, since the de Sitter modes are related to
the Minkowski ones through a conformal transformation. This is indeed the case,
and we prove it by analysing the massless case of the conformally coupled 
scalar field.

In the conformally coupled massless case we have $\mu^2 = 1/4$, as given by
\eqref{eq:kg_def_M}, and thus $\nu = 1/2$ is the order of the Hankel functions
$Z_\nu(z) = \Hank{1}{1/2}(z)$. The explicit form of this Hankel function is
given in the appendix by \eqref{eq:hankel_1_2}. In order to evaluate 
the coefficients $\tilde{A}(p, t), \tilde{B}(p, t)$, we must compute the
derivative of this function:
\begin{equation*}
	\Hank{1}{1/2}^\prime(z) = \left(1 + \frac{i}{2z}\right) 
	\sqrt{\frac{2}{\pi z}} e^{iz},\qquad
	\Hank{2}{1/2}^\prime(z) = \left(1 - \frac{i}{2z}\right) 
	\sqrt{\frac{2}{\pi z}} e^{-iz}
\end{equation*}
Substituting in \eqref{eq:kg_junction_solAB} we arrive at
\begin{subequations}\label{eq:kg_massless_conf_beta}
\begin{gather}
	\tilde{A}(p, t) = i e^{-ipt - ip/\omega} ,\qquad
	\tilde{B}(p, t) = 0,\\
	\beta(p) = 0.
\end{gather}
\end{subequations}
If the field is not conformally coupled, the result is non-zero.
\end{subsection}
\begin{subsection}{Graphical analysis}\label{eq:kg_bogo_fig}
In this subsection we illustrate the exact analytical solution for $\bsq$ 
obtained by squaring \eqref{eq:kg_beta_explicit}. We anticipate some of the
results of the next section when analysing the figures.

The two different regimes corresponding to $\mu > 0$ (hyperbolic) and 
$\mu < 0$ (trigonometric) require two sets of graphs because of the 
difference in order of magnitude in the middle and low momentum regions.
\begin{figure}[!ht]
	\centering
	\includegraphics[width=0.8\linewidth]{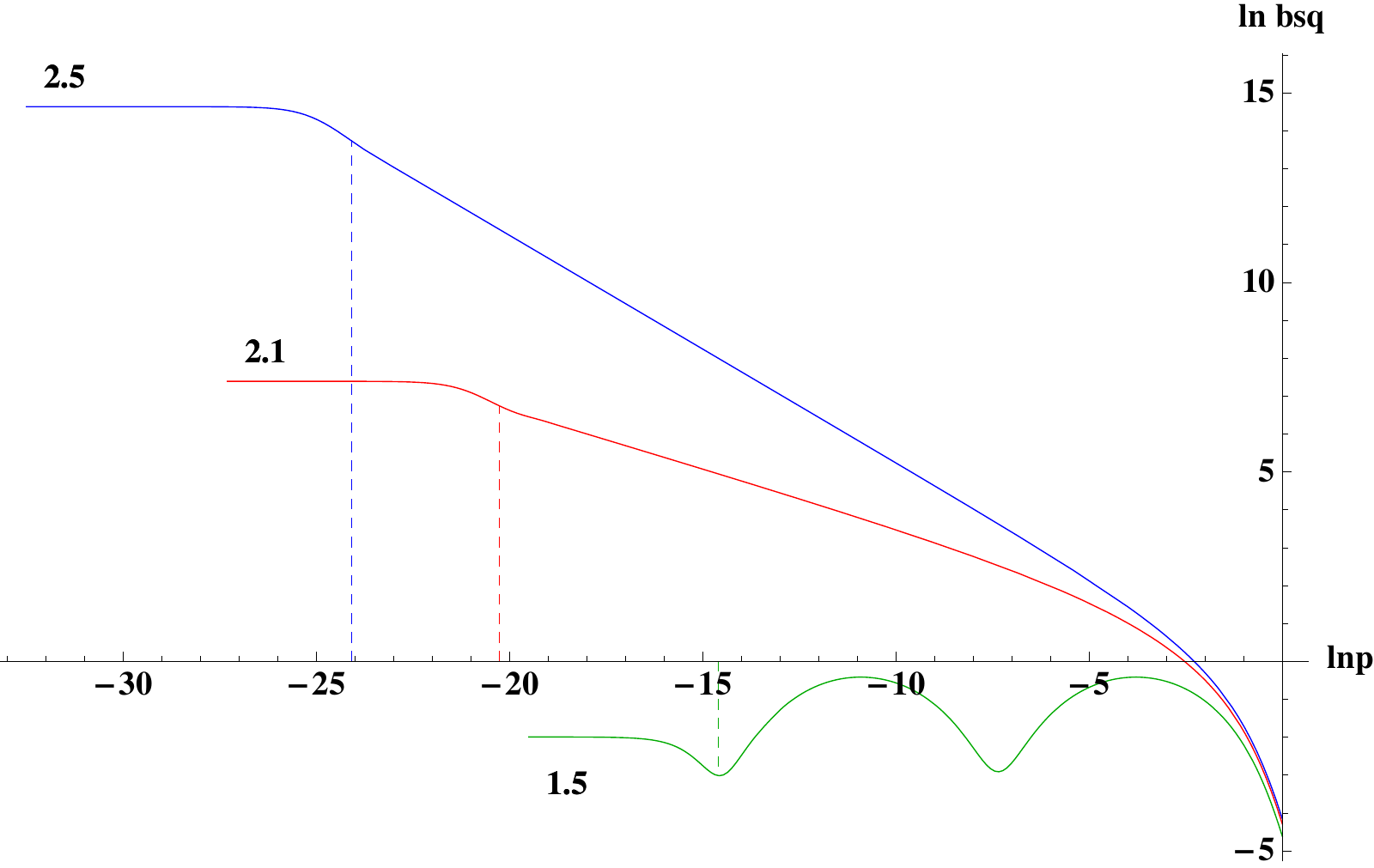}
	\caption{\small log-log plot of $\ln \bsq$ against $\ln p$ at 
	$m = 1$, $\tin = 0$, $\tout = 10$ and $\omega=\colthree{2.5}{2.1}{1.5}$}
	\label{fig:kg_ohyp}
\end{figure}
The hyperbolic $\mu > 0$ regime is shown in {\em log-log} plots of $\ln \bsq$ 
against $\ln p$ in \autoref{fig:kg_ohyp} and \autoref{fig:kg_thyp}. The first
figure presents three curves for different values of the expansion parameter
while the second uses different $\tout$. The values for the other parameters
involved are $m = 1$ and $\tin = 0$ with $\tout = 10$ for the first figure and
$\omega = 1.0$ for the second. We draw the conclusion that the value of
$\bsq$ in the region $q \ll \omega$ (delimited by the dashed line) increases
exponentially with both $\omega$ and $\tout$ (more precisely, with the 
expansion time), and is independent of $p$. In \autoref{sec:kg_asympt_low} we 
prove that this value is proportional to $\sinh{\nu \omega \Delta t}$
\eqref{eq:kg_bsqs}. The oscillations of the 
green curve $\omega = 1.5$ in the former figure are characteristic to the 
trigonometric regime \eqref{eq:kg_middle_sq_nmu}, while the 
declining lines are characteristic to the factor $(2\omega/p)^{2\nu}$ of
the hyperbolic regime \eqref{eq:kg_middle_sq_pmu_final}. From the latter 
figure we conclude that the
middle region $p \ll \omega \ll q$ is independent on the expansion time 
$\tout$, as shown in \autoref{sec:kg_asympt_middle}.
\begin{figure}[!ht]
	\centering
	\includegraphics[width=0.8\linewidth]{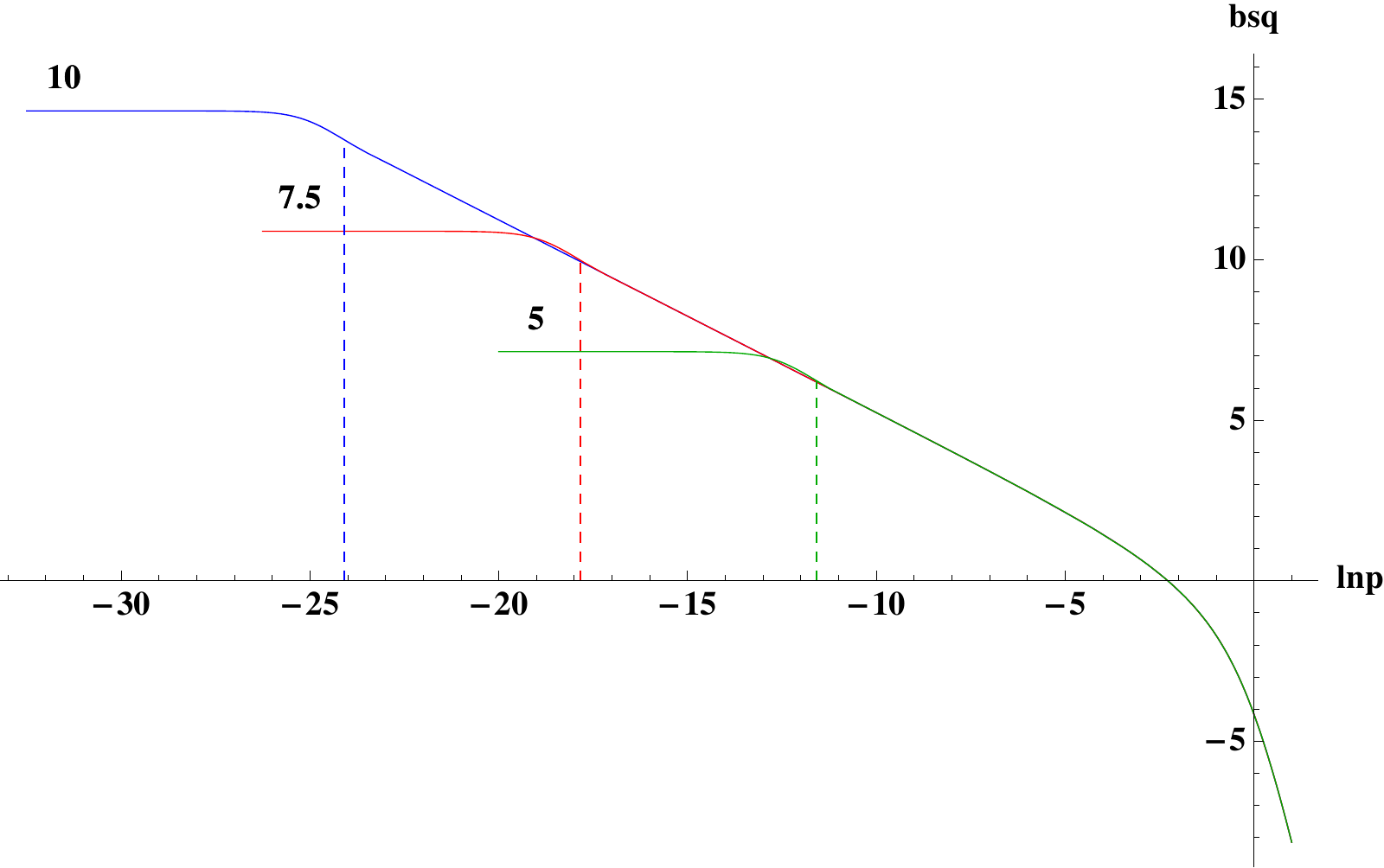}
	\caption{\small log-log plot $\tout=\colthree{10.0}{7.5}{5.0}$}
	\label{fig:kg_thyp}
\end{figure}

\begin{figure}[!ht]
	\centering
	\includegraphics[width=0.8\linewidth]{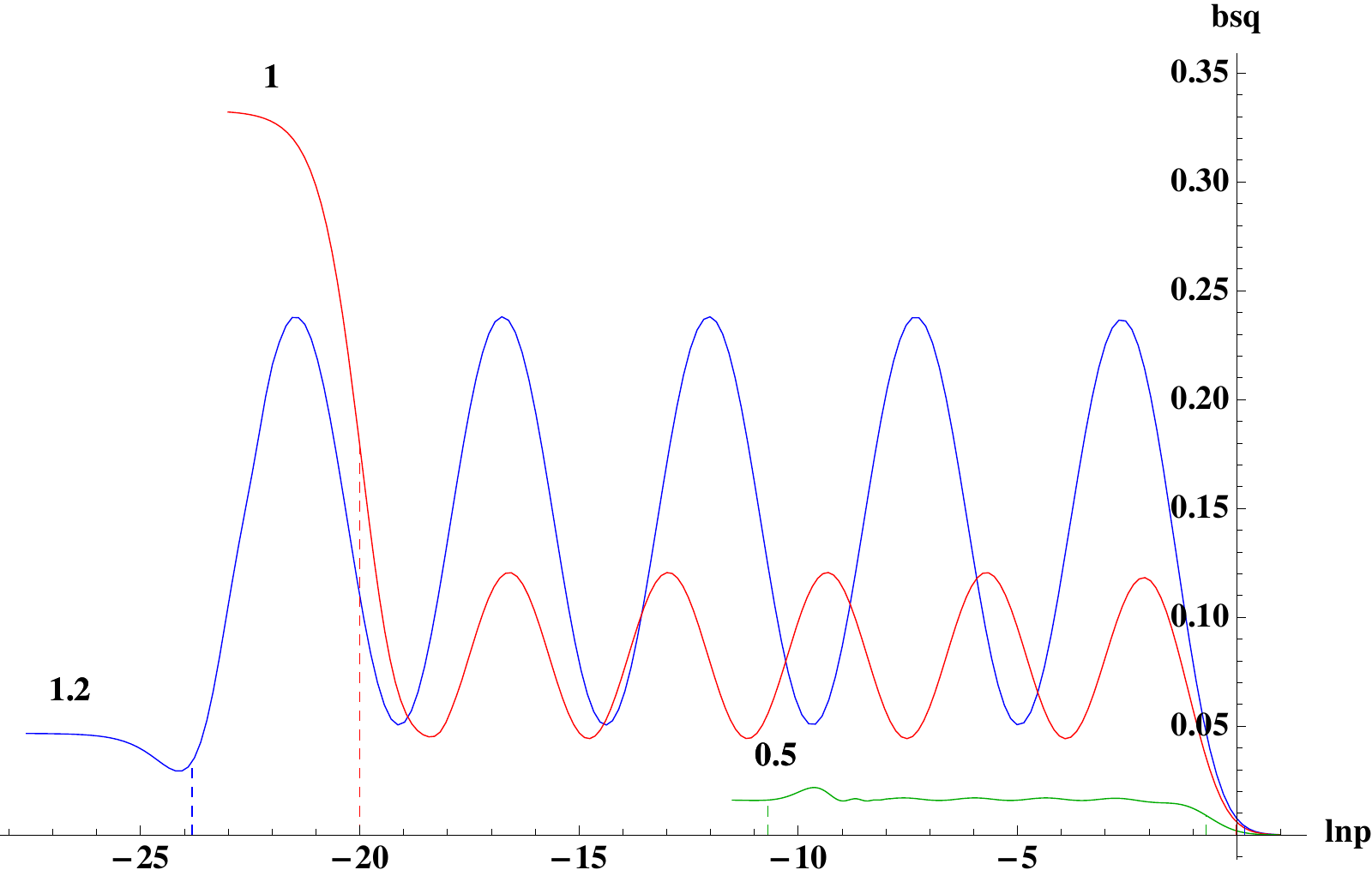}
	\caption{\small lin-log plot $\omega=\colthree{1.2}{1.0}{0.5}$}
	\label{fig:kg_otrig}
\end{figure}
Similar figures represent the trigonometric regime, \autoref{fig:kg_otrig}
for different expansion factors and \autoref{fig:kg_ttrig} for three
different values of $\tout$. The parameters
used are $m = 1$, $\tin = 0$, $\tout = 20$ for the former and $\omega = 1.0$ 
for the latter figure. The second figure confirms the result that the middle 
region is independent of the expansion time. Unlike the hyperbolic case, 
the value of $\bsq$ in the region $q\ll \omega$ oscillates
according to a factor $\sin{\nu \omega \Delta t}$ given in \eqref{eq:kg_bsqs}.
In the middle region we observ oscillations of constant amplitude about a constant 
value, which we find to be $m/2\omega\nu(\coth{\pi\nu} - \nu\omega/m)$. Both 
this value and the amplitude of the oscillations decrease as $\omega$
decreases ($\nu$ increases), the oscillations having a dampening factor 
proportional to $(\sinh{\pi\nu})^{-1}$ (see \eqref{eq:kg_middle_oscil_ampl}).
\begin{figure}[!ht]
	\centering
	\includegraphics[width=0.8\linewidth]{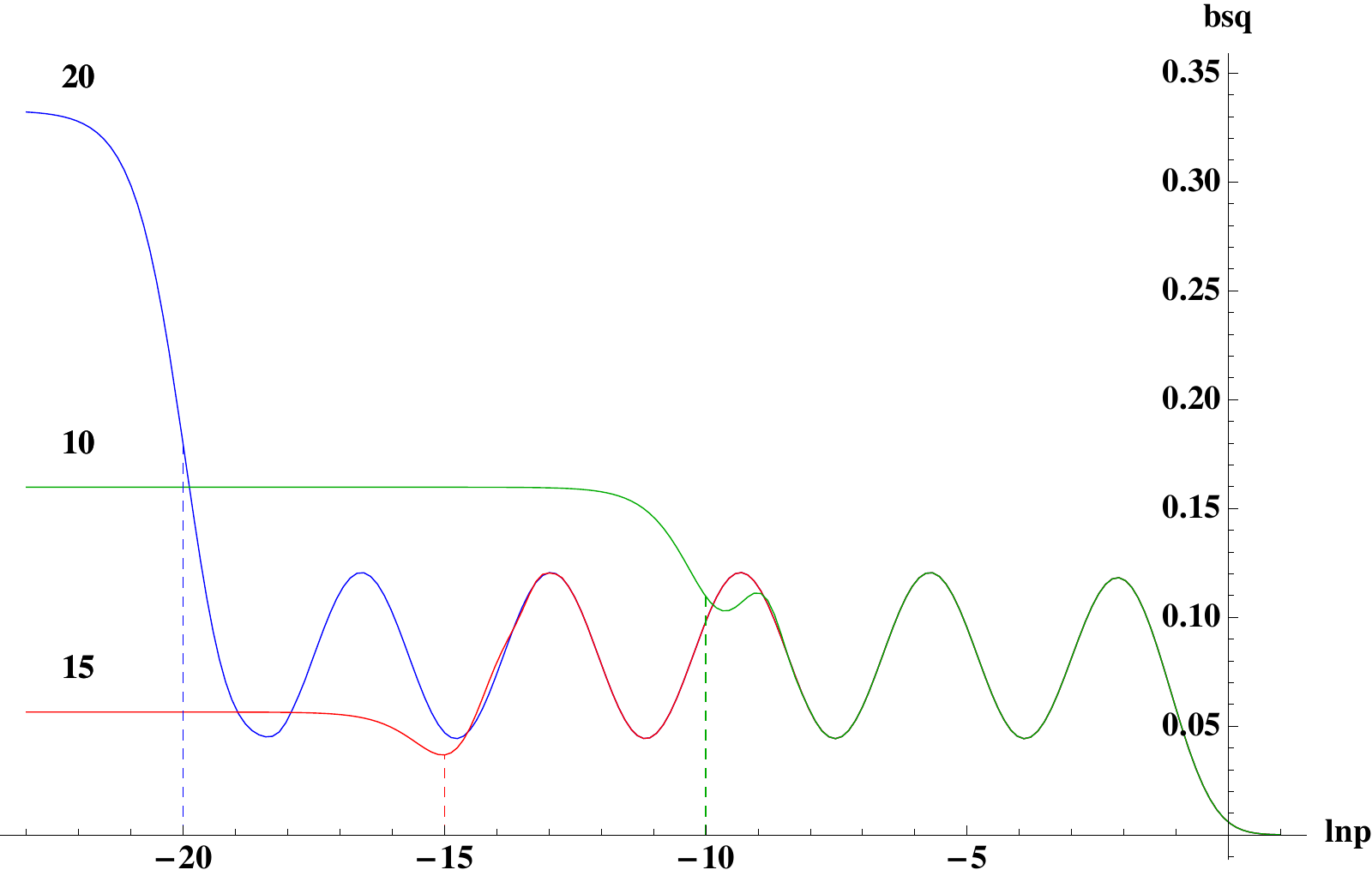}
	\caption{\small lin-log plot $\tout=\colthree{20.0}{15.0}{10.0}$}
	\label{fig:kg_ttrig}
\end{figure}

The fast decline in the ultraviolet region is given by a factor $1/p^6$. 
We illustrate this behaviour in
\autoref{fig:kg_aslarge_p6} where we represent $p^6 \bsq$ against $\ln p$ for
different values of $\omega$, using $m = 1, \tin = 0$ and $\tout = 10$. The
exact solution is checked against the asymptotic value $m^4\omega^2/16$, and
we observe an excellent agreement for sufficiently large $p$.
\begin{figure}[!ht]
	\centering
	\includegraphics[width=0.8\linewidth]{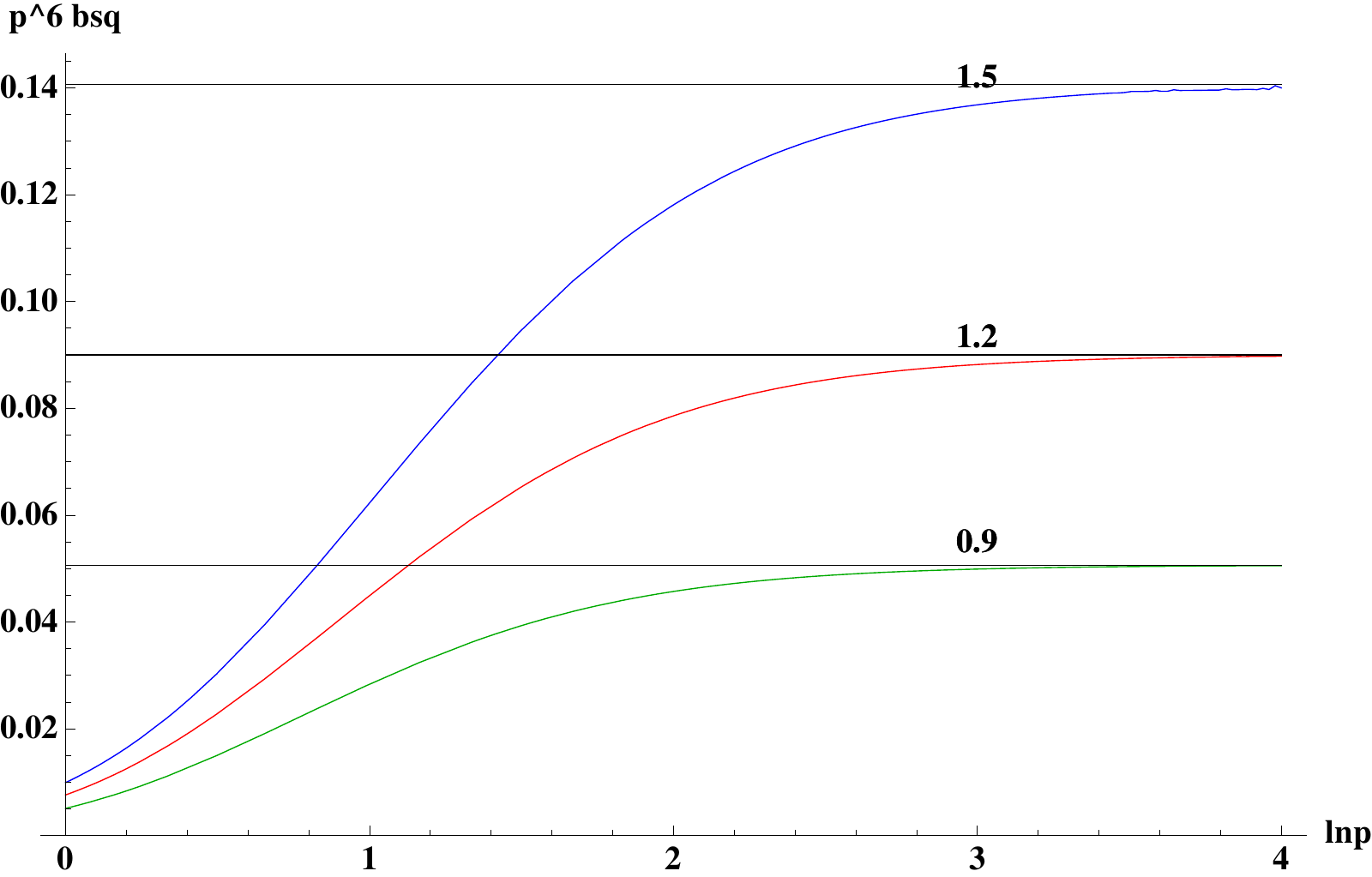}
	\caption{\small lin-log plot $\omega=\colthree{1.5}{1.2}{0.9}$}
	\label{fig:kg_aslarge_p6}
\end{figure}
\end{subsection}
\end{section}
\begin{section}[Asymptotic analysis]{Asymptotic analysis of the particle density}
\label{sec:kg_asympt}
Before delving into any algebra presented in this chapter, we recommend the 
reader to carefully read through \autoref{app:hankel} for insight in Hankel 
functions and their asymptotic forms. We recall some of the
notation used throughout this section:
\begin{gather*}
	M = m + 12\omega (\xi - 1/6),\qquad 
	\mu^2 = 1/4 - M^2,\qquad
	\nu = \sqrt{\abs{\mu}}q,\\
	Z_\nu\pparant \text{ is given in \eqref{eq:kg_Z_def}},\qquad
	\beta(p) \text{ is given in \eqref{eq:kg_betaAB}},\qquad
	Z_i(q,p) \text{ defined in \eqref{eq:kg_beta_Z}},\\
	\Gamma(p,q) = (2\omega/\sqrt{pq})^{\nu} \Gamma(\nu),\qquad
	\Gamma_\pm(p,q) = (2\omega/\sqrt{pq})^{\pm i\nu} \Gamma(\pm i\nu)
\end{gather*}
\begin{subsection}{Large momentum \texorpdfstring{$p \gg \omega$}{}}
\label{sec:kg_asympt_large}
The first trial of our theory of particle creation is to have a sufficiently
fast decay in the ultraviolet region for the particle number density. Since 
$n_p \sim p^2 \abs{\beta(p)}^2$, we require $\beta(p)$ to decay faster than
$p^{3/2}$. Furthermore, the energy density must also go to $0$, therefore we
require $\beta(p)$ to decay faster than $p^2$.

Although a conformally coupled scalar field satisfies the above 
requirements, a non-conformally coupled scalar field (and the spinorial field)
have a leading term of order $1/p^2$, which makes the energy density infinite.
Nevertheless, the total number of particles is finite and approaches a 
constant value for increasing expansion times.

We point out in this subsection that the behaviour of
$\abs{\beta}^2$ in the ultraviolet region is, for a conformally coupled scalar 
field, of order $1/p^6$, while for any other coupling it is of order
$1/p^4$.

First we shall perform a quick analysis on the behaviour of the coefficients
$\tilde{A}$ and $\tilde{B}$ defined by \eqref{eq:kg_junction_solAB_tilde} 
for large values of $p$. In order to estimate their order of magnitude, 
we will use the zeroth order approximation
of the Hankel functions \eqref{eq:bessel_all_as_largez}:
\begin{subequations}\label{eq:kg_large_AB}
\begin{align}
	\abs{\tilde{A}(p,t)}^2&\xrightarrow[p\rightarrow \infty]{} \frac{E_p}{4p}
	\left(\left(1 + \frac{p}{E_p}\right)^2 + \frac{\omega^2}{4E_p^2}\right)
	\label{eq:kg_large_A},\\
	\abs{\tilde{B}(p,t)}^2&\xrightarrow[p\rightarrow \infty]{} \frac{E_p}{4p}
	\left(\left(1 - \frac{p}{E_p}\right)^2 + \frac{\omega^2}{4E_p^2}\right)
	\label{eq:kg_large_B}.
\end{align}
\end{subequations}
We see that $A(p,t)$ approaches unity as $p\rightarrow E_p$, while $B$ is at 
most 
of the order of $p^{-1}$. In order to accurately capture the behaviour of the
$\beta$ coefficient for large $p$, we must expand up to order $1/p^3$. 
We shall use Hankel's expansion given in the appendix 
\eqref{eq:bessel_all_as_hankel_largez}. We note that the terms $Z_i$ given
by \eqref{eq:kg_beta_Z} take the following form:
\begin{equation}
	Z_i(q, p) = \frac{4\omega}{\pi\sqrt{pq}}\left(
	\cos\qmpparant C_i(q, p) + \sin\qmpparant S_i(q, p)
	\right).
	\label{eq:kg_large_CS_def}
\end{equation}
For brevity, we shall omit the arguments $(q, p)$ from the functions $C_i, S_i$,
which evaluate to:
\begin{subequations}\label{eq:kg_large_CS}
\begin{align}
	C_1 &= \phantom{-}Q_{q} P_{p} - P_{q} Q_{p}, &
	S_1 &= \phantom{-}P_{q} P_{p} + Q_{q} Q_{p}, 
	\label{eq:kg_large_CS1}\\
	C_2 &= \phantom{-}R_{q} P_{p} + S_{q} Q_{p}, &
	S_2 &= \phantom{-}R_{q} Q_{p} - S_{q} P_{p},
	\label{eq:kg_large_CS2}\\
	C_3 &= -P_{q} R_{p} - Q_{q} S_{p}, &
	S_3 &= -P_{q} S_{p} + Q_{q} R_{p}, 
	\label{eq:kg_large_CS3}\\
	C_4 &= -R_{q} S_{p} + S_{q} R_{p}, &
	S_4 &= \phantom{-}R_{q} R_{p} + S_{q} S_{p} .
	\label{eq:kg_large_CS4}
\end{align}
\end{subequations}
The polynomials $Q, P, R, S$ are given in the appendix by
\eqref{eq:P_hankel},\eqref{eq:Q_hankel},\eqref{eq:R_hankel},\eqref{eq:S_hankel}.
We have used the notation $P_q = P(\sqrt{\mu},q/\omega)$. Note that 
the order of the Hankel functions in $Z_{\nu}$ is imaginary for negative $\mu$,
in accord with our definition \eqref{eq:kg_Z_def}. However, the polynomials
are written using $\mu$, and not $\sqrt{\mu}$, thus $C_i$ and $S_i$ are all 
real functions.
Because the Minkowski energies $E_p, E_q$ can be expanded in a power 
series as
\begin{equation*}
	\frac{1}{E_p} = \frac{1}{p}
	\left(1 - \frac{m^2}{2p^2} + \mathcal{O}(p^{-4})\right),
\end{equation*}
it would not be correct to speak of orders of $p$ when analysing $\beta$. Instead, 
since the terms coming from the Hankel functions are of the form $\omega/p$,
we will discuss the coefficient in orders of $\omega$, such that
\begin{equation}
	\beta(p) = -\sqrt{\frac{E_p E_q}{4 pq}} e^{-i (E_q \tin + E_p \tout)} 
	\sum_{n = 0}^{\infty} \Omega^{(n)}(q, p) \omega^n,
	\label{eq:kg_large_Omegan_def}
\end{equation}
with $\sqrt{\mu}$ being treated as independent of $\omega$. 
The coefficients $\Omega^{(n)}$, up to order $3$, are given by
\begin{subequations}\label{eq:kg_large_Omegan}
\begin{align}
	\Omega^{(0)}(q, p) &= \frac{2}{m^2} \mathcal{C}(q, p)
	+ \mathcal{O}(p^{-4}),\label{eq:kg_large_Omega0}\\
	\omega\Omega^{(1)}(q, p) &= \frac{i \omega m^2 (4\mu - 1)}{16}
	\left(\frac{1}{p} - \frac{1}{q}\right) \mathcal{C}(q, p)
	+ \mathcal{O}(p^{-5}),\label{eq:kg_large_Omega1}\\
	\omega^2 \Omega^{(2)}(q, p) &= \frac{\omega^2 (4\mu - 1)}{8} \mathcal{C}(q, p)
	+ \mathcal{O}(p^{-4}),\label{eq:kg_large_Omega2}\\
	\omega^3 \Omega^{(3)}(q, p) &= \frac{i \omega^3 (4\mu - 1)^2}{64}
	\left(\frac{1}{p} - \frac{1}{q}\right) \mathcal{C}(q, p) -
	\frac{i\omega^3 (4\mu - 1)}{8} \times\nonumber\\
	&\left(
	\cos\qmpparant\left(\frac{1}{p^3} - \frac{1}{q^3}\right) - 
	i \sin\qmpparant\left(\frac{1}{p^3} + \frac{1}{q^3}\right)\right)
	+ \mathcal{O}(p^{-5}).
	\label{eq:kg_large_Omega3}
\end{align}
The reccurent term is
\begin{align}
	\mathcal{C}(q, p) = 
	\cos\qmpparant\left(\frac{1}{p^2} - \frac{1}{q^2}\right) - 
	i \sin\qmpparant\left(\frac{1}{p^2} + \frac{1}{q^2}\right).
	\label{eq:kg_large_Omegarecurrent}
\end{align}
\end{subequations}
The term $\mu - 1/4 = -m^2 / \omega^2 - 12(\xi - 1/6)$ magically appears in every
term of the expansion, since all higher orders of the polynomials $P,Q,R,S$ 
share a $4\mu - 1$ in their numerator. In the conformally coupled massles case,
this term is zero, and thus $\beta$ is 0, as expected (no particle creation for
a conformal field). 

What is indeed extraordinary is that the leading term in the conformally 
coupled case $\xi = 1/6$ is of order $p^{-3}$, while in any other coupling 
(including the minimal coupling $\xi = 0$), it is of order
$p^{-2}$:
\begin{multline}
	\beta(p) = \omega^2\frac{E_p E_q}{4 pq} e^{-i (E_q \tin + E_p \tout)} 
	\bigg\{ 12(\xi - 1/6) (\mathcal{C}(q, p) + \mathcal{O}(p^{-3})) -\\
	i \omega\left(\frac{m^2}{\omega^2} + 12(\xi - 1/6) \right) \left(
	\cos\qmpparant\left(\frac{1}{p^3} - \frac{1}{q^3}\right) - 
	i \sin\qmpparant\left(\frac{1}{p^3} + \frac{1}{q^3}\right)\right)\\
	 + \mathcal{O}(p^{-4})
	\bigg\}.
	\label{eq:kg_large}
\end{multline}
The square of this coefficient, in the minimally coupled case, is
\begin{equation}
	\abs{\beta(p)}^2_{\xi=0} \xrightarrow[p\gg \omega]{}
	\frac{\omega^4}{2 p^4} e^{-2\omega(\tout-\tin)}
	\left(\cosh{2\omega \Delta t} - \cos\frac{2(q-p)}{\omega} \right).
	\label{eq:kg_betasq_large_minimal}
\end{equation}
In the conformally coupled case we have
\begin{equation}
	\abs{\beta(p)}^2_{\xi=1/6} \xrightarrow[p\gg \omega]{}
	\frac{m^4\omega^2}{8 p^6} e^{-3\omega(\tout-\tin)}
	\left(\cosh{3\omega \Delta t} - \cos\frac{2(q-p)}{\omega} \right).
	\label{eq:kg_betasq_large_conf}
\end{equation}
With $\Delta t = \tout - \tin$ being the total time of expansion.
In the limit of large expansion times, we can neglect the cosine term while
the hyperbolic cosine can be replaced by $1/2$:
\begin{align}
	\bsq_{\xi=0} &\xrightarrow[\tout \gg \tin]{} 
	\frac{\omega^4}{4 p^4},
	\label{eq:kg_betasq_large_minimal_larget}\\
	\bsq_{\xi=1/6} &\xrightarrow[\tout \gg \tin]{} 
	\frac{m^4\omega^2}{16 p^6}.
	\label{eq:kg_betasq_large_conf_larget}
\end{align}
For future reference we shall denote the coefficient of the conformally 
coupled scalar field by
\begin{equation}
	\bsq_l \equiv \frac{m^4\omega^2}{16 p^6}.
	\label{eq:kg_bsql}
\end{equation}
A short note about different junction conditions: if we choose not to impose
continuity on $e^{\omega t} f_{\vp}^{\rm dS}$, but on a different 
function $e^{\omega t (1 - \alpha)} f_{\vp}^{\rm dS}$, then the 
$\beta$ coefficient will have a leading term of order $p^{-1}$,
\begin{equation}
	\Omega^{(1)}(q, p) = i \alpha\omega 
	\left(\cos{\qmpparant}\left(\frac{1}{p} - \frac{1}{q}\right) - 
	i \sin\qmpparant\left(\frac{1}{p} + \frac{1}{q}\right)\right),
	\label{eq:kg_large_Omega1_junc}
\end{equation}
which has the effect of producing an infinite number of particles per unit 
volume of space, since the spectral density of particles approaches a constant
in the ultraviolet region:
\begin{equation}
	n_p(x) \xrightarrow[\tout \gg \tin]{}  \frac{\alpha^2 \omega^2}{8\pi^2}.
\end{equation}
We note that this result is independent of the choice of the coupling $\xi$ 
with the gravitational field.
\end{subsection}
\begin{subsection}{Low momentum \texorpdfstring{$q \ll \omega$}{}}
\label{sec:kg_asympt_low}
In this section we point out that the particle number density 
approaches a constant value in the infrared region, meaning low energy
modes are equally populated by the expansion of space.

In order to investigate the infrared asymptotic behaviour of $\beta$ we need
to consider the approximation $q \ll \omega$. We will approximate the Hankel 
functions in \eqref{eq:kg_beta_Z} with their asymptotic forms for small
values of the argument given by \eqref{eq:bessel_all_as_smallz}, and their
derivatives with \eqref{eq:hankelp_as_smallz}. Note that since the order
of the functions is $\sqrt{\mu}$ with an unbounded $\mu$, we cannot restrict
ourselves to the term $2/z$ since there is a significant contribution 
comming from $z/2$ if the power $\sqrt{\mu}$ is not real.

In this approximation, the real terms $Z_i$ given by \eqref{eq:kg_beta_Z} 
read
\begin{subequations}\label{eq:kg_small_Z}
\begin{align}
	Z_1 &= \phantom{-}\frac{4}{\pi\snu}\sinh{\snu\omega\Delta t}, &
	Z_2 &= \phantom{-}\frac{4 \omega}{q\pi}\cosh{\snu\omega\Delta t},\\
	Z_4 &= -\frac{4\snu\omega^2}{\pi p q}\sinh{\snu\omega\Delta t},&
	Z_3 &= -\frac{4 \omega}{p\pi}\cosh{\snu\omega\Delta t}. 
\end{align}
\end{subequations}
With this we evaluate the $\beta$ coefficient to
\begin{multline}
	\beta(p) \xrightarrow[q\ll \omega]{} -\frac{\sinh{\snu\omega\Delta t}}{2\snu}
	e^{-i (E_q \tin + E_p \tout)} \times\\
	\left\{\frac{1}{2}\left(\sqrt{\frac{E_q}{E_p}}+ 
	\sqrt{\frac{E_p}{E_q}}\right) + 
	\snu \coth{\snu\omega\Delta t} 
	\left(\sqrt{\frac{E_p}{E_q}} - \sqrt{\frac{E_q}{E_p}}\right) + 
	\right.\\\left.
	i \frac{\sqrt{E_p E_q}}{\omega} \left(-1 + 
	\frac{m^2}{E_pE_q} + \frac{12\omega^2}{E_pE_q}\left(\xi - 1/6\right)
	\right)\right\}.
	\label{eq:kg_small}
\end{multline}
Since we have not retained any powers of $p$ in the above approximation, we 
shall replace $E_p$ and $E_q$ with $m$. Higher order terms will
not contribute correctly, because we have suppressed potentially balancing 
terms. Using $1/4 - \mu = m^2/\omega^2 + 12(\xi - 1/6)$
we arrive at the following form for the $\beta$ coefficient:
\begin{equation}
	\beta(p) \xrightarrow[q\ll \omega]{} -\frac{\sinh{\snu\omega\Delta t}}{2\snu}
	e^{-i (E_q \tin + E_p \tout)} \left(1 - 12 i \left(\xi - 1/6\right)\right).
	\label{eq:kg_small_smallp}
\end{equation}	
Again, in the conformally coupled case the production of particles is at a 
minimum. We give the two values for the square of $\beta$ in the minimally 
and conformally coupled cases:
\begin{align}
	\bsq_{\xi=0} &\xrightarrow[q\ll \omega]{} 
	\frac{5\sinh^2{\snu\omega\Delta t}}{16\mu},
	\label{eq:kg_betasq_small_minimal}\\
	\bsq_{\xi=1/6} &\xrightarrow[q\ll \omega]{} 
	\frac{\sinh^2{\snu\omega\Delta t}}{4\mu}.
	\label{eq:kg_betasq_small_conf}
\end{align}
We identify two different regimes: if $\mu = 1/4 - m^2/\omega^2 - 12\xi + 2$
is pozitive (small mass or large expansion rate), $\beta$ increases 
exponentially as the expansion rate increases. In the case
$\mu < 0$, the hyperbolic functions turn into trigonometric ones, and 
$\abs{\beta(p)}^2$ oscillates as the expansion time increases. 
The massless case is not correctly captured by this approximation, since
we have replaced $E_p$ and $E_q$ by $m$, and neglected the difference 
$E_p - E_q$. If we neglect terms of order $p$ in \eqref{eq:kg_small} we obtain 
\begin{equation}
	\beta(p)_{\substack{\xi=1/6\\m = 0}} \xrightarrow[q\ll \omega]{} 0.
\end{equation}
The equality is not exact (as shown in \autoref{sec:kg_massless})
because of an inaccurate treatment of higher order
terms in the approximation used.
We note that if the massless field is not conformally coupled, $\beta(p)$ is 
infinite for $p\rightarrow 0$ because of the coefficient of $\xi-1/6$:
\begin{equation}
	\beta(p) \xrightarrow[\substack{q\ll \omega\\m = 0}]{} 
	-\frac{\sinh{\snu\omega\Delta t}}{2\snu} e^{-i (E_q \tin + E_p \tout)} 
	\frac{12 i \omega}{\sqrt{qp}} \left(\xi - 1/6\right) + \mathcal{O}(p^0).
	\label{eq:kg_small_nonc_m0}
\end{equation}
We denote the asymptotic form of $\bsq$ in the conformally coupled case by 
\begin{equation}
	\bsq_s = \frac{\sinh^2{\snu\omega\Delta t}}{4\mu}.
	\label{eq:kg_bsqs}
\end{equation}
\end{subsection}
\begin{subsection}{Middle region
\texorpdfstring{$p \ll \omega \ll q$}{}}
\label{sec:kg_asympt_middle}
The ubiquitous thermal spectrum of the particle density created by the
de Sitter space is not recovered. Not only does it not emerge for large 
$\omega$, but even when $\omega$ is sufficiently
small, there is a polynomial correction to the thermal Bose-Einstein 
distribution, which is dominant for large masses (small expansion parameter).

In the middle region we shall use the small argument approximation for
the Hankel functions of argument $p/\omega$, and the large argument 
approximation for those of $q/\omega$. We shall use first order 
approximations, and we shall substitute directly in the expressions for the 
$A,B$ coefficients \eqref{eq:kg_junction_solAB}, because the asymmetry of the 
approximations for $p$ and $q$ makes the use of the general formula for $\beta$
\eqref{eq:kg_beta_explicit} unnecessarily cumbersome.

In \autoref{sec:kg_asympt_large} we have investigated the behaviour of the 
coefficient $A$ and $B$ for large values of the argument, and have concluded
that the $B(q,t)$ coefficient drops like $1/q$ \eqref{eq:kg_large_AB}. 
Since the domain of interest is for increasing values of $p$ (and $q$), we 
shall not delve on accurate 
representations of the small momentum domain, and instead use the approximation
$B(q, t) \simeq 0$. Therefore $\beta$ reduces to
\begin{equation*}
	\beta(p) \xrightarrow[p \ll \omega \ll q]{} 
	\tilde{A}(q,\tin)\tilde{B}(p,\tout).
\end{equation*}
Substituting \eqref{eq:kg_large_A} for $\tilde{A}(q,\tin)$, and computing 
$\tilde{B}(p, \tout)$ by substituting \eqref{eq:bessel_all_as_smallz} and
\eqref{eq:hankelp_as_smallz} for the Hankel functions and their derivatives,
we arrive at
\begin{multline}
	\beta(p) \xrightarrow[p\ll\omega\ll q]{}
	-\sqrt{\frac{\pi E_p E_q}{8\omega q}} e^{-i(E_q \tin + E_p \tout)} 
	e^{-i \frac{q}{\omega} + \frac{i\pi\snu}{2} + \frac{i\pi}{4}} 
	\left\{ \left(\frac{2\omega}{p}\right)^{\snu} \frac{\Gamma(\snu)}{i\pi}
	\left(1 - \frac{i\nu\omega}{E_p} + \frac{i\omega}{2E_p}\right) + 
	\right.\\\left.
	\left(\frac{p}{2\omega}\right)^{\snu} 
	\frac{1+i\cot{\pi\snu}}{\snu \Gamma(\snu)}
	\left(1 + \frac{i\nu\omega}{E_p} + \frac{i\omega}{2E_p}\right)\right\}.
	\label{eq:kg_middle_snu}
\end{multline}
In order to compute $\bsq$ we must consider separately the cases $\mu > 0$ 
and $\mu < 0$. If $\mu > 0$ (thus $\snu$ is real), we have $\snu = \nu$ and 
the dominant 
term is $1/p^\nu$. Although we can discard the higher order terms for the
purpose of this analysis, they are not negligible near the region 
$p\sim \omega$,
important for the analysis of the total number of created particles. Moreover,
the free term uncovers a thermal factor corresponding to an imaginary energy. 
We write the square of \eqref{eq:kg_middle_snu} in this approximation:
\begin{multline}
	\abs{\beta(p)}^2_{\mu>0} \xrightarrow[p \ll \omega \ll q]{}
	\frac{E_pE_q}{2\omega\nu q}\left(\cot{\pi\nu} \left(\frac{1}{2} + 
	\frac{M^2}{2E_p^2}\right) + \frac{\nu\omega}{E_p}\right) +
	\frac{\omega E_q}{8\pi E_p q} \Bigg\{\\
	\Gamma(p,p)
	\left(\frac{E_p^2}{\omega^2} + \left(\frac{1}{2}-\nu\right)^2\right) + 
	\frac{\pi^2}{\nu^2 \sin^2{\pi\nu} \Gamma^2(p,p)}
	\left(\frac{E_p^2}{\omega^2} + \left(\frac{1}{2}+\nu\right)^2\right)\Bigg\}.
	\label{eq:kg_middle_sq_pmu}
\end{multline}
We have used the notation
\begin{equation}
	\Gamma(q,p) = \left(\frac{2\omega}{\sqrt{qp}}\right)^{\nu} \Gamma(\nu).
	\label{eq:kg_gamma_def}
\end{equation}
In the region where $E_p \sim m$, and for a conformally coupled scalar field
($M=m$), the first term takes the form
\begin{equation*}
	-\frac{m}{2\omega\nu}\left(\cot{\pi\nu} + \frac{\nu\omega}{m}\right).
\end{equation*}
Assuming that a Taylor series expansion for $\nu$ about $\omega/m = 0$ 
makes sens (which it doesn't, since the case $\mu > 0$ implies 
$m/\omega < 1/2$), and ignoring the apparent complex (i.e. non-real) nature
of the result, this term gives a thermal factor (plus a polynomial
correction):
\begin{equation}
	\frac{1}{e^{2\pi i \nu} - 1} + \frac{\omega^2}{16m^2} - 
	\frac{\omega^2}{8m^2} \frac{1}{e^{2\pi i \nu} - 1} + \mathcal{O}\left(
	\frac{\omega^4}{m^4}\right).
	\label{eq:kg_middle_pmu_pseudothermal_factor}
\end{equation}
We note that this factor contributes significant corrections for 
$\nu \lesssim 1/4$, and in the region $p \sim \omega$, therefore we shall 
include it in the final form:
\begin{multline}
	\bsq_{\substack{\xi=1/6\\\mu>0}} \xrightarrow[p\ll\omega\ll q]{} \bsq_m = 
	-\frac{m}{2\omega\nu}\left(\cot{\pi\nu} + \frac{\omega\nu}{m}\right)+
	\frac{\omega}{8\pi m} \Bigg\{\\
	\Gamma^2(p,p)
	\left(\frac{m^2}{\omega^2} + \left(\frac{1}{2}-\nu\right)^2\right) + 
	\frac{\pi^2}{\nu^2 \sin^2{\pi\nu} \Gamma^2(p,p)}
	\left(\frac{m^2}{\omega^2} + \left(\frac{1}{2}+\nu\right)^2\right)\Bigg\}.
	\label{eq:kg_middle_sq_pmu_final}	
\end{multline}

Let us turn to the case $\mu < 0$, $\snu = i \nu$. In the square of $\beta$ we
shall encounter an approximately constat term which will give the thermal 
character about which a second term oscillates, whose amplitude decreases as
the expansion factor decreases, or the mass increases.
Using
\begin{equation*}
	\abs{\Gamma(i\nu)}^2 = \frac{\pi}{\nu \sinh{\pi\nu}},
\end{equation*}
the constant term writes
\begin{equation*}
	T_c = \frac{E_p}{4\nu\omega}\left(\coth{\pi\nu}\left(1 + \frac{M^2}{E_p^2}\right)
	- \frac{2\nu\omega}{E_p} + \dots\right).
\end{equation*}
while the oscillating term is
\begin{equation*}
	T_o = \frac{E_p}{8\pi\omega}\left\{\left(1 - \frac{M^2}{E_p^2} + 
	\frac{\omega^2}{2E_p^2}\right)\left(\Gamma_+^2(p,p) + \Gamma_-^2(p,p)\right)-
	\frac{i \omega^2 \nu}{E_p^2} 
	\left(\Gamma_+^2(p,p) - \Gamma_-^2(p,p)\right)\right\}.
\end{equation*}
We have used the notation
\begin{equation}
	\Gamma_{\pm}(p,q) = \left(\frac{2\omega}{\sqrt{pq}}\right)^{\pm i\nu} \Gamma(\pm i\nu)
	\,,\,
	\Gamma_{+}^*(p,q) = \Gamma_{-}(p,q) \,,\,
	\abs{\Gamma_{+}(p,q)}^2 = \frac{\pi}{\nu \sinh{\pi \nu}}.
	\label{eq:kg_gammapm_def}
\end{equation}
In the conformally coupled case, we make the approximation $E_p \simeq M = m$,
and obtain:
\begin{align*}
	T_c &= \frac{m}{2\nu\omega}\left(\coth{\pi\nu} - \frac{\nu\omega}{m}\right),\\
	T_o &= \frac{\omega}{16\pi m}\left\{
	\left(\Gamma_+^2(p,p) + \Gamma_-^2(p,p)\right) - 
	2 i \nu\left(\Gamma_+^2(p,p) - \Gamma_-^2(p,p)\right)\right\}.
\end{align*}
For large masses (or small expansion factors), we can obtain a term resembling
the Bose-Einstein distribution function, by expanding $\nu$ in a power series
about $\omega/m = 0$:
\begin{equation}
	T_c = \frac{1}{e^{2\pi\nu} - 1}
	+ \frac{\omega^2}{16m^2} + \mathcal{O}\left(\frac{\omega^4}{m^4}\right).
	\label{eq:kg_middle_nmu_pseudothermal}
\end{equation}
However, this distribution does not have the characteristic of a Bose-Einstein
decaying exponential, since for large $m$ the polynomial term is dominant.

In order to determine the amplitude of the oscillatory term $T_o$ we define 
$\theta_{\Gamma}$ such that
\begin{equation}
	\Gamma_\pm(p,p)^2 = \frac{\pi}{\nu \sinh{\pi \nu}} e^{2i \theta_{\Gamma}},
\end{equation}
since $\abs{\Gamma_+(p,p)}^2 = \pi/\nu \sinh{\pi \nu}$. In this notation,
the oscillatory term writes
\begin{equation}
	T_o = \frac{1}{4 \nu\sinh{\pi\nu}}\left(
	\frac{1}{\sqrt{1+4\nu^2}}\cos{2\theta_\Gamma} + \frac{2\nu}{\sqrt{1 + 4\nu^2}}
	\sin{2\theta_\Gamma}\right).
	\label{eq:kg_middle_oscil_ampl}
\end{equation}
This term exponentially approaches $0$ with increasing $\nu$, but becomes large
at $\nu \rightarrow 0$ (the result is not accurate for $\nu = 0$ since the 
expansion we used for low arguments is not valid for Hankel functions of 
order $0$). We shall keep the oscillatory term in the final form of $\bsq$:
\begin{multline}
	\qquad\bsq_{\substack{\xi=1/6\\\mu<0}}\xrightarrow[p\ll\omega\ll q]{}\bsq_m=
	\frac{m}{2\omega\nu} \left(\coth{\pi\nu} - \frac{\nu\omega}{m}\right) +\\
	\frac{\omega}{8\pi m}\left\{\frac{1}{2}
	\left(\Gamma_+^2(p,p) + \Gamma_-^2(p,p)\right) - 
	i \nu\left(\Gamma_+^2(p,p) - \Gamma_-^2(p,p)\right)\right\}.\qquad
	\label{eq:kg_middle_sq_nmu}
\end{multline}
The behaviour of $\bsq$ in the middle region $p\ll \omega \ll q$ is unrelated
to $\tin, \tout$.
\end{subsection}
\begin{subsection}{The number of created particles}\label{sec:kg_asympt_N}
The asymptotic analysis of the previous section enables us to approximate 
$\bsq$ to a good degree of accuracy on the entire domain $p\in [0,\infty)$.
However, even though the approximation approaches the form of the exact 
solution, the function $p^2 \bsq$ is notably different because the relevant
region of integration is near $p\sim \omega$, where none of the above 
approximations are valid.

The difficult part in obtaining integrals of $\bsq$ is choosing the right
ranges for the asymptotic forms. The delicate areas are the borders of 
the asymptotic regions, namely $q \sim \omega$ and $p \sim \omega$. The 
solution is to define
\begin{equation}
	\bsq_{as} = \begin{cases}
		\bsq_s & q < \omega,\\
		\bsq_m & \omega < q \text{ and } p < p_\nu,\\
		\bsq_l & p > p_\nu.
	\end{cases}
	\label{eq:kg_bsqas}
\end{equation}
The choice for the first branch is natural, since $\bsq_s$ is constant (see 
\eqref{eq:kg_bsqs}), and $\bsq_m$ is either decreasing (in the hyperbolic case,
see \eqref{eq:kg_middle_sq_pmu_final}) or oscillating about a constant value 
(in the trigonometric case, see \eqref{eq:kg_middle_sq_nmu}), therefore there
is no risc of having the asymptotic form increase too much. Inevitably, there
will be a region where this approximation is not accurate. On the other hand,
the choice of the second point $p_\nu$ is not straightforward. Although the
asymptotic form for large arguments $\bsq_l$ given by \eqref{eq:kg_bsql} is
monotonic and decreasing, it approaches infinity as $p$ approaches $0$, while
$\bsq_m$ goes to large values when $p > \omega$. In order to solve this 
difficulty we choose $p_\nu$ such that 
\begin{equation}
	\bsq_m(p_\nu) = \bsq_l(p_\nu).
	\label{eq:kg_pnudef}
\end{equation}
We have to analyze both the hyperbolic and the trigonometric case. 
Unfortunately, the complex behaviour of $\bsq_m$ outside the scope of their
definition ($p \sim \omega$) makes the equation \eqref{eq:kg_pnudef} 
unsolvable. In the hyperbolic case we choose, by trial and error,
$p_\nu = \sqrt{2m\omega/3}$. For the trigonometric case we consider the constant
term minus the amplitude of the oscillations (since $\bsq$ is decreasing below
this value at a fast rate), and arrive at
\begin{equation}
	p_\nu = \begin{cases}
		\sqrt{2m\omega/3} & \mu > 0,\\
		{\displaystyle \sqrt{\frac{m\omega}{2}} \nu^{1/6} (\coth{\pi\nu} - 
		\frac{\omega\nu}{m} - \frac{\omega}{2m\sinh{\pi\nu}})^{-1/6}}
		& \mu < 0.
	\end{cases}
	\label{eq:kg_pnusol}
\end{equation}
We write the particle number space density as the integral of the particle 
number density with magnitude $p$ given in \autoref{eq:kg_bogo_np}, which we
split according to the piecewise definition of our asymptotic form 
\eqref{eq:kg_bsqas}:
\begin{gather*}
	n(\vx) = \frac{1}{2\pi^2} \left(I_s + I_m + I_l\right),\\
	I_s = \int_0^{q=\omega} dp\, p^2 \bsq_s ,\qquad
	I_m = \int_{q=\omega}^{p=p_\nu} dp\, p^2 \bsq_m ,\qquad
	I_l = \int_{p_\nu}^{\infty} p^2 \bsq_l.
\end{gather*}
We have used $q = \omega$ as a shorthand for $p = \omega e^{-\omega \Delta t}$.
$I_s$ and $I_l$ evaluate to
\begin{align}
	I_s = \frac{\sinh^2{\sqrt{\mu}\omega\Delta t}}{12\mu} \omega^3 
	e^{-3\omega\Delta t} ,\qquad
	I_l = \frac{m^4 \omega^2}{48} \frac{1}{p_\nu^3}.
	\label{eq:kg_N_Isl}
\end{align}
In the middle region we need to integrate $\Gamma^{\pm 2}(p, p)$ (defined 
by \eqref{eq:kg_gamma_def} for the hyperbolic case) and 
$\Gamma^2_{\pm}(p,p)$ (defined by \eqref{eq:kg_gammapm_def} for the 
trigonometric case), for which we find the results:
\begin{align*}
	\int_{q=\omega}^{p=p_\nu} dp\, p^2 \Gamma^{\pm 2}(p,p) &= 
	\frac{1}{3\mp 2\nu} \left(p_\nu^3 \Gamma^{\pm 2}(p_\nu, p_\nu) - 
	\omega^3 e^{-3\omega \Delta t} \Gamma^{\pm 2}(\omega e^{-\omega \Delta t},
	\omega e^{-\omega \Delta t})\right),\\
	\int_{q=\omega}^{p=p_\nu} dp\, p^2 \Gamma^2_{\pm}(p,p) &= 
	\frac{1}{3\mp 2i \nu} \left(p_\nu^3 \Gamma^2_{\pm}(p_\nu, p_\nu) - 
	\omega^3 e^{-3\omega \Delta t} \Gamma^2_{\pm}(\omega e^{-\omega \Delta t},
	\omega e^{-\omega \Delta t})\right).
\end{align*}
After neglecting the terms $e^{-\omega \Delta t}$, we arrive at
\begin{multline}
	n^{\text{hyp}}(\vx) = \frac{1}{2\pi^2} \Bigg\{ 
	\frac{m^4 \omega^2}{48 p_\nu^3} - 
	\frac{m p_\nu^3}{6\omega\nu} \left(\cot{\pi\nu + \omega\nu/m}\right)+\\
	\frac{\omega p_\nu^3}{8\pi m} \left[
	\frac{\Gamma^2(p_\nu,p_\nu)}{3 - 2\nu} \left(m^2/\omega^2 + 
	\left(1/2 - \nu\right)^2\right) + \frac{\pi^2}{\nu^2 \sin^2{\pi\nu}}
	\frac{\Gamma^{-2}(p_\nu, p_\nu)}{3 + 2\nu} \left(m^2/\omega^2 + 
	\left(1/2 + \nu\right)^2\right)\right] \Bigg\},\\	
	\shoveleft{n^{\text{trig}}(\vx) = \frac{1}{2\pi^2} \Bigg\{ 
	\frac{m^4 \omega^2}{48 p_\nu^3} + 
	\frac{m p_\nu^3}{6\omega\nu} \left(\coth{\pi\nu} - \omega\nu/m\right)+}\\
	\frac{\omega p_\nu^3}{8\pi m} \left[
	\Gamma^2_+(p_\nu,p_\nu)\frac{1/2 - i\nu}{3 - 2i\nu} +
	\Gamma^2_-(p_\nu,p_\nu)\frac{1/2 + i\nu}{3 + 2i\nu}\right] \Bigg\}.
	\label{eq:kg_N_as}
\end{multline}
We conclude that if the time interval 
$\Delta t = \tout - \tin$ is sufficiently large the particle number density
approaches a constant value. In order to understand the dependence on 
$m$ and $\omega$, which is highly dependent on the choice of $p_\nu$, we 
take the extreme cases $m/\omega \rightarrow 0$ (hyperbolic regime), and
$\omega/m \rightarrow 0$ (trigonometric regime).

In the first case we can approximate $\nu = 1/2$ and get
\begin{equation}
	n^{\text{hyp}}(\vx) \xrightarrow[m/\omega \rightarrow 0]{} 
	\frac{1}{2\pi^2} \left\{ \frac{m \omega^2}{36} - 
	\frac{\sqrt{6}}{27} (m\omega)^{3/2} + \frac{\omega m^2}{12} + 
	\frac{\sqrt{6}}{64} m^2 (m\omega)^{1/2}\right\}.
	\label{eq:kg_N_largeo}
\end{equation}
The leading term is quadratic in $\omega^2$, but it becomes dominant only for
large $\omega/m$. 
Since $m$ appears in each of the above terms, we conclude that there is
no particle production in the masless case:
\begin{equation}
	n^{\text{hyp}}(\vx) \xrightarrow[m = 0]{} 0.
\end{equation}
The asymptotic $m$ dependence can be investigated by letting 
$\omega/m \rightarrow 0$ (trigonometric regime). In this limit we can use:
\begin{gather*}
	\nu \simeq \frac{m}{\omega}\left(1 - \frac{\omega^2}{8m^2}\right),\qquad
	\frac{1}{\nu} \simeq \frac{\omega}{m}\left(1 + \frac{\omega^2}{8m^2}\right),
	\qquad \coth{\pi\nu} \simeq 1,\\
	\sinh{\pi\nu} \simeq e^{\pi\nu},\qquad
	\Gamma_{\pm}^2(p_\nu,p_\nu) = \frac{1}{\nu\sinh{\pi\nu}} 
	e^{\pm 2i \theta_{\Gamma}} \simeq 0\\
	p_{\nu} \simeq \sqrt{\frac{m\omega}{2}} (1/\nu - \omega/m)^{-1/6} \simeq m
\end{gather*}
With which we get
\begin{equation}
	n^{\text{trig}}(\vx) \xrightarrow[\omega/m \rightarrow 0]{} 
	\frac{m \omega^2}{48 \pi^2}. \label{eq:kg_N_largem}
\end{equation}
This dependency is similar to that for large $\omega/m$, up to a factor $3/2$.
We emphasize again that these results depend strongly on the choice of $p_\nu$,
which we have chosen rather empirically in the hyperbolic case. This term 
becomes quickly dominant and reproduces remarkably well the exact result. 
From the above formula we can conclude that there is no particle production for 
$\omega = 0$:
\begin{equation}
	n^{\text{trig}}(\vx) \xrightarrow[\omega = 0]{} 0.
\end{equation}
\end{subsection}
\begin{subsection}{Graphical comparison to the exact solution}\label{sec:kg_asympt_fig}
The piecewise definition of the asymptotic form of $\bsq$ \eqref{eq:kg_bsqas}
can be used to approximate $\bsq$ on the entire domain $p\in[0,\infty)$. This
approximate form can be used for the computation of the particle number
density $n(\vx)$. However, we expect some differences to occur because of the
inaccuracy of the approximation near the delimiters $q \sim\omega$ and
$p\sim\omega$.

\begin{figure}[ht]
	\centering
	\includegraphics[width=0.8\linewidth]{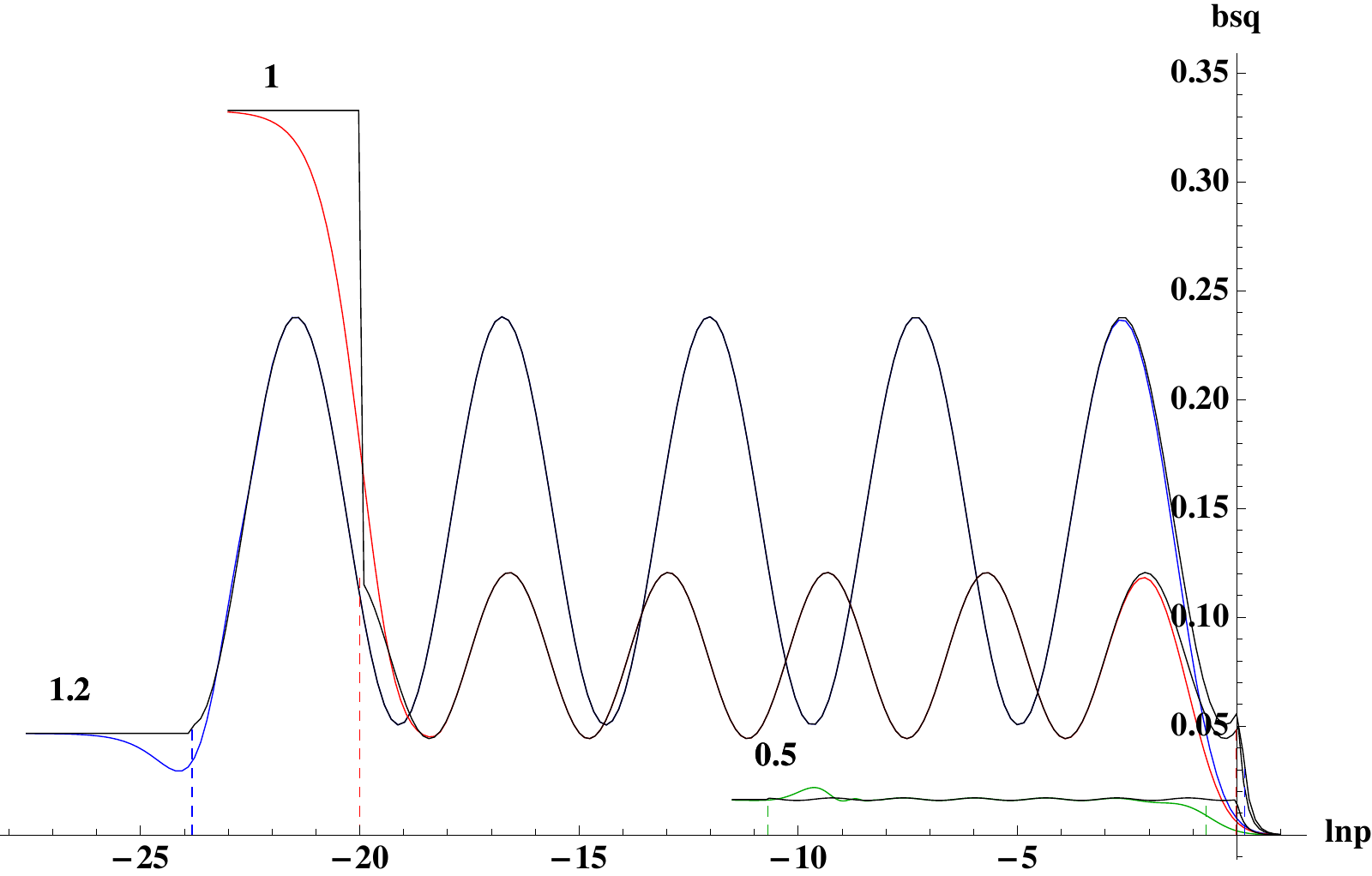}
	\caption{\small lin-log plot $\omega=\colthree{1.2}{1.0}{0.5}$}
	\label{fig:kg_asotrig}
\end{figure}
The overlap between the analytical solution and the corresponding asymptotic
forms is given in \autoref{fig:kg_asotrig} (compare to \autoref{fig:kg_otrig})
and \autoref{fig:kg_asohyp}(compare to \autoref{fig:kg_ohyp}). The most 
striking disagreement is near $q \lesssim \omega$ (before the dashed lines),
where the asymptotic form for low arguments is no longer accurate. There is 
another disagreement near $p \sim \omega$, where the asymptotic forms 
corresponding to large $p$ and small $p$ respectivelly start to increase (the 
former like $1/p^6$), while $\bsq$ enters a decline.
\begin{figure}[ht]
	\centering
	\includegraphics[width=0.8\linewidth]{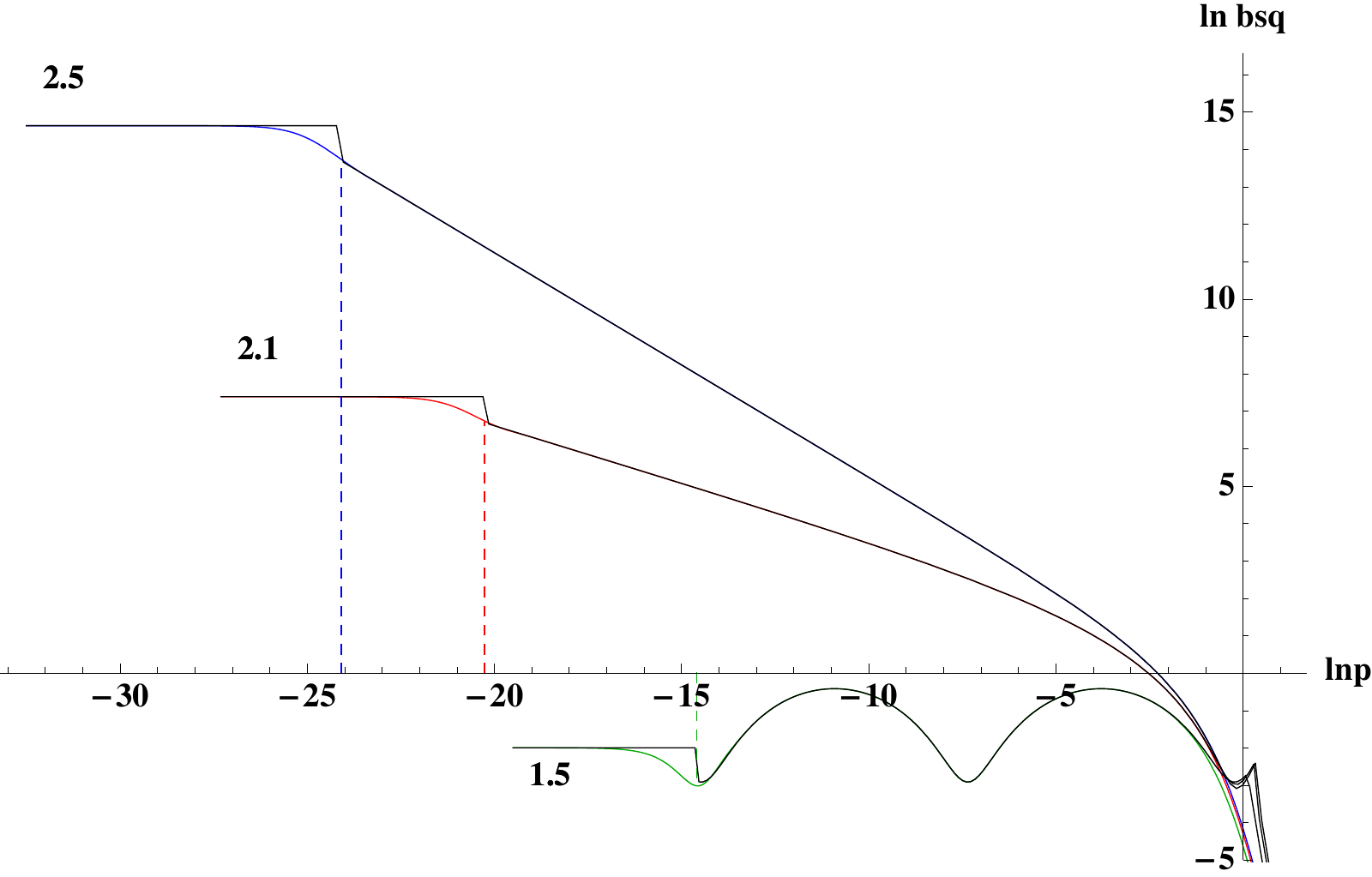}
	\caption{\small log-log plot $\omega=\colthree{2.5}{2.1}{1.5}$}
	\label{fig:kg_asohyp}
\end{figure}

\begin{figure}[!ht]
  \begin{minipage}[b]{0.47\linewidth} 
    \centering
    \includegraphics[width=\linewidth]{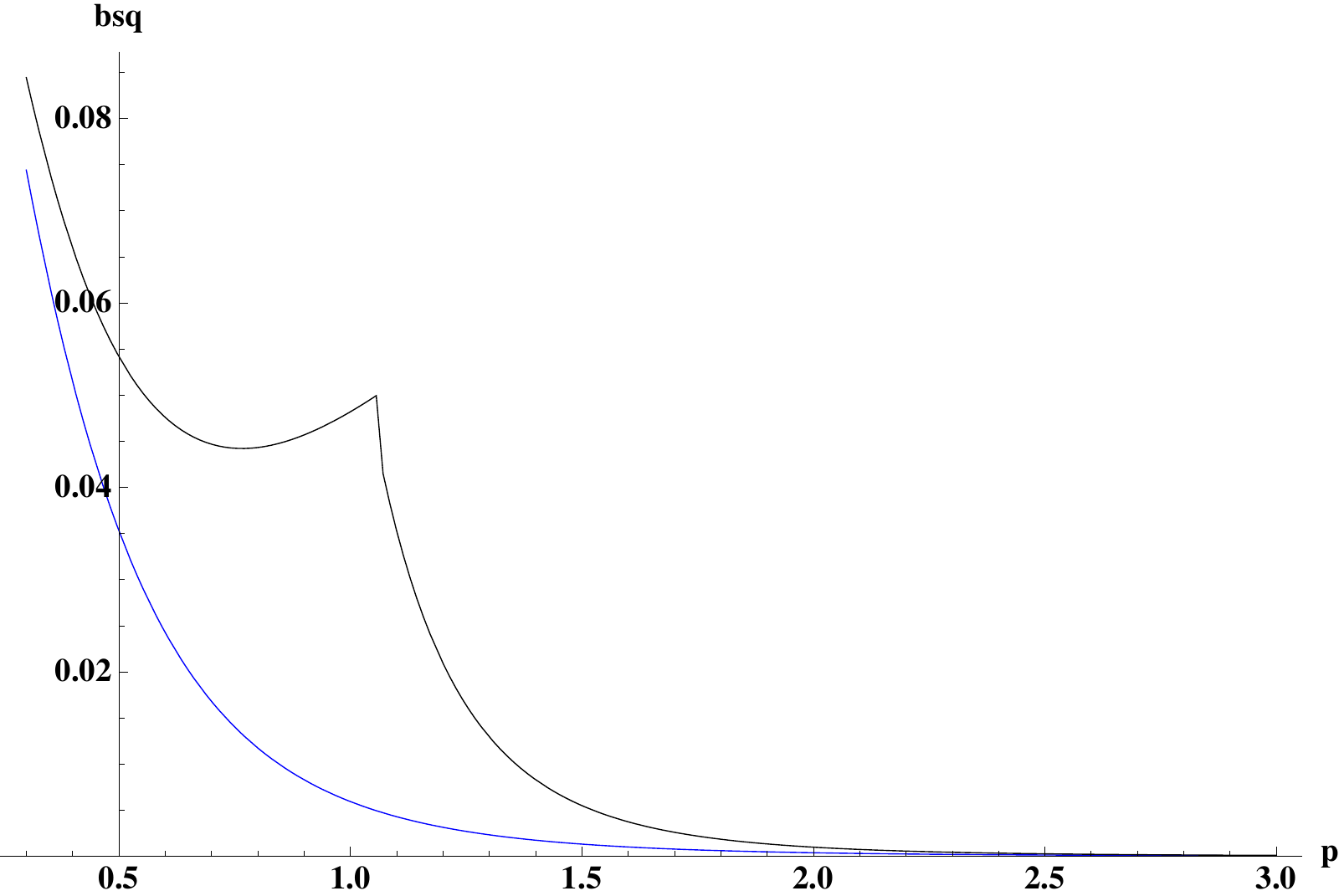}
    \caption{\small lin-lin plot $m=\omega=1$}
    \label{fig:kg_as_bsq}
  \end{minipage}
  \hspace{0.03\linewidth} 
  \begin{minipage}[b]{0.47\linewidth}
    \centering
    \includegraphics[width=\linewidth]{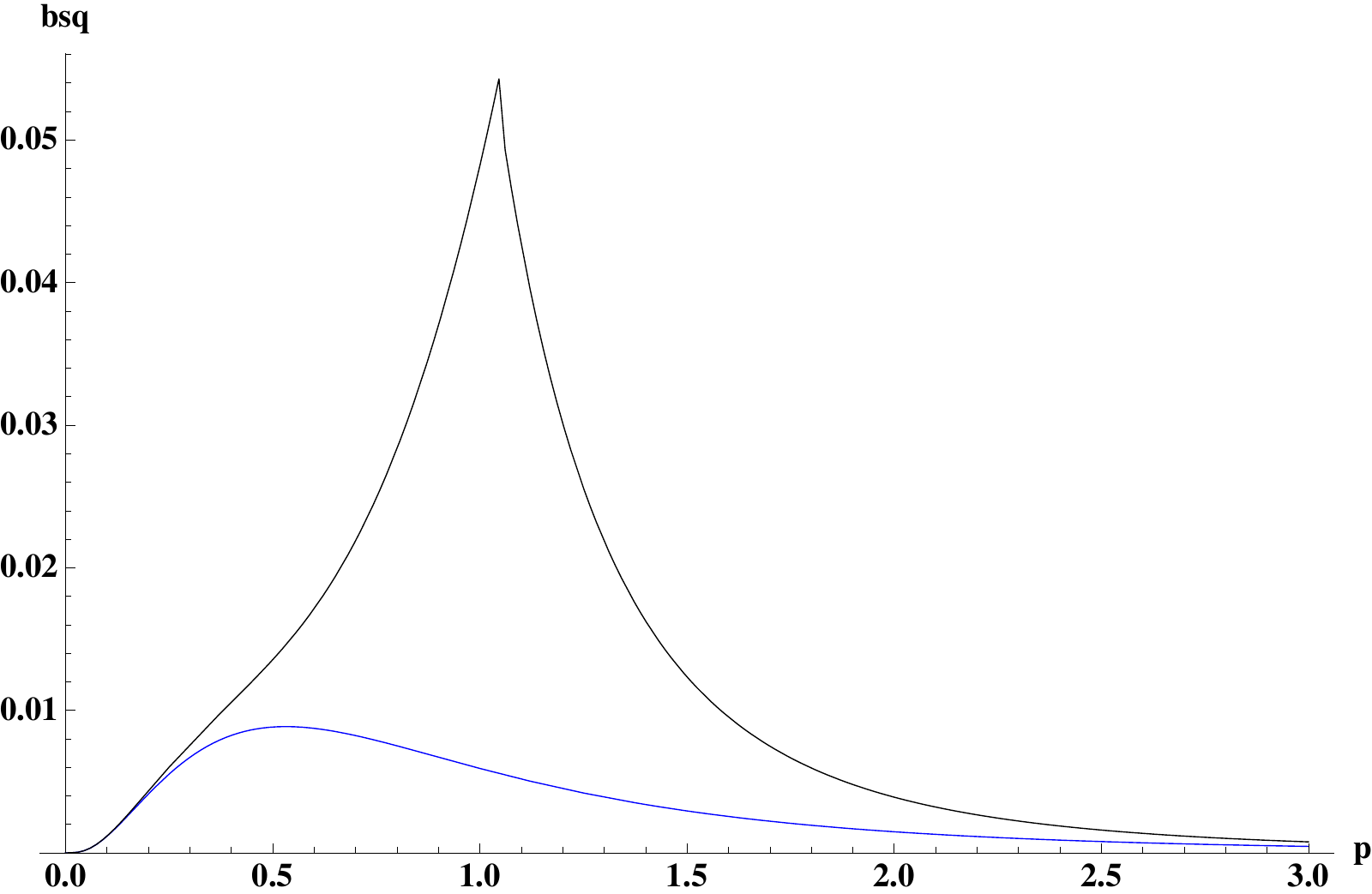}
    \caption{\small lin-lin plot of $p^2 \bsq$ against $p$ for 
    $m=\omega=1$}
    \label{fig:kg_as_p2bsq}
  \end{minipage}
\end{figure}
The inaccuracy of the asymptotic forms is depicted in \autoref{fig:kg_as_bsq} 
at the level of $\bsq$ and in \autoref{fig:kg_as_p2bsq} at the level of
$p^2 \bsq$. 

The resulting total number of particles,
\begin{equation*}
	n(\vx) = \int_{0}^{\infty} \frac{p^2}{2\pi^2} \bsq,
\end{equation*}
is a function of $m$ and $\omega$, irrespective of the expansion time (in the 
limit $q\gg p$). Our asymptotic analysis provides a good approximation of
this number, up to an almost constant proportionality factor caused by the 
difference between the exact solution and the asymptotic form depicted in 
\autoref{fig:kg_as_p2bsq}. This can be seen by the plot in 
\autoref{fig:kg_lnNdiff_mass}, where we show the
result of the numerical integration $\ln n(\vx)$ as a function of $m$, at 
$\tin = 0$, $\tout = 10$ and $\omega = 1.0$, compared with our asymptotic form,
from which we approximate this factor by $3.5$. In subsequent plots we shall 
divide the asymptotic results by this factor. There is a very small 
discontinuity occuring at $\mu = 0$ because of the discontinuity in $p_\nu$
and because of the piecewise definition we have employed.
\begin{figure}[!ht]
	\centering
	\includegraphics[width=0.8\linewidth]{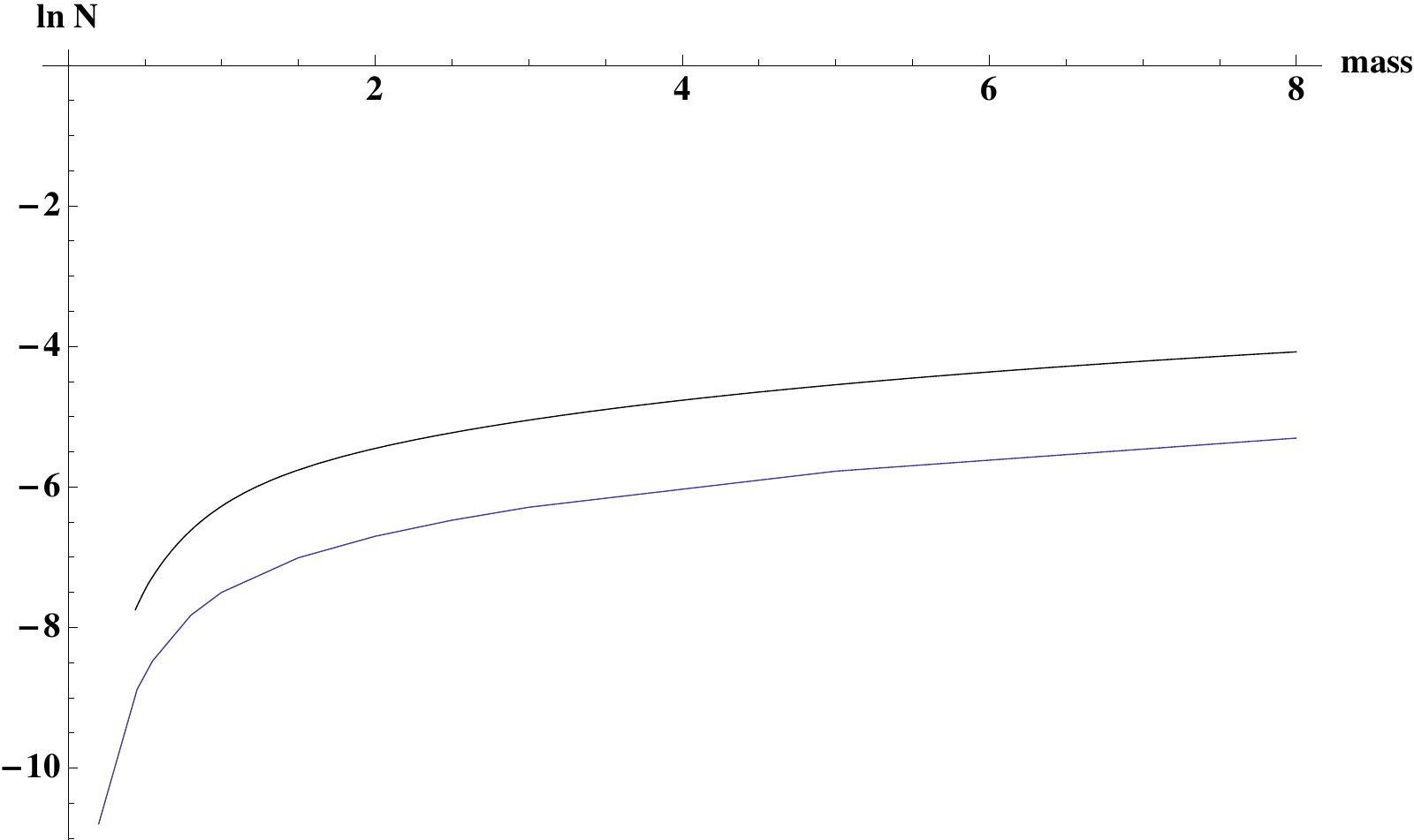}
	\caption{\small $\ln n(\vx)$ as a function of the particle mass $m$ 
	at $\omega=1$}
	\label{fig:kg_lnNdiff_mass}
\end{figure}

\begin{figure}[!ht]
	\centering
	\includegraphics[width=0.8\linewidth]{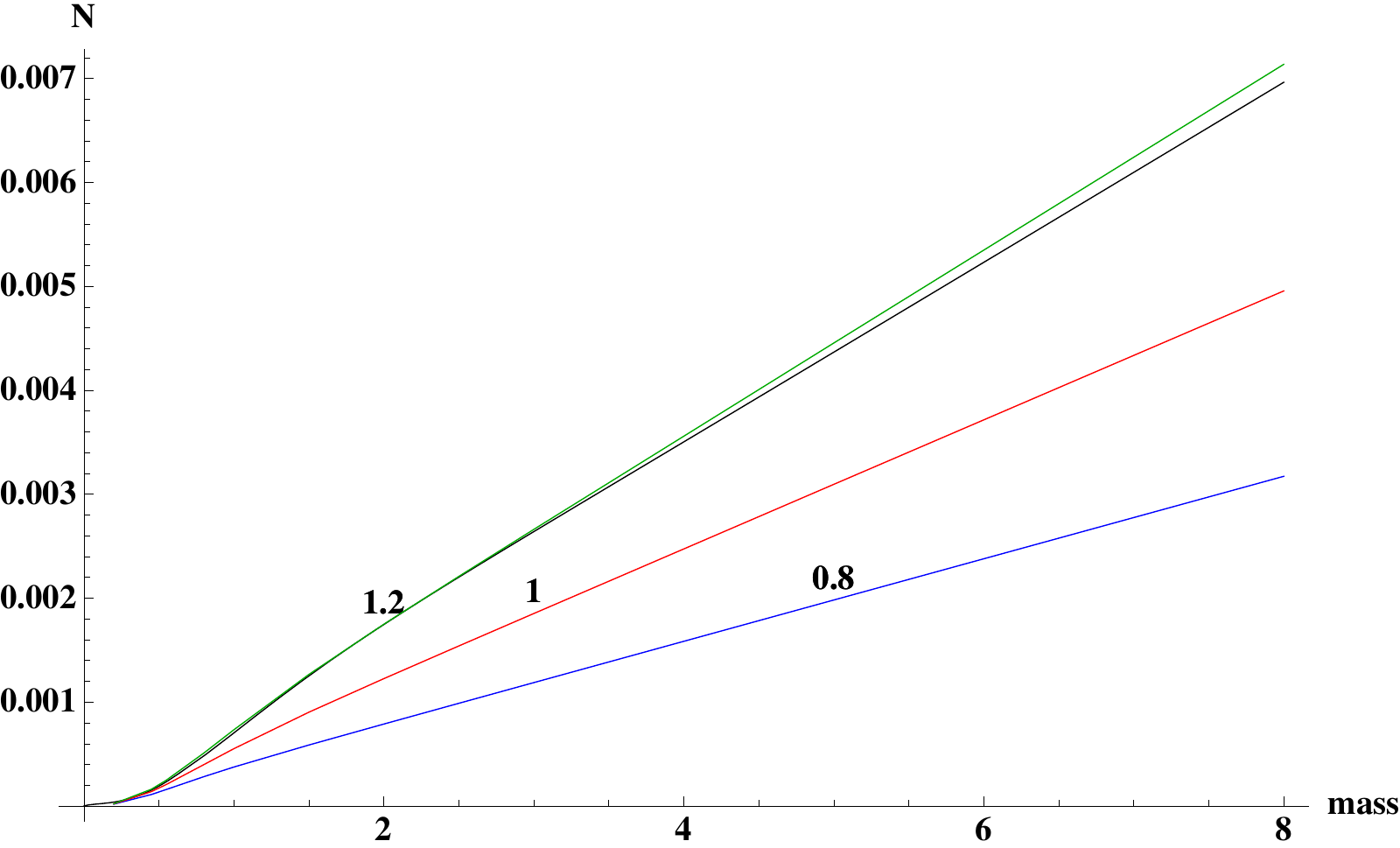}
	\caption{\small $n(\vx)$ as a function of the particle mass $m$ 
	for $\omega=\colthree{0.8}{1.0}{1.2}$ (asymptotic solution is divided by 
	$3.5$)}
	\label{fig:kg_Nm}
\end{figure}
It is remarkable that we obtain a very good agreement with the exact solution.
In \autoref{fig:kg_Nm} we show the particle number density
$n(\vx)$ as a function of the particle mass for different $\omega$, and in
\autoref{fig:kg_No} we plot against $\omega$ for different $m$. The other 
parameters are $\tin = 0$ and $\tout = 10$.
\begin{figure}[!ht]
	\centering
	\includegraphics[width=0.8\linewidth]{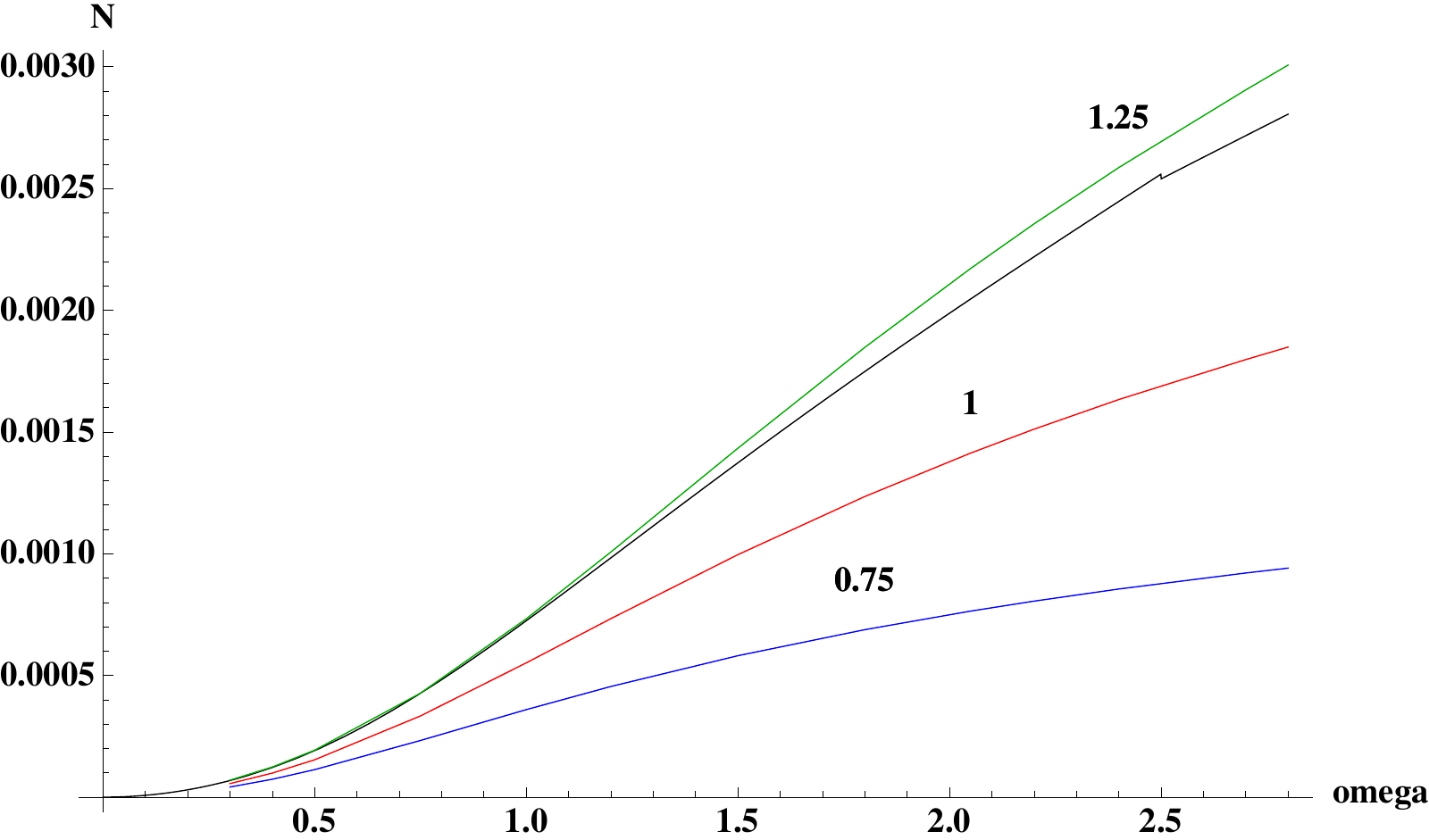}
	\caption{\small $n(\vx)$ as a function of the expansion factor $\omega$ 
	for $m=\colthree{0.75}{1.00}{1.25}$ (asymptotic solution is divided by 
	$3.5$)}
	\label{fig:kg_No}
\end{figure}

Finally, the energy density $\mathcal{E}(\vx)$ is shown in \autoref{fig:kg_Em}
as a function of $m$ and in \autoref{fig:kg_E0} with respect to $\omega$.
\begin{figure}[!ht]
  \begin{minipage}[b]{0.47\linewidth}
    \centering
    \centering
    \includegraphics[width=\linewidth]{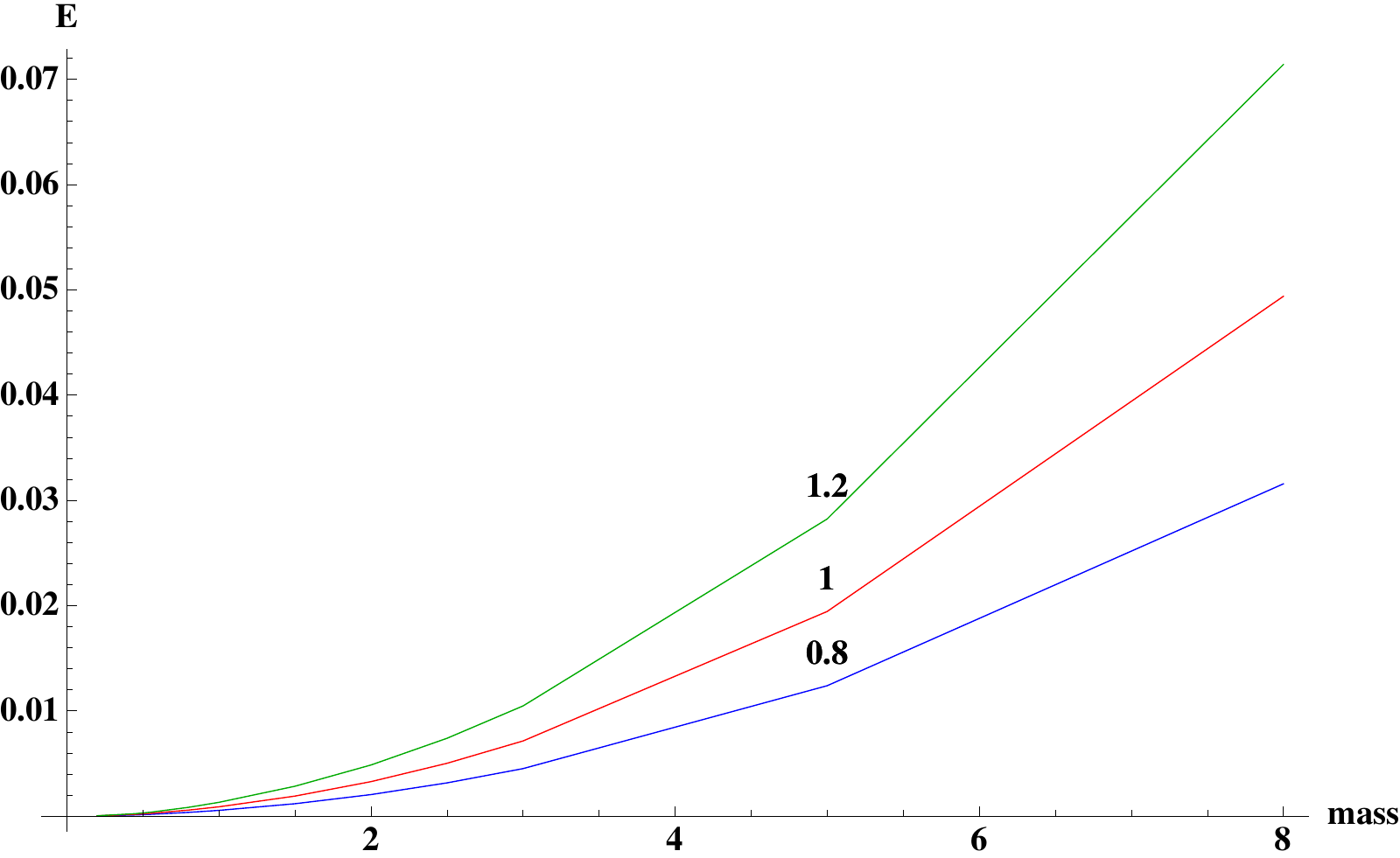}
			\caption{\small $\mathcal{E}(\vx)$ as a function of the mass $m$
			for $\omega=\colthree{0.8}{1.0}{1.2}$}
    \label{fig:kg_Em}
  \end{minipage}
  \hspace{0.03\linewidth} 
  \begin{minipage}[b]{0.47\linewidth} 
    \includegraphics[width=\linewidth]{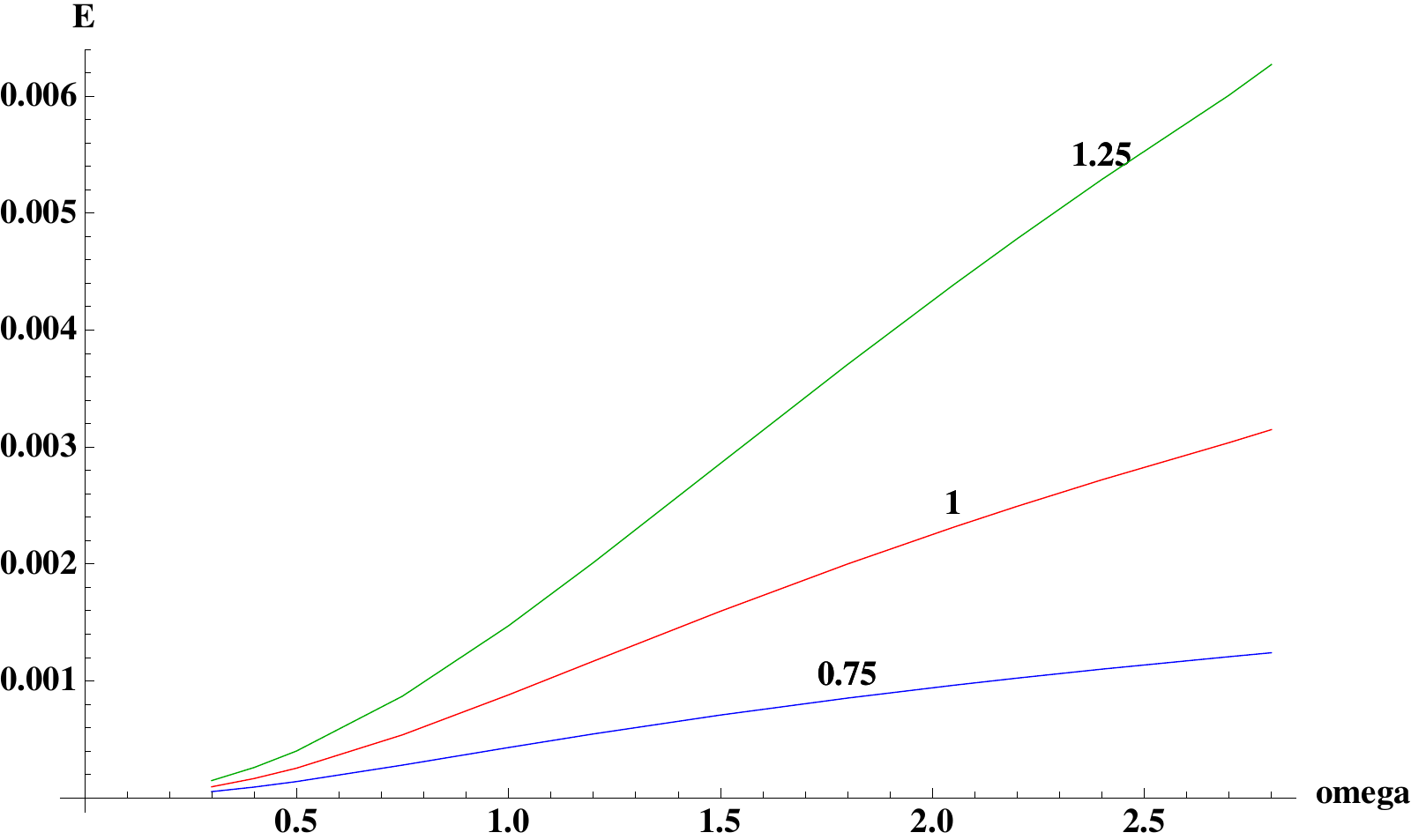}
			\caption{\small $\mathcal{E}(\vx)$ as a function of the expansion factor
			$\omega$ for $m=\colthree{0.75}{1.00}{1.25}$}
    \label{fig:kg_E0}
  \end{minipage}
\end{figure}

We conclude that the particle number density does not depend on the expansion 
time $\tout - \tin$ as long as the latter is sufficiently large 
($\omega \Delta t > 1$). There is linear increase with the particle 
mass \eqref{eq:kg_N_largem}, and a quadratic increase with the expansion factor
\eqref{eq:kg_N_largeo}. The energy density $\mathcal{E}(\vx)$
exhibits a similar increase with the expansion factor, but the $m$ dependence
is more pronounced, and looks quadratic.
\end{subsection}
\end{section}

\chapter{Creation of spinorial particles}\label{chap:dirac}
\begin{section}{Bogoliubov coefficients}\label{chap:dirac_bogo}
We follow the procedure of \autoref{sec:kg_bogo} for the continuation of
Minkowski modes from the {\em out} ({\em in}) region through the de Sitter
expansion phase in the {\em in} ({\em out}) region.
\begin{subsection}{de Sitter \texorpdfstring{{\em in} and {\em out}}{in and out} modes}
\label{sec:junction_dirac}
The Dirac equation, being a first order differential equation, requires only
one initial value equation to completely determine a mode. However, since the 
modes are spinors, there are actually two independent equations, one for the 
upper spinor and one for the lower one. The junction equation is
\begin{equation}
	\Uout_{\vp,\lambda}(\tout, \vx) = U^{\Mout}_{\vp,\lambda}(\tout, \vx),
	\label{eq:dirac_outdef}
\end{equation}
with $\Uout$ given as a linear combination of de Sitter solutions, with
the coefficients determined by
\begin{equation}
	A(p, \tout) \DSol{U}{dS}{\vpout}{\lambda}(\tout, \vx) + 
	B(p, \tout) \DSol{V}{dS}{-\vpout}{\lambda}(\tout, \vx) = 
	\DSol{U}{\Mout}{\vp}{\lambda}(\tout, \vx).
\end{equation}
The de Sitter spinors can be read from \eqref{eq:dirac_UVsol_frw} and the
Minkowski ones from \eqref{eq:dirac_UVsol_out}. Using the connection between
the Pauli spinors \eqref{eq:eta_mvn_xi}, we can cast the above relation in
matrix form:
\begin{multline}
	\qquad\begin{pmatrix}
		ie^{\pi k/2} \Hank{1}{\nu_-}\pparant& -e^{-\pi k/2} \Hank{2}{\nu_-}\pparant\smallskip\\
		ie^{-\pi k/2} \Hank{1}{\nu_+}\pparant& e^{\pi k/2} \Hank{2}{\nu_+}\pparant
	\end{pmatrix}
	\begin{pmatrix}
		A(p, \tout) \smallskip\\
		B(p, \tout)
	\end{pmatrix} =\\
	  \sqrt{\frac{2\omega}{\pi p}} e^{-i E(p) \tout + \frac{3}{2}\omega \tout}
	\begin{pmatrix}
		\sqrt{1 + m/E} \smallskip\\
		\sqrt{1 - m/E}	
	\end{pmatrix}.
	\qquad\label{eq:dirac_junction_matrixeq}
\end{multline}
The determinant of the matrix in the LHS can be computed using the
identity \eqref{eq:hank_idcota}, and we find
\begin{subequations}\label{eq:dirac_junction_solAB}
\begin{align}
  A(p, \tout) =& -i \sqrt{\frac{\pi p}{8\omega}} e^{-i E(p) \tout + \frac{3}{2} \omega \tout}\nonumber\\
  &\qquad\times
  \left\{\sqrt{1 + \frac{m}{E}} e^{\pi k/2} \Hank{2}{\nu_+}\pparant + 
  \sqrt{1 - \frac{m}{E}} e^{-\pi k/2} \Hank{2}{\nu_-}\pparant\right\}
  \label{eq:dirac_junction_solA},\\
  B(p, \tout) =& -\phantom{i} \sqrt{\frac{\pi p}{8\omega}} e^{-i E(p) \tout + \frac{3}{2} \omega \tout}\nonumber\\
  &\qquad\times
  \left\{\sqrt{1 + \frac{m}{E}} e^{-\pi k/2} \Hank{1}{\nu_+}\pparant - 
  \sqrt{1 - \frac{m}{E}} e^{\pi k/2} \Hank{1}{\nu_-}\pparant\right\}.
  \label{eq:dirac_junction_solB}
\end{align}
\end{subequations}
Using the same identity, we arrive at the normalization relation
\begin{equation}
	\abs{A(p, \tout)}^2 + \abs{B(p, \tout)}^2 = e^{3 \omega \tout},
	\label{eq:dirac_AB_norm}
\end{equation}
and thus the modes are orthonormal throughout all space, both in the
expansion phase, with respect to the de Sitter scalar product 
\eqref{eq:dirac_scprod_dS_frw} and on the Minkowski region with respect to the 
scalar product \eqref{eq:dirac_scprod_mink}:
\begin{equation*}
	\scprod{U^{\rm f}_{\vpp,\lambdap}}{U^{\rm f}_{\vp,\lambda}} = 
	\delta_\llp\delta^3(\vp - \vpp).
\end{equation*}
Note the sign difference between \eqref{eq:dirac_AB_norm} and the scalar case
\eqref{eq:kg_AB_norm}. In analogy with the scalar case, we define a new
pair of coefficients normalized to unity:
\begin{equation}
	\tilde{A}(p, \tout) = e^{-\frac{3}{2} \omega t} A(p, \tout)\quad,\quad
	\tilde{B}(p, \tout) = e^{-\frac{3}{2} \omega t} B(p, \tout).
	\label{eq:dirac_junction_solAB_tilde}
\end{equation}
The Bogoliubov coefficients will be determined in the following section.
\end{subsection}
\begin{subsection}{Mode mixing and density of created particles}\label{sec:bogo_dirac}
In this section we apply the general theory of Bogoliubov transformation 
outlined in \autoref{sec:particle_prod_dirac}.

The difference between fermionic and bosonic fields has already been 
pointed out by the different normalizations of the Bogoliubov 
coefficients (compare \eqref{eq:kg_bogo_orthonormij} to
\eqref{eq:dirac_bogo_orthonormij}). While the $-$ sign appeared for the scalar
field from scalar products of the form $\scprod{f^*_\vp}{f^*_\vpp}$, the 
polarized solutions to the Dirac equation have a peculiar behaviour to 
rotations of the label $\vp$ in the sense that 
$U_{-(-\vp),\lambda} = -U_{\vp,\lambda}$. This produces the $-$ signs required 
for the theory to be valid.

Similar to the scalar case, we have determined coefficients $A$ and $B$ such 
that
\begin{subequations}
\begin{equation}
	\DSol{U}{out}{\vp}{\lambda}(t, \vx) = 
	A(p, \tout) \DSol{U}{dS}{\vpout}{\lambda}(t, \vx) + 
	B(p, \tout) \DSol{V}{dS}{-\vpout}{\lambda}(t, \vx).
\end{equation}
The same procedure applies in defining {\em in} modes, except when expressing
$\DSol{V}{in}{-\vp}{\lambda}$:
\begin{align}
	\DSol{U}{in}{\vp}{\lambda}(t, \vx) &= 
	A(p, \tin) \DSol{U}{dS}{\vpin}{\lambda}(t, \vx) + 
	B(p, \tin) \DSol{V}{dS}{-\vpin}{\lambda}(t, \vx), \\
	\DSol{V}{in}{-\vp}{\lambda}(t, \vx) &= 
	-B^*(p, \tin) \DSol{U}{dS}{\vpin}{\lambda}(t, \vx) + 
	A^*(p, \tin) \DSol{V}{dS}{-\vpin}{\lambda}(t, \vx).
\end{align}
\end{subequations}
The $-$ sign next to $B^*$ appeared because we did a full rotation of the 
label of the spinor $V$. The effect of such a rotation manifests on the level
of the Pauli 2-component spinors $\xi$ and $\eta$ that make up the Dirac 
spinors $U$ and $V$, as discussed in \autoref{chap:pauli_spinors}:
\begin{align*}
	U_{-(-\vp), \lambda}(t, \vx) &= - U_{\vp, \lambda}(t, \vx), \\
	\vp=(p, \theta, \varphi)\quad&,\quad -\vp=(p, \pi - \theta, \varphi + \pi)
\end{align*}	
The full rotation has the effect of adding $2\pi$ to the angle $\varphi$, 
which produces the $-$ sign. The Bogoliubov coefficients 
\eqref{eq:dirac_bogo_inout} are defined in 
a simlar way to the scalar case:
\begin{equation}
	\DSol{U}{out}{\vp}{\lambda}(t, \vx) = \sum_{\lambdap} \int d^3p^\prime 
	\left\{\alpha_\llp(\vp, \vpp) \DSol{U}{in}{\vpp}{\lambdap}(t, \vx) + 
	\beta_\llp(\vp, -\vpp) \DSol{V}{in}{-\vpp}{\lambdap}(t, \vx)\right\}.
\end{equation}	
The $\beta$ coefficients come from scalar products between Dirac spinors:
\begin{equation}
	\beta_\llp(\vpp, -\vp) = \scprod{\DSol{V}{i}{-\vq}{\lambda}}
	{\DSol{U}{f}{\vpp}{\lambdap}}.
	\label{eq:dirac_betappdef_scprod}
\end{equation}
It follows that
\begin{equation}
	\beta_\llp(\vpp, -(-\vp)) = - \beta_\llp(\vpp, \vp) ,\qquad
	\beta_\llp(-(-\vpp), \vp)) = - \beta_\llp(\vpp, \vp),
	\label{eq:dirac_betapp_mmp}
\end{equation}
and furthermore, since $\beta_\llp(\vpp, -\vp) = \beta_\llp(-\vpp, -(-\vp))$,
we have
\begin{equation}
	\beta_\llp(\vpp, -\vp) = - \beta_\llp(-\vpp, \vp).
	\label{eq:dirac_betapp_pmp_mpp}
\end{equation}
The Bogoliubov coefficients are expressed in terms of reduced coefficients,
\begin{subequations}
\begin{align}
	\alpha_\llp(\vp, \vpp) &= e^{\frac{3}{2}\omega (\tout - \tin)} 
	\delta_\llp \delta^3(\vpp - \vq) \alpha(p),
	\label{eq:dirac_alphadef}\\
	\beta_\llp(\vp, -\vpp) &= e^{\frac{3}{2}\omega (\tout - \tin)} 
	\delta_\llp \delta^3(\vpp - \vq) \beta(p),
	\label{eq:dirac_betadef}
\end{align}
explicitly given by
\begin{align}
	\alpha(p) &= \tilde{A}^*(q, \tin) \tilde{A}(p, \tout) + 
	\tilde{B}^*(q, \tin) \tilde{B}(p, \tout),
	\label{eq:dirac_alphaAB}\\
	\beta(p) &= \tilde{A}(q, \tin) \tilde{B}(p, \tout) - 
	\tilde{B}(q, \tin) \tilde{A}(p, \tout),
	\label{eq:dirac_betaAB}
\end{align}
such that the following normalization condition is obeyed:
\begin{equation}
	\abs{\alpha(p)}^2 + \abs{\beta(p)}^2 = 1.
	\label{eq:dirac_orthonorm}
\end{equation}
\end{subequations}
This states that the particle number density, which is proportional to 
$\abs{\beta(p)}^2$, cannot exceed~1. This is not the case for the scalar 
field, where the Bogoliubov coefficients can be arbitrarily large.
	
In order to understand the orthogonality relation 
\eqref{eq:dirac_orthonormij}, we must make use yet again of the spinorial 
characteristics of the Dirac solutions. The condition reduces to
\begin{equation}
	\beta_{\llp}(\vpp, \vq) + \beta_{\llp}(\vp, \vq^\prime) = 0.
\end{equation}
The first term is proportional to
\begin{equation}
	q^2\delta(q - q^\prime) \delta(\pi - \theta - \theta^\prime)
	\delta(\varphi^\prime - \varphi - \pi),
\end{equation}
while the second term is proportional to
\begin{equation}
	q^2\delta(q - q^\prime) \delta(\pi - \theta - \theta^\prime)
	\delta(\varphi - \varphi^\prime - \pi).
\end{equation}
There is a difference between the $\varphi$ delta functions: If the argument
of the first delta function is $0$, then the argument of the second one is 
$-2\pi$, and thus one of the spinors gets shifted by $2\pi$ in the second
term. Similar considerations apply if the second delta has null argument. 
Therefore, a $-$ sign will accompany one and only one of the 
$\beta$ coefficients above, and thus the orthogonality relations are
automatically satisfied.

We can readily express the {\em out} one-particle operators with the 
coefficients introduced in this chapter. In order to see that our results do
indeed follow from the general discussion in \autoref{sec:particle_prod}, 
we will use the expressions \eqref{eq:dirac_bogo_bd}
\begin{subequations}
\begin{align}
	\bout(\vp, \lambda) &= e^{\frac{3}{2}\omega(\tout - \tin)} \left(
	\alpha^*(p) \bin(\vq, \lambda) + \beta^*(p) \ddin(-\vq, \lambda)\right),\\
	\ddout(\vp, \lambda) &= e^{\frac{3}{2}\omega(\tout - \tin)} \left(
	-\beta(p) \bin(-\vq, \lambda) + \alpha(p) \ddin(\vq, \lambda)\right).
\end{align}
\end{subequations}	
The expectation value of these operators in the {\em in} vacuum state is
\begin{subequations}
\begin{align}
	\braket{0_{\rm in} | \dout(\vp) \bout(\vpp) | 0_{\rm in}} &=
	\alpha^*(p) \beta^*(p) \delta^3(\vp + \vpp),\\
	\braket{0_{\rm in} | \bdout(\vp) \ddout(\vpp) | 0_{\rm in}} &=
	-\beta(p) \alpha(p) \delta^3(\vp + \vpp),\\
	\braket{0_{\rm in} | \bdout(\vp) \bout(\vpp) | 0_{\rm in}} &=
	\abs{\beta(p)}^2\delta^3(\vp - \vpp),\\
	\braket{0_{\rm in} | \ddout(\vp) \dout(\vpp) | 0_{\rm in}} &=
	\abs{\beta(p)}^2 \delta^3(\vp - \vpp).
\end{align}
\end{subequations}
With these expectation values we can evaluate the particle number density
\eqref{eq:dirac_bogo_pnumberi}
\begin{equation}
	n_{\vp} = 2 \sum_{\lambda} \abs{\beta(p)}^2 \delta^3(\vp - \vp).
	\label{eq:dirac_bogo_pnumberp_inf}
\end{equation}	
There is an extra factor of $4$ compared to the scalar case 
\eqref{eq:kg_bogo_pnumberp_inf}, comming from the two different polarizations
and the two particle types (particle and antiparticle). The 
volumic particle density is 
\begin{equation}
	n_{\vp}(\vx) = \frac{4}{(2\pi)^3} \abs{\beta(p)}^2.
\end{equation}
In order to evaluate the number of particles with the magnitude of the 
momentum $p$, we integrate away the spherical coordinates and arrive at
\begin{equation}
	n_p(\vx) = \frac{2 p^2}{\pi^2} \abs{\beta(p)}^2.
\end{equation}
The expectation value for the stress-tensor is given by:
\begin{subequations}\label{eq:dirac_Tmunu_beta}
\begin{align}
	\braket{0_{\rm in} | :T^{\rm out}_{00}(x): | 0_{\rm in}} &=\mathcal{E}(\vx)=
	\int \frac{d^3p}{2\pi^3} E(p) \bsq,
	\label{eq:dirac_T00_beta}\\
	\braket{0_{\rm in} | :T^{\rm out}_{ij}(x): | 0_{\rm in}} &= \delta_{ij}
	\int \frac{d^3p}{2\pi^3} \frac{\vp^2}{3 E(p)} 
	\left\{\bsq + \frac{m}{2ip} 
	\left(\alpha^*(p) \beta^*(p) e^{-2i E t} - \alpha(p) \beta(p) e^{2i E t}\right)
	\right\}.
	\label{eq:dirac_Tij_beta}
\end{align}
\end{subequations}
The ${}^{\rm out}$ label indicates that the normal ordering has been done
with respect to the ${\rm out}$ vacuum. 
We shall refer to the spectral energy density through the following notation:
\begin{equation}
	\mathcal{E}_p = \frac{2 p^2}{\pi^2} E(p) \bsq.
	\label{eq:dirac_Epdef}
\end{equation}
By substituting \eqref{eq:dirac_junction_solAB} for $A$ and $B$ in the formula
for $\beta(p)$ \eqref{eq:dirac_betaAB}, we arrive at the following expression:
\begin{multline}
	\beta(p) = i \frac{\pi}{8\omega} \sqrt{q p} e^{-i(E_q \tin + E_p \tout)} \times\\
	\left\{\epsilon^+_q \epsilon^+_p H_1(q,p) + 
	\epsilon^-_q \epsilon^-_p H_1^*(q,p) -
	\epsilon^+_q \epsilon^-_p H_2(q,p) +
	\epsilon^-_q \epsilon^+_p H_2^*(q,p)\right\},
	\label{eq:dirac_beta_explicit}
\end{multline}
where we have used $\epsilon^{\pm}_q = \sqrt{1 \pm m/E_q}, 
E_q = \sqrt{m^2 + \vq^2}$ being the Minkowski energy of a particle of 
mass $m$ and momentum $\vq$, and 
\begin{subequations}\label{eq:dirac_beta_H}
\begin{align}
	H_1(q, p) &= \phantom{e^{\pi k}} \Hank{2}{\nu_+}(q)\Hank{1}{\nu_+}(p) - 
	\phantom{e^{-\pi k}} \Hank{1}{\nu_+}(q)\Hank{2}{\nu_+}(p)
	\label{eq:dirac_beta_H1},\\
	H_2(q, p) &= e^{\pi k} \Hank{2}{\nu_+}(q)\Hank{1}{\nu_-}(p) + 
	e^{- \pi k} \Hank{1}{\nu_+}(q)\Hank{2}{\nu_-}(p)
	\label{eq:dirac_beta_H2}.
\end{align}
\end{subequations}
with $k = m/\omega$ and $\nu_{\pm} = 1/2 \pm ik$. An important conclusion can
be drawn from \eqref{eq:dirac_beta_explicit}: except for the leading phase 
factor (which has no effect on $\bsq$), there is no dependency on $\tin$ and
$\tout$ independently, only on $\Delta t = \tout - \tin$ (through $q$).
We shall take advantage of this by setting $\tin = 0$.

In the squared form of $\beta$, the square-root terms $\epsilon^\pm_p$ 
simplify according to the following realtions:
\begin{equation}
	\epsilon^+_p \epsilon^-_p = p/E_p ,\qquad
	\epsilon^\pm_p \epsilon^\pm_p = 1 \pm m/E_p.
\end{equation}
By using the identity \eqref{eq:hank_idcota}, we arrive at the result
\begin{equation*}
	\abs{H_1(q, p}^2 + \abs{H_2(q, p)}^2 = \frac{16 \omega^2}{\pi^2 pq},
\end{equation*}
with which we evaluate $\abs{\beta}^2$ to
\begin{equation}
	\abs{\beta(p)}^2 = \frac{1}{2} + \frac{\pi^2 q p}{32 E_q E_p}
	\left\{k^2 \mathcal{H}_1(q,p) + k \frac{q}{\omega} \mathcal{H}_2(q,p) - 
	k\frac{p}{\omega}\mathcal{H}_3(q,p) + 
	\frac{qp}{\omega^2} \mathcal{H}_4(q,p)\right\}.
	\label{eq:dirac_betasq_explicit}
\end{equation}
$\mathcal{H}_i(q,p)$ are expressed using the functions $H_i(q,p)$ defined in 
\eqref{eq:dirac_beta_H}:
\begin{subequations}\label{eq:dirac_betasq_Hrond}
\begin{gather}
	\mathcal{H}_1 = \abs{H_1}^2 - \abs{H_2}^2 ,\qquad
	\mathcal{H}_2 = H_1 H_2 + H_1^* H_2^*,\qquad
	\mathcal{H}_3 = H^*_1 H_2 + H_1 H_2^*,\\
	\mathcal{H}_4 = \frac{1}{2} \left(H_1^2 + H_1^{*\, 2} - 
	(H_2^2 + H_2^{*\, 2})\right).
\end{gather}
\end{subequations}
Every function $\mathcal{H}_i, H_i$ is understood to take the arguments
$(q, p)$. We emphasize that $\bsq$ only depends on 
$\Delta t = \tout - \tin$, and from \eqref{eq:dirac_betaAB} it follows that
$\bsq$ vanishes for $\tin = \tout$.
\end{subsection}
\begin{subsection}{Production of massless Dirac particles}\label{sec:dirac_massless}
We have shown that there is no particle production for a conformally 
coupled massless scalar field in \autoref{sec:kg_massless}. In this section
we shall prove the same result for the massless fermionic field.

The orders $\nu_{\pm}$ of the Hankel functions reduce to $1/2$ for $m = 0$. 
The computation of the coefficients $\tilde{A}, \tilde{B}$ is 
straightforward from \eqref{eq:dirac_junction_solAB}:
\begin{subequations}\label{eq:dirac_massless_conf_beta}
\begin{gather}
	\tilde{A}(p, t) = -i \sqrt{\frac{\pi p}{2\omega}} e^{-ipt} 
	\Hank{2}{1/2}\pparant ,\qquad	\tilde{B}(p, t) = 0,\\
	\beta(p) = 0.
\end{gather}
\end{subequations}
\end{subsection}
\begin{subsection}{Graphical analysis}
Before embarking for the asymptotic analysis of the analytical solution 
\eqref{eq:kg_beta_explicit}, we take a short survey of its form. When 
commenting the spectra, we shall use some results in anticipation.

\begin{figure}[h!t]
	\centering
	\includegraphics[width=0.8\linewidth]{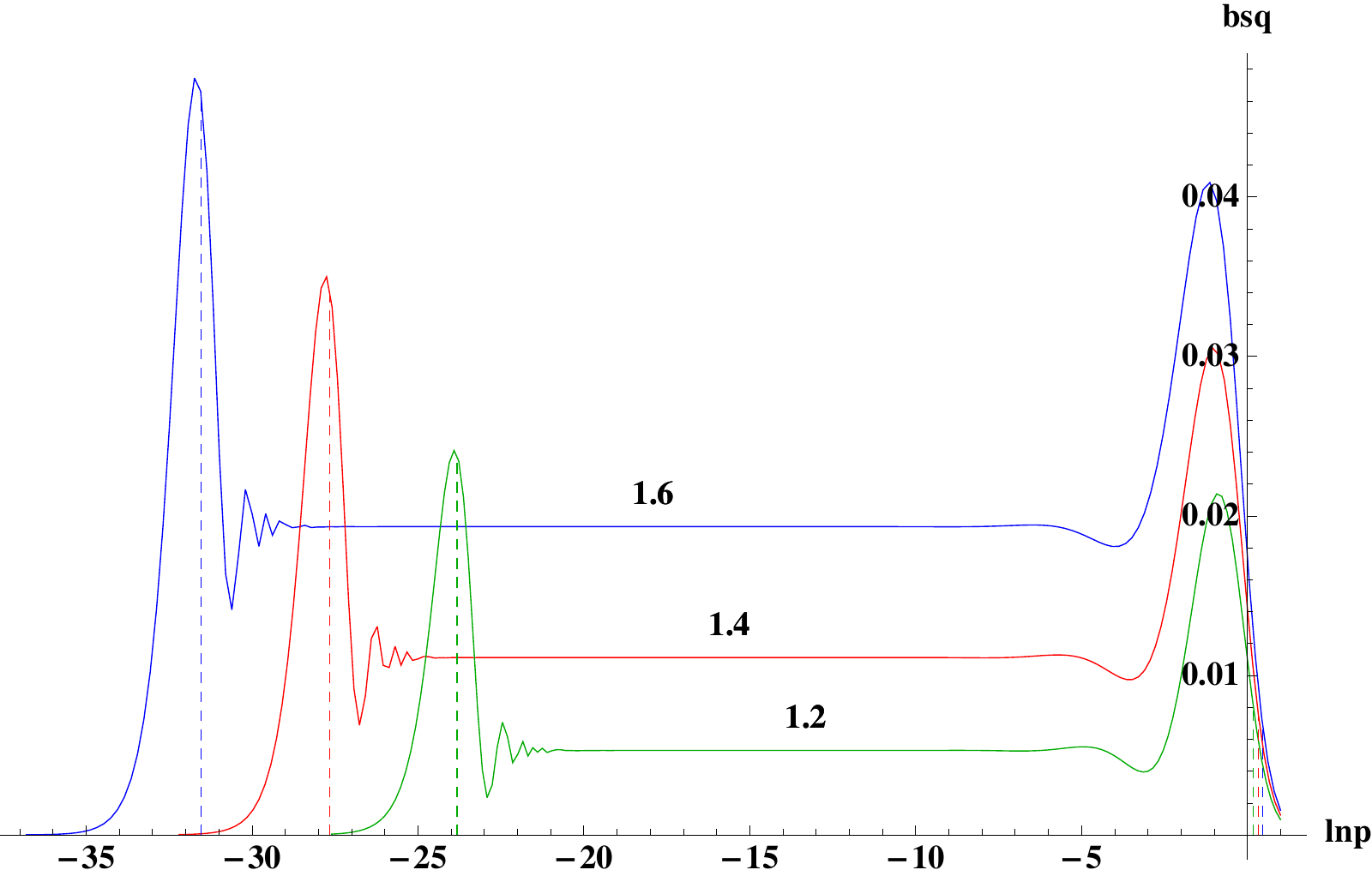}
	\caption{\small The exact solution $\bsq$ plotted against $\ln p$ for
	$m=1$, $\tin = 0$, $\tout = 20$ and $\omega=\colthree{1.6}{1.4}{1.2}$.}
	\label{fig:dirac_o}
\end{figure}
The exact form of $\bsq$ is depicted in \autoref{fig:dirac_o} and 
\autoref{fig:dirac_t1}. The first figure shows $\bsq$ for three 
values of $\omega$, while the second uses three values of $\tout$. The rest of 
the parameters have the values $m = 1$, $\tin = 0$, $\tout = 10$ and 
$\omega = 1.5$. 

These pictures show three regions of interest, corresponding to three 
different regimes:
\begin{enumerate}[(a)]
\item $q \ll \omega$: $\bsq$ goes to $0$ as $pq$, as we shall prove in 
\eqref{eq:dirac_beta_small}
\item $p \gg \omega$: $\bsq$ goes to $0$ as 
$1/p^4$, (see \eqref{eq:dirac_bsql})
\item $p \ll \omega \ll q$: $\bsq$ is remarcably well described by a 
constant value which we find to be a Fermi-Dirac distribution function, 
$(1+e^{2\pi m/\omega})^{-1}$ \eqref{eq:dirac_middle_as_o0}.
\end{enumerate}
These regions are delimited by two resonances which occur roughly at 
$q \sim \omega$ and $p \sim \omega$.

\autoref{fig:dirac_t1} shows three plots of different $\tout$, having the 
same parameters $m = 1$, $\tin = 0$ and $\omega = 1.5$. The middle plateau
and the second resonance overlap. This independence on the expansion time 
$\Delta t = \tout - \tin$ will be uncovered by the asymptotic analysis.
\begin{figure}[h!t]
	\centering
	\includegraphics[width=0.8\linewidth]{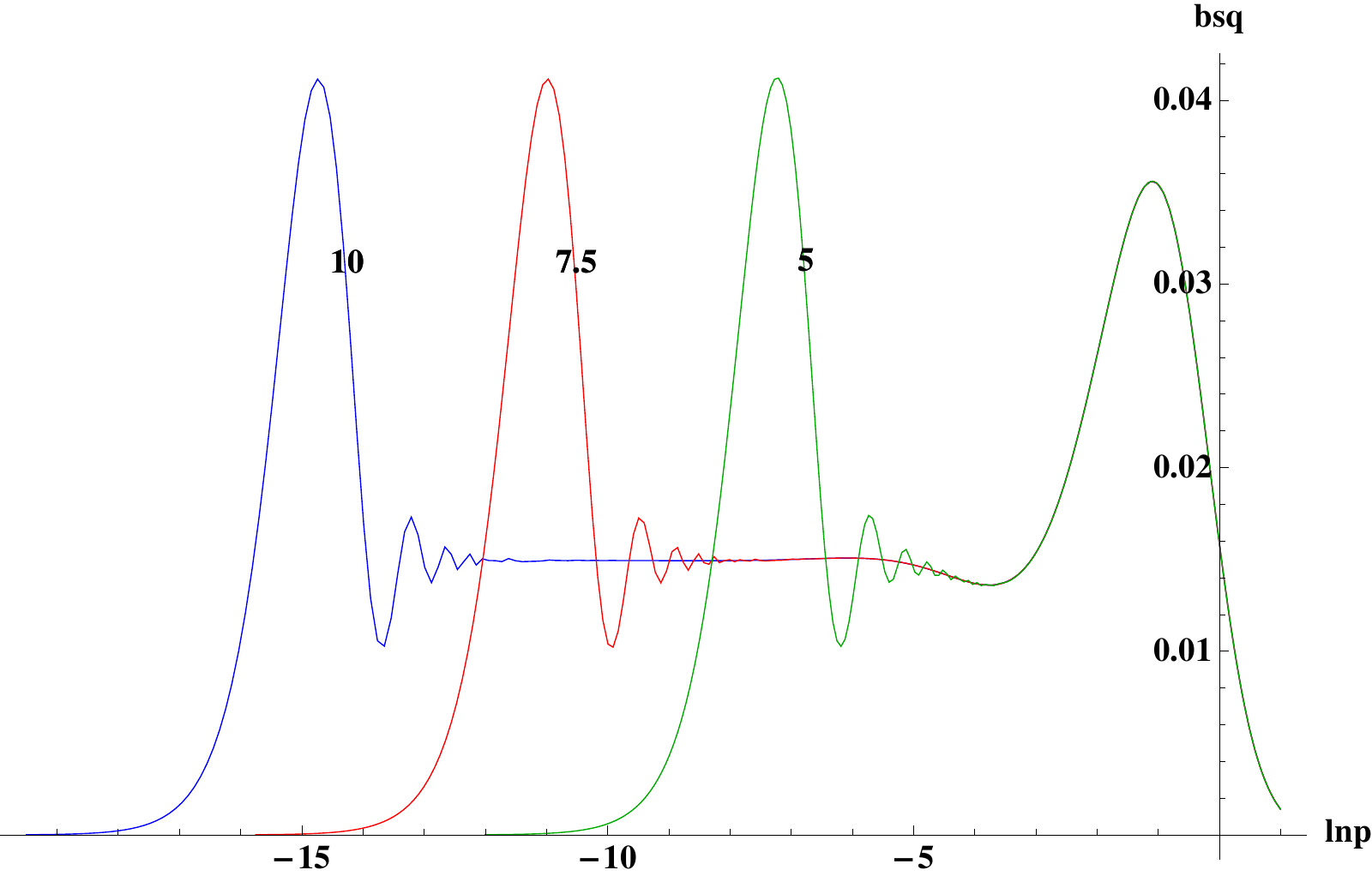}
	\caption{\small $\bsq$ plotted against $\ln p$ for $m=1$, $\tin = 0$, 
	$\omega = 1.5$ and $\tout=\colthree{10}{7.5}{5.0}$.}
	\label{fig:dirac_t1}
\end{figure}

The large momentum region is investigated in \autoref{fig:dirac_as_p4bsq},
where we show that $p^4 \bsq$ approaches the value $m^2 \omega^2/16$, indicated
by black lines. This behaviour guarantees a finite number of particles per unit 
volume, since $n_p \sim p^2 \bsq$. However, the volumic density of the energy 
$\mathcal{E}(\vx)$ \eqref{eq:dirac_T00_beta} has a logarithmic divergence. 
The spectral energy density $\mathcal{E}_p$ is plotted in
\autoref{fig:dirac_energydens}.
\begin{figure}[h!t]
  \begin{minipage}[b]{0.47\linewidth}
    \centering
    \includegraphics[width=\linewidth]{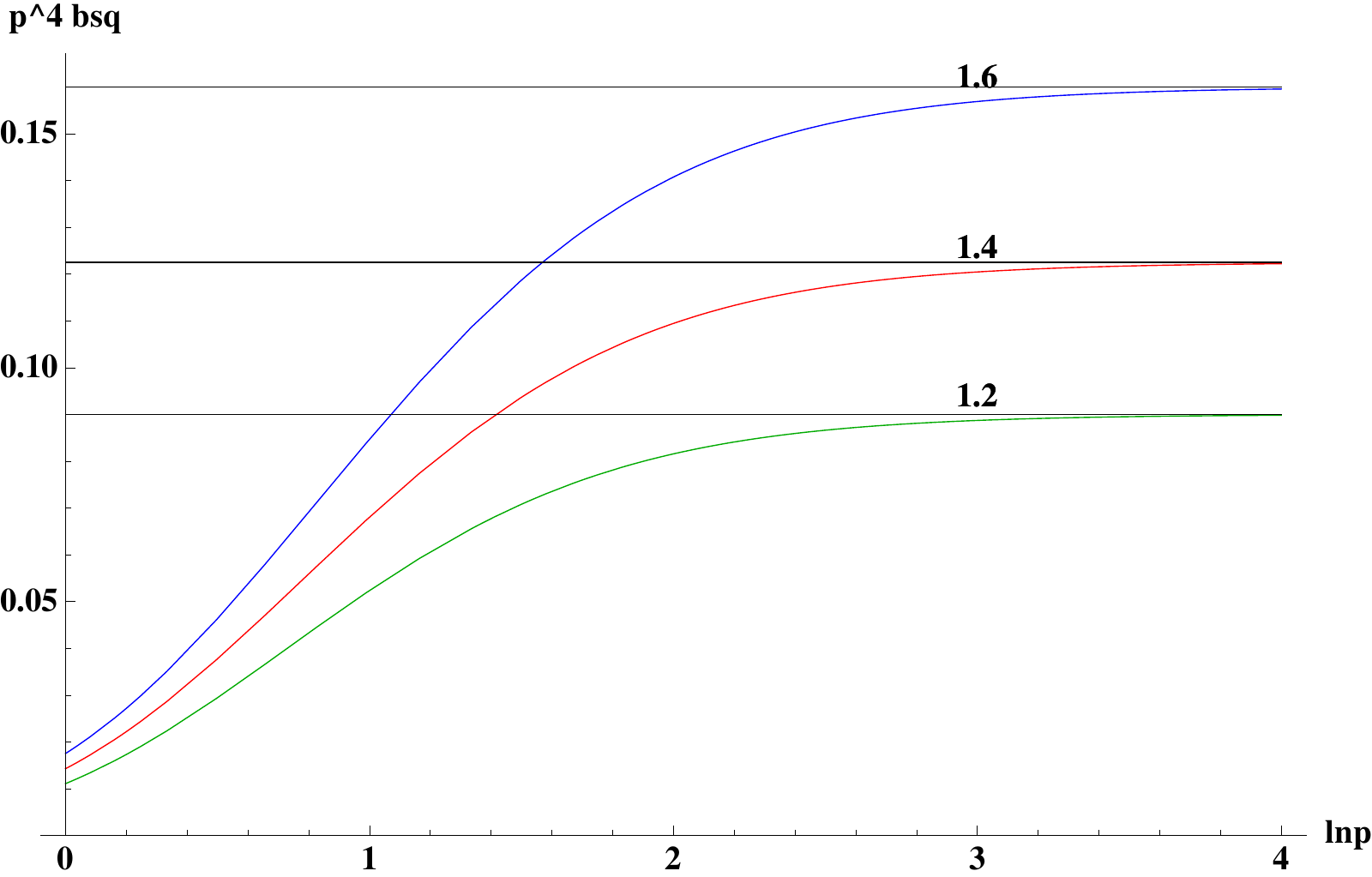}
    \caption{\small The particle number density $n_p = p^2 \bsq$ is represented 
    against $\ln p$ for $m=1$,$\tin=0$, $\tout=10$, $\omega=\colthree{1.6}{1.4}{1.2}$.}
    \label{fig:dirac_as_p4bsq}
  \end{minipage}
  \hspace{0.03\linewidth} 
  \begin{minipage}[b]{0.47\linewidth} 
    \centering
    \includegraphics[width=\linewidth]{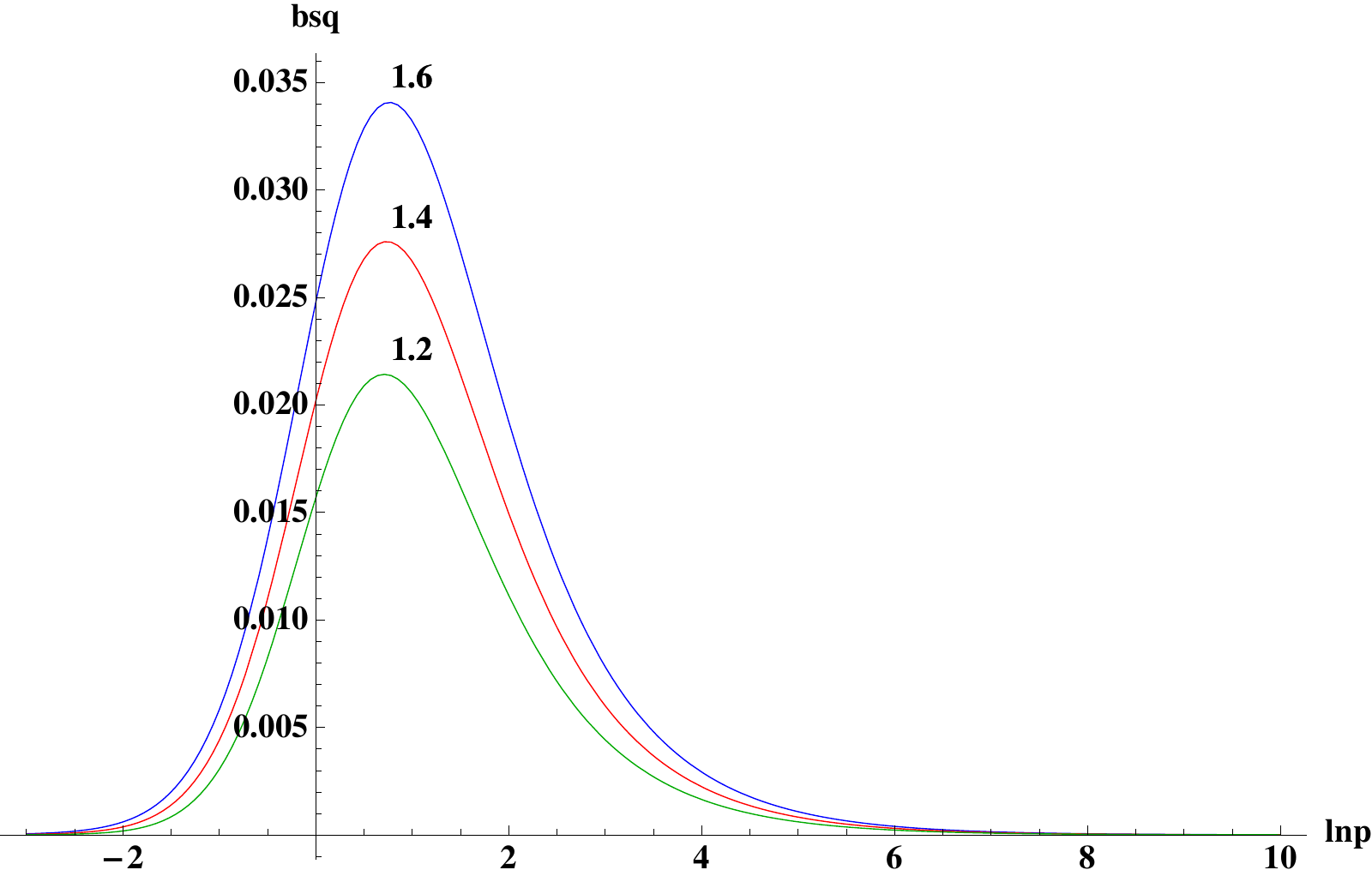}
    \caption{\small the energy density $p^2 E_p \bsq$ is plotted against 
    $\ln p$ for $m=1$, $\tin = 0$, $\tout = 10$, 
    $\omega=\colthree{1.6}{1.4}{1.2}$.}
    \label{fig:dirac_energydens}
  \end{minipage}
\end{figure}
\end{subsection}
\end{section}
\begin{section}[Asymptotic analysis]{Asymptotic analysis of the particle density}
\label{sec:dirac_asympt}
The analysis mainly involves the use of approximation formulas for the Hankel
functions, given in \autoref{app:hankel}. Some of the more frequent notation
used in this section are given below.
\begin{gather*}
	k = m/\omega,\qquad \nu_{\pm} = 1/2 \pm i k,\qquad 
	\epsilon^{\pm}_p = (1 \pm m/E_p)^{1/2}\\
	\Gamma_{\pm}(q,p) \text{ is defined in \eqref{eq:def_gammapm}},\quad
	\beta(p) \text{ is given in \eqref{eq:dirac_betaAB}},\quad
	H_i(q, p) \text{ are given by \eqref{eq:dirac_beta_H}}
\end{gather*}
We assume the reader has gone through \autoref{sec:kg_asympt}, and omit
explanatory text that would otherwise be repeated.
\begin{subsection}{Large momentum \texorpdfstring{$p \gg \omega$}{}}
\label{sec:asympt_dirac_large}
We find that $\bsq$ drops like $1/p^4$, which gives a logarithmic divergency of
the energy of the created particles. This divergency might be explained by the 
sudden transition between the Minkowski and de Sitter phases.
The analysis is done using the same method used in the scalar case: we 
substitute Hankel's expansion \eqref{eq:bessel_all_as_hankel_largez} for the 
Hankel functions in the terms \eqref{eq:dirac_beta_H}, for which we employ a 
similar notation:
\begin{gather}
	H_i(q, p) = \frac{4\omega}{\pi\sqrt{pq}}\left(
	\cos\qmpparant C_i(q, p) + \sin\qmpparant S_i(q, p)
	\right)
	\label{eq:dirac_large_CS_def}
\end{gather}
with the functions $C_i, P_i$ given by
\begin{subequations}\label{eq:dirac_large_CS}
\begin{align}
	C_1 &= \phantom{-}i\left(P^+_{q} Q^+_{p} - Q^+_{q} P^+_{p}\right) &
	C_2 &= P^+_{q} P^-_{p} + Q^+_{q} Q^+_{p}
	\label{eq:dirac_large_CS1}\\
	S_1 &= -i\left(P^+_{q} P^+_{p} + Q^+_{q} Q^+_{p}\right) &
	S_2 &= P^+_{q} Q^-_{p} - Q^+_{q} P^-_{p}
	\label{eq:dirac_large_CS2}
\end{align}
\end{subequations}
The polynomials $Q, P$ are given by \eqref{eq:P_hankel},\eqref{eq:Q_hankel}.
The notation $P^\pm_p$ stands for $P(\nu_{\pm}, p/\omega)$.
Contrary to the scalar case, the $C_i, S_i$ functions are not real.
We proceed in a similar fashion and express $\beta$ as a power series in
$\omega$, keeping $\nu_{\pm}$ independent:
\begin{equation}
	\beta(p) = \frac{i}{2} e^{-i (E_q \tin + E_p \tout)} 
	\sum_{n = 0}^{\infty} \Omega^{(n)}(q, p) \omega^n
	\label{eq:dirac_large_Omegan_def}
\end{equation}
We shall retain terms up to $\omega^2$:
\begin{align*}
	\Omega^{(0)}(q, p) &= \cos\qmpparant \mathcal{E}^{-++-}_-
	- i \sin\qmpparant \mathcal{E}^{++--}_-
	\\
	\Omega^{(1)}(q, p) &= -\frac{ikm}{2} \left(\frac{1}{p} - \frac{1}{q}\right)
	\left(\cos\qmpparant \mathcal{E}^{++--}_- 
	- i \sin\qmpparant \mathcal{E}^{-++-}_- \right)\\
	&-\frac{m}{2} \mathcal{E}^{++--}_+ \left(
	\cos\qmpparant \left(\frac{1}{p} - \frac{1}{q}\right) -
	i \sin\qmpparant\left(\frac{1}{p} + \frac{1}{q}\right)\right)\\
	\Omega^{(2)}(q, p) &= -\frac{ik(k^2 + 1)}{4} \left\{
	\cos\qmpparant\left(\frac{ik}{2} 
	\left(\frac{1}{p} - \frac{1}{q}\right)^2 \mathcal{E}^{++--}_-
	-\left(\frac{1}{p^2} - \frac{1}{q^2}\right) \mathcal{E}^{+--+}\right)
	\right.\\
	&\left.+ i \sin\qmpparant \left(\frac{1}{p^2} + \frac{1}{q^2}\right)
	\left(\frac{ik}{2} \mathcal{E}^{++--}_- + \mathcal{E}^{++--}_+\right)\right\}
	\\
	&+\frac{ik^2}{4qp} \sin\qmpparant \left(
	(1 - k^2) \mathcal{E}^{++--}_- + 2ik \mathcal{E}^{++--}_+\right)
\end{align*}
The terms $\mathcal{E}^{++--}_{-}$ stand for
\begin{equation}
	\mathcal{E}^{rslm}_{\pm} = 
	\epsilon^r_q \epsilon^s_p \pm \epsilon^l_q \epsilon^m_p
	\quad,\quad r,s,l,m \in \{-,+\}
\end{equation}
We have used $4\nu_{\pm}^2 - 1 = 4ik(ik\pm 1)$, etc.
The terms $\epsilon^{\pm}_p$ can be expressed in powers of $p$ as
\begin{equation*}
	\epsilon^{\pm}_p = 1 \pm \frac{m}{2E_p} + \frac{m^2}{8E_p^2} + 
	\mathcal{O}(p^{-3}),
\end{equation*}
and so the $\Omega$ terms simplify to
\begin{subequations}\label{eq:dirac_large_Omegan}
\begin{align}
	\Omega^{(0)}(q, p) &= m \mathcal{C}_-(q,p) + \mathcal{O}(p^{-3})
	\label{eq:dirac_large_Omega0}\\
	\omega \Omega^{(1)}(q, p) &= -\frac{ikm^2}{2} \left(\frac{1}{p} - \frac{1}{q}\right)
	\mathcal{C}_+(q,p) - m \mathcal{C}_-(q,p) + \mathcal{O}(p^{-3})
	\label{eq:dirac_large_Omega1}\\
	\omega^2 \Omega^{(2)}(q, p) &= -\frac{im \omega(k^2 + 1)}{2} 
	\left(\frac{1}{p} - \frac{1}{q}\right) \mathcal{C}_+(q,p) + 
	\frac{m\omega}{qp} \sin\qmpparant + \mathcal{O}(p^{-3})
	\label{eq:dirac_large_Omega2}
\end{align}
The recurrent terms are
\begin{equation}
	\mathcal{C}_{\pm}(q, p) = 
	\cos\qmpparant\left(\frac{1}{p} \pm \frac{1}{q}\right) - 
	i \sin\qmpparant\left(\frac{1}{p} \mp \frac{1}{q}\right)
	\label{eq:dirac_large_Omegarecurrent}
\end{equation}
\end{subequations}
We have used $k = m/\omega$. Summing the contributions, we are left with 
a leading term of order $1/p^2$, and thus we express $\beta$ as
\begin{multline}
	\beta(p) = - \frac{m\omega}{4} e^{-i (E_q \tin + E_p \tout)}\times\\
	\left\{ \cos\qmpparant\left(\frac{1}{p^2} - \frac{1}{q^2}\right)
	- i \sin\qmpparant\left(\frac{1}{p^2} + \frac{1}{q^2}\right)
	+\mathcal{O}(p^{-3})\right\}
	\label{eq:dirac_beta_large}
\end{multline}
All higher order terms contain $4\nu_{\pm}^2 - 1$,
which is proportional to the mass of the field, and thus we confirm the 
result that there is no particle creation in the massless case (derived in
\autoref{sec:kg_massless}). 
For the case of sufficiently large expansion time $\Delta t = \tout - \tin$ 
we approximate the square of $\beta$ with $\bsq_l$ defined by
\begin{equation}
	\bsq_l \xrightarrow[\tout \gg \tin]{} \frac{m^2\omega^2}{16p^4}
	\label{eq:dirac_bsql}
\end{equation}
Although this dependency guarantees a finite total number of particles, this
is not so with the total energy, which diverges.
\end{subsection}
\begin{subsection}{Low momentum \texorpdfstring{$q \ll \omega$}{}}
\label{sec:dirac_asympt_low}
While a constant number of scalar particles are created in the low momentum 
region, the number of spinorial particles approaches $0$ as $p^2$.

In the limit $p\ll 0, q\ll 0$ we shall use the approximation 
\eqref{eq:hankelnu_as_smallz} for the Hankel functions appearing in
\eqref{eq:dirac_beta_H}. The computation of $H_2$ is more demanding since the
Hankel functions involved have different orders $\nu_\pm$. Nevertheless, it
requires no special tricks, and the result is:
\begin{align}
	H_1(q, p) &= \frac{4 \sinh{\omega \Delta t \nu_+}}{i \pi \nu_+}\\
	H_2(q, p) &= \frac{2}{\pi \cosh{\pi k}} \sinh(\pi k - im\Delta t) \left(
	\sqrt{\frac{4\omega^2}{pq}} - \frac{1}{\abs{\nu_+}^2} 
	\sqrt{\frac{qp}{4\omega^2}} \right) + \nonumber\\
	&\qquad\frac{2i}{\pi^2} \cosh{\frac{1}{2}\omega \Delta t} \left(
	\Gamma_+^2(q, p)/\nu_- - \Gamma_-^2(q, p)/\nu_+\right)
\end{align}
The square of $\beta$ is needlessly cumbersome to compute. We shall 
approximate the square roots appearing in \eqref{eq:dirac_beta_explicit} 
through $\epsilon^\pm_p, \epsilon^\pm_q$ by their series expansion about
$p/m = 0$, ignoring terms of order $\mathcal{O}(p^2)$:
\begin{align*}
	\epsilon_q^- \epsilon_p^- &\sim \frac{pq}{2m^2} \simeq 0 &
	\epsilon_q^+ \epsilon_p^+ &\simeq \sqrt{2} \\
	\epsilon_q^- \epsilon_p^+ &\simeq \frac{q}{m} &
	\epsilon_q^+ \epsilon_p^- &\simeq \frac{p}{m}
\end{align*}
Substituting the above in the expression for $\beta$, we arrive at
\begin{multline}
	\beta(p) \xrightarrow[q\ll 0]{} e^{-i(E_q \tin + E_p \tout)} 
	\frac{\sqrt{qp}}{m} \left\{
	\frac{\sinh{\omega \Delta t \nu_+}}{\nu_+} + \right.\\
	\left.\frac{i}{2k \cosh{\pi k}} 
	\left( e^{\omega \Delta t / 2} \sinh(\pi k + i m\Delta t) - 
	e^{- \omega \Delta t / 2} \sinh(\pi k - i m\Delta t) \right) 
	+ \mathcal{O}(p^2) \right\}
	\label{eq:dirac_beta_small}
\end{multline}
We have dropped a term of order $\mathcal{O}(p^2)$. Therefore $\bsq$ goes to
$0$ as $p^2$ for $q \ll \omega$. 
\end{subsection}
\begin{subsection}{Middle region \texorpdfstring{$p \ll \omega \ll q$}{}}
\label{sec:dirac_asympt_middle}
The thermal spectrum of particle density not recovered in the scalar case
emerges for the Dirac field in the middle region, subject to the constraint
$p \ll m$. The flat plateau is given by a Fermi-Dirac distribution law of 
temperature $T = \omega / 2\pi$, for the energy $E = m$.

Applying the same reasoning outlined in \autoref{sec:kg_asympt_middle}, we 
use the first order approximation for large arguments of the Hankel functions
entering in the coefficients $A,B$ of argument $q$, and thus arrive at
\begin{equation}
	\tilde{A}(q, \tin) \simeq -i E^{-i  E_q \tin - i q/\omega + 3i \pi/4}
	\quad,\quad \tilde{B}(q, \tin) \simeq 0
\end{equation}
Thus the $\beta$ coefficient reduces to 
\begin{equation}
	\beta(p) \xrightarrow[p\ll\omega\ll q]{} \tilde{B}(p, \tout)
\end{equation}
For a quick derivation of the thermal behaviour of $\abs{\beta(p)}^2$, we may
proceed by using only the term of order $p^{-1/2}$ in the expansion
\eqref{eq:hankelnu1_as_smallz}, by which $\tilde{B}$ evaluates to
\begin{equation*}
	\tilde{B}(p, \tout) \xrightarrow[p\ll\omega\ll q]{} \frac{i}{\sqrt{4\pi}}
	e^{-i E_p \tout} \left(\epsilon^+_p e^{-\pi k/2} \Gamma_+(p,p) - 
	\epsilon^-_p e^{\pi k/2} \Gamma_-(p,p)\right)
\end{equation*}
and the square is
\begin{equation*}	
	\abs{\tilde{B}(p, \tout)}^2 \xrightarrow[p\ll\omega\ll q]{}
	\frac{1}{2}\left(1 - \frac{m}{E_p} \tanh{\pi k}\right) - 
	\frac{p}{4\pi E_p} \left(\Gamma_+^2(p,p) + \Gamma_-^2(p,p)\right)
\end{equation*}
The second term is at most $\pi^2/\cosh^2{\pi k}$, and is negligible for 
large enough mass or small enough expansion factor, but it becomes important 
as we depart from these conditions. However, this is not the ``right'' 
first order correction to $\abs{\beta(p)}^2$, as we shall point out 
later in this section. The term $m/E_p$ can be expanded about $p/m = 0$, and
we arrive at 
\begin{equation}
	\abs{\beta(p)}^2 \xrightarrow[p\ll\omega\ll q]{} 
	\frac{1}{e^{2\pi m/\omega} + 1} + \mathcal{O}\left(p/m\right)
	\label{eq:dirac_middle_as_o0}
\end{equation}
This resembles a Fermi-Dirac distribution function, if we let the energy
equal $m$ (just as in our approximation), and interpret $\omega/2\pi$ as
the temperature. This function approximates remarcably well the middle region
where the condition $p\ll\omega\ll q$ is valid. However, the extreme regions
of this plateau present two maxima that are highly pronounced when $\omega$
is small. In an attempt to approach this behaviour we consider both terms
in the expansion \eqref{eq:hankelnu1_as_smallz}. Note that in this 
approximation the identity \eqref{eq:hank_idcota} is no longer valid.
It is more convenient to first square $\tilde{B}(p,\tout)$ and then apply
the expansion of the Hankel functions:
\begin{equation*}
	\abs{B(p,\tout)} = \frac{\pi p}{8\omega} \left\{ \mathcal{H}_1(p) -
	\frac{m}{E_p} \mathcal{H}_2(p) - \frac{p}{E_p} \mathcal{H}_3(p)\right\},
\end{equation*}
with the terms $\mathcal{H}_i$ given by
\begin{align*}
	\mathcal{H}_1(p) &= e^{\pi k} \Hank{2}{\nu_+}\Hank{1}{\nu_-} + 
	e^{- \pi k} \Hank{2}{\nu_-} \Hank{1}{\nu_+}\\
	\mathcal{H}_2(p) &= e^{\pi k} \Hank{2}{\nu_+}\Hank{1}{\nu_-} - 
	e^{- \pi k} \Hank{2}{\nu_-} \Hank{1}{\nu_+}\\
	\mathcal{H}_3(p) &= \phantom{e^{\pi k}}\Hank{2}{\nu_+}\Hank{1}{\nu_+} + 
	\phantom{e^{-\pi k}}\Hank{2}{\nu_-} \Hank{1}{\nu_-}
\end{align*}
Every Hankel function is understood to take the argument $p/\omega$. 
These functions evaluate to
\begin{align*}
	\mathcal{H}_1(p) &= \frac{4\omega}{\pi p} + 
	\frac{p}{\omega \pi \abs{\nu_+}^2}\\
	\mathcal{H}_2(p) &= -\frac{4\omega}{\pi p} \tanh{\pi k} + \frac{2}{i\pi^2}
	\left(\Gamma^2_+(p,p)/\nu_- - \Gamma^2_-/\nu_+\right) + 
	\frac{p}{\pi \omega \abs{\nu_+}^2} \tanh{\pi k}
	+ \mathcal{O}(p^2)\\
	\mathcal{H}_3(p) &= 
	\frac{2\omega}{\pi^2 p} \left(\Gamma^2_+(p,p) + \Gamma^2_-(p,p)\right) + 
	\frac{4k}{\pi\abs{\nu_+}^2} \tanh{\pi k} + 
	\frac{p}{2\omega \pi^2} 
	\left(\Gamma^2_+(p,p)/\nu_-^2 + \Gamma^2_-(p,p)/\nu_+^2\right)
\end{align*}
Since we are also interested in the behaviour of $\bsq$ near $p\sim\omega$,
we cannot replace $E_p$ by $m$, and thus we settle to:
\begin{multline}
	\bsq \xrightarrow[p\ll\omega\ll q]{} \bsq_m = 
	\frac{1}{2}\left(1 - \frac{m}{E_p} \tanh{\pi k}\right) -
	\frac{p}{8\pi E_p}\left(\Gamma^2_+(p,p)/\nu_- + \Gamma^2_-(p,p)/\nu_+\right)+\\
	\frac{p^2}{8\omega^2\abs{\nu_+}^2}\left(1-\frac{3m}{E_p}\tanh{\pi k}\right) -
	\frac{p^3}{16\pi\omega^2 E_p}\left(\Gamma^2_+(p,p)/\nu^2_- + 
	\Gamma^2_-(p,p)/\nu^2_+\right)+	\dots
	\label{eq:dirac_middle_as}
\end{multline}
Note however that this expansion is not correct in the third order, but 
nevertheless it approaches the form of the exact solution better than
if we would have omitted this contribution. There is no reference to the
times $\tin, \tout$, which means the form of $\bsq$ in the middle region is
independent of the expansion time.
\end{subsection}
\begin{subsection}{Resonances}\label{sec:dirac_asympt_max}
There are three distinct regions of interest, for both the spinorial and 
the scalar case, separated by points situated roughly at $p \sim \omega$ 
and $q \sim \omega$. We have studied the asymptotic behaviour in these
three regions in the preceding sections. This section is devoted to the 
analytical investigation of the pronounced maxima of produced fermions which
occur at these points.

The analysis starts with the explicit form of $\bsq$, given in 
\autoref{eq:dirac_betasq_explicit}. The extrema can be identified by setting
$d\bsq/dp = 0$. Since the functions $\mH_i$ 
(defined by \eqref{eq:dirac_betasq_Hrond}) employed in the 
expression for $\abs{\beta}^2$ have an explicit dependence on $q$ and $p$,
we write the total derivative with respect to $p$ as
\begin{equation*}
	\frac{d}{dp} \mH_i(q, p) = \left(\partialdiv{\mH_i}{p}\right)_q+
	\left(\partialdiv{\mH_i}{q}\right)_p
\end{equation*}
First we compute the derivatives of the coefficients of $\mH_i$:
\begin{subequations}\label{eq:dirac_max_ddp}
\begin{align}
	\frac{d}{dp}p &= 1 & \frac{d}{dp}q &= \frac{q}{p} &
	\frac{d}{dp}qp &= 2q\\
	\frac{d}{dp}\frac{p}{E_p} &= \frac{m^2}{E_p^3} &
	\frac{d}{dp}\frac{q}{E_q} &= \frac{q m^2}{p E_q^3} &
	\frac{d}{dp} \frac{qp}{E_qE_p} &= \frac{m^2 q}{E_q E_p}
	\left(\frac{1}{E_p^2} + \frac{1}{E_q^2}\right)
\end{align}
\end{subequations}
Next, we compute the derivatives of $H_i(q, p)$ (defined in 
\eqref{eq:dirac_beta_H}):
\begin{subequations}\label{eq:dirac_dH}
\begin{align}
	\frac{d}{dp} H_1(q,p) &= \frac{i}{\omega p}\left(p H_2 - q H_2^*\right)
	- \frac{2ik}{p} H_1 - \frac{1}{p} H_1\\
	\frac{d}{dp} H_2(q,p) &= \frac{i}{\omega p}\left(p H_1 + q H_1^*\right)
	- \frac{1}{p} H_2
\end{align}
\end{subequations}
With these we compute the derivatives of $\mH_i$:
\begin{subequations}\label{eq:dirac_dHrond}
\begin{align}
	\frac{d}{dp} \mH_1(q,p) &= \frac{2i}{\omega}
	\left(H_1^* H_2 - H_1 H_2^*\right) - \frac{2i q}{\omega p} 
	\left(H_1^* H_2^* - H_1 H_2\right) - \frac{2}{p} \mH_1\\
	\frac{d}{dp} \mH_2(q,p) &= \frac{i}{\omega}\left(
	H_1^2 - H_1^{*\, 2} + H_2^2 - H_2^{*\, 2}\right) + \frac{2ik}{p} 
	\left(H_1^* H_2^* - H_1 H_2\right) - \frac{2}{p} \mH_2\\
	\frac{d}{dp} \mH_3(q,p) &= -\frac{iq}{\omega p}\left(
	H_1^2 - H_1^{*\, 2} - H_2^2 + H_2^{*\, 2}\right) + \frac{2ik}{p} 
	\left(H_1^* H_2 - H_1 H_2^*\right) - \frac{2}{p} \mH_3\\
	\frac{d}{dp} \mH_4(q,p) &= -\frac{2 ik}{p}\left(
	H_1^2 - H_1^{*\, 2}\right) - \frac{2}{p} \mH_4
\end{align}
These derivatives satisfy the remarcable property
\begin{multline}
	k^2 \frac{d}{dp} \mH_1(q,p) + 
	k \frac{q}{\omega} \frac{d}{dp} \mH_2(q,p) - 
	k\frac{p}{\omega} \frac{d}{dp} \mH_3(q,p) + 
	\frac{qp}{\omega^2} \frac{d}{dp} \mH_4(q,p) = \\
	-\frac{2}{p}
	\left\{k^2 \mH_1(q,p) + k \frac{q}{\omega} \mH_2(q,p) -
	k\frac{p}{\omega}\mH_3(q,p) + 
	\frac{qp}{\omega^2} \mH_4(q,p)\right\}
\end{multline}
\end{subequations}
Using these results, the derivative of $\abs{\beta}^2$ yields:
\begin{multline}
	\frac{d}{dp} \abs{\beta(p)}^2 = \frac{\pi^2 q}{32 E_q E_p} 
	\left\{-k^2 \left(\frac{q^2}{E_q^2} + \frac{p^2}{E_p^2}\right) \mH_1+
	\right.\\\left.
	\left(1 - \frac{q^2}{E_q^2} - \frac{p^2}{E_p^2}\right) 
	\left(\frac{kq}{\omega}\mH_2 - \frac{kp}{\omega}\mH_3\right) + 
	qpk^2 \left(\frac{1}{E_q^2} + \frac{1}{E_p^2}\right) \mH_4\right\}
	\label{eq:dirac_dbetasq}
\end{multline}
In the following subsections we shall work in the asymptotic cases 
$q \sim \omega \gg p $ and $p \sim \omega \ll q$, near the points where
we hope to uncover the observed maxima.
\begin{subsubsection}{First maximum \texorpdfstring{$q \sim \omega \gg p$}{(small p)}}
\label{sec:asympt_maxlow}
We shall approximate the Hankel functions of argument $p/\omega$ with the first
term in \eqref{eq:hankelnu_as_smallz}. In this approximation, all $\mH_i$ 
functions are of order $1/p$, allowing us to eliminate terms with coefficients
of order $p$ in \eqref{eq:dirac_dbetasq}:
\begin{equation}
	\frac{d}{dp} \abs{\beta(p)}^2 \xrightarrow[p\rightarrow 0]{} 
	\frac{\pi^2 k^2 q^2}{32 E_q^3} 
	\left(-\frac{q}{m} \mH_1 + \mH_2\right)
	\label{eq:dirac_dbetasq_smallp}	
\end{equation}
The equation which determines the point of extremum is
\begin{equation}
	\frac{q}{m} \mH_1(q, p) = \mH_2(q, p)
	\label{eq:dirac_maxlow_eq}
\end{equation}
The functions $\mH_i$ take the form
\begin{subequations}
\begin{multline*}
	\mH_1(q,p)\xrightarrow[p\rightarrow 0]{} -\frac{4\omega}{\pi p \cosh{\pi k}}
	\left\{\sinh{\pi k} \left(e^{\pi k} \Hank{2}{\nu_+} \Hank{1}{\nu_-} - 
	e^{-\pi k} \Hank{2}{\nu_-} \Hank{1}{\nu_+}\right) - 
	\right.\\\left.
	\left(\Hank{2}{\nu_+} \Hank{2}{\nu_-} +
	\Hank{1}{\nu_+} \Hank{1}{\nu_-}\right)\right\}
\end{multline*}
\begin{multline*}
	\mH_2(q,p)\xrightarrow[p\rightarrow 0]{} -\frac{4\omega}{\pi p \cosh{\pi k}}
	\left\{\sinh{\pi k} \left(\Hank{2}{\nu_+} \Hank{1}{\nu_+} +
	\Hank{2}{\nu_-} \Hank{1}{\nu_-}\right) + \phantom{\frac{1}{2}}
	\right.\\\left.
	\frac{1}{2} e^{\pi k} \left(\Hank{2}{\nu_+}^2 + \Hank{1}{\nu_-}^2\right) -
	\frac{1}{2} e^{-\pi k}\left(\Hank{2}{\nu_-}^2 + \Hank{1}{\nu_+}^2\right)
	\right\}
\end{multline*}
\end{subequations}
The Hankel functions in the right-hand-side of the above expressions take the
argument $q/\omega$. These forms are rather cumbersome to work with. 
Nevertheless, we find that the results obtained for large $\omega$ 
($\nu_{\pm} \rightarrow 1/2$) give reasonably good predictions. In this limit,
equation \eqref{eq:dirac_maxlow_eq} takes the form
\begin{equation*}
	\sinh{\pi k} \left(1 - \frac{q}{m} \sinh{\pi k}\right) = 
	\left(\frac{q}{m} + \sinh{\pi k} \right) \cos{\frac{2q}{\omega}}	
\end{equation*}
In the limit of large $\omega$, we can substitute $\sinh{\pi k}$ with
$\pi k$, and thus we arrive at
\begin{equation}
	p_{\text{max}} = \frac{\pi \omega}{4} e^{-\omega (\tout - \tin)}
	\label{eq:dirac_maxlow_p}
\end{equation}
This formula becomes a good prediction only for $\omega > m$, but the 
predicted value falls farther away as we depart from this condition. Note
that this value depends on the difference $\tout - \tin$.
\end{subsubsection}
\begin{subsubsection}{Second maximum \texorpdfstring{$p \sim \omega \ll q$}{(large q)}}
\label{sec:asympt_maxhigh}
Although the position of the maximum situated at $q\sim \omega$ is given by
a simple relation \eqref{eq:dirac_maxlow_p}, we were unable to find a similar 
one for the maximum at $p\sim \omega$. However, we can extract several 
conclusions out of the asymptotic form of the equation which gives the 
point. In the region $p \sim \omega$, we can consider $q$ as infinitely large,
and approximate the Hankel functions of argument $q/\omega$ with the asymptotic
forms \eqref{eq:bessel_all_as_largez}. In this approximation, all $\mH_i$ 
functions are of order $1/q$, and thus we shall retain only the highest order 
in $q$ from \eqref{eq:dirac_dbetasq}:
\begin{equation}
	\frac{d}{dp} \abs{\beta(p)}^2 \xrightarrow[q\rightarrow \infty]{} 
	\frac{\pi^2 k^2 q p}{32 E_p^3} 
	\left(-\frac{p}{m} \mH_2 + \mH_4\right)
	\label{eq:dirac_dbetasq_largeq}	
\end{equation}
The equation which determines the point of extremum is
\begin{equation}
	\frac{p}{m} \mH_2(q, p) = \mH_4(q, p)
	\label{eq:dirac_maxlarge_eq}
\end{equation}
This is very similar to equation \eqref{eq:dirac_maxlow_eq}, which determines
the point of maximum at $q\sim \omega$. We proceed in a similar fashion, and
evaluate the functions $\mH_i$:
\begin{subequations}\label{eq:dirac_maxlarge_Hrond}
\begin{align}
	\mH_2(q,p)&\xrightarrow[q\rightarrow \infty]{} -\frac{4\omega}{\pi q}
	\left(e^{\pi k} \Hank{2}{\nu_+} \Hank{1}{\nu_-} - 
	e^{-\pi k} \Hank{2}{\nu_-} \Hank{1}{\nu_+}\right)\\
	\mH_4(q,p)&\xrightarrow[q\rightarrow \infty]{} -\frac{4\omega}{\pi q}
	\left(\Hank{2}{\nu_+} \Hank{1}{\nu_+} +	
	\Hank{2}{\nu_-} \Hank{1}{\nu_-}\right)
\end{align}
\end{subequations}
It is understood that the Hankel functions take the argument $p/\omega$.
This gives accurate results as long as we have corresponding numerical software
to solve the equation \eqref{eq:dirac_maxlarge_eq}. 

Taking the limit $\omega \rightarrow \infty$ (with $m$ finite) simplifies 
\eqref{eq:dirac_maxlarge_Hrond} to
\begin{equation}
	\mH_2(q,p)\xrightarrow[\substack{q\rightarrow \infty\\m \ll \omega}]{} 
	-\frac{16\omega^2}{\pi^ qp} \sinh{\pi k} \quad,\quad
	\mH_4(q,p)\xrightarrow[\substack{q\rightarrow \infty\\m \ll \omega}]{} 
	-\frac{16\omega^2}{\pi^ qp}
\end{equation}
This predicts the maximum at
\begin{gather}
	p_{\text{max}} = \frac{\omega}{\pi} 
	\label{eq:dirac_maxlarge_p_o0}
\end{gather}
This result is not accurate, since for increasing $\omega$ it 
predicts a shift to the right for the point of maximum, which contravenes
numerical results by which we find that the point actually shifts to the
left. However, there is one important conclusion to be drawn: the position
of the maximum does not depend on $\tin$ nor on $\tout$, as long as 
$\tout - \tin$ is sufficiently large to regard $q/\omega$ as approaching
infinity.

A further approximation can be done by assuming that $p/\omega < 1$ (which
is not true for small masses). In this case we substitute 
\eqref{eq:hankelnu_as_smallz} in \eqref{eq:dirac_maxlarge_Hrond} and arrive
at
\begin{align*}
	\mH_2 &\xrightarrow[\substack{q\rightarrow \infty\\p \ll \omega}]{} 
	- \frac{4 \omega}{\pi^2 q}\left\{ 
	\left(\frac{4\omega}{p} - \frac{p}{\omega \abs{\nu_+}^2}\right) \tanh{\pi k}+
	\frac{2i}{\pi}\left(\frac{\Gamma_+^2(p,p)}{\nu_-} - 
	\frac{\Gamma_-^2(p,p)}{\nu_+}\right) \right\}\\
	\mH_4 &\xrightarrow[\substack{q\rightarrow \infty\\p \ll \omega}]{} 
	- \frac{4 \omega}{\pi^3 q}\left\{ 
	\frac{2\omega}{p}\left(\Gamma_+^2(p,p) + \Gamma_-^2(p,p)\right) + 
	\frac{p}{2\omega}\left(\frac{\Gamma_+^2(p,p)}{\nu_-^2} + 
	\frac{\Gamma_-^2(p,p)}{\nu_+^2}\right) + 
	\frac{4\pi k}{\abs{\nu_+}^2} \tanh{\pi k} \right\}
\end{align*}
Substituting in \eqref{eq:dirac_dbetasq_largeq}, the derivative of 
$\abs{\beta}^2$ follows:
\begin{multline}
	\frac{d}{dp} \abs{\beta(p)}^2 
	\xrightarrow[\substack{q\rightarrow \infty\\p \ll \omega}]{}
	\frac{m^2}{8 E_p^3} \left\{ -\frac{2}{\pi} 
	\left(\Gamma_+^2(p,p) + \Gamma_-^2(p,p)\right) + 
	\frac{\tanh{\pi k}}{k \abs{\nu_+}^2} \frac{p}{\omega} + 
	\right.\\\left.
	\left(\frac{i}{k\pi}\left(\frac{\Gamma_+^2(p,p)}{\nu_-^2} - 
	\frac{\Gamma_-^2(p,p)}{\nu_+^2}\right) + 
	\frac{3}{2\pi} \left(\frac{\Gamma_+^2(p,p)}{\nu_-^2} +
	\frac{\Gamma_-^2(p,p)}{\nu_+^2}\right)\right) \frac{p^2}{\omega^2} - 
	\frac{\tanh{\pi k}}{k \abs{\nu_+}^2} \frac{p^3}{\omega^3}\right\}
	\label{eq:dirac_dbetasq_largeq_smallp}		
\end{multline}
Analytically, nothing much is gained because of the presence of the functions
$\Gamma_\pm(p,p)$ (defined by \eqref{eq:def_gammapm}). However, if the ratio
$\omega/m$ is not less than $1/2$, we obtain good predictions for the 
position of the maximum. If $m$ is increased, the point of maximum becomes
larger than $\omega$, and the approximation $p \ll \omega$ is no longer valid.
This region is not accessible by this method.
\end{subsubsection}
\end{subsection}
\begin{subsection}{The number of created particles}\label{sec:dirac_asympt_N}
Because the asymptotic form for the middle region $p\ll\omega\ll q$ contains
$E_p$, which cannot be replaced by a series expansion in powers of $p$. Such
an expansion departs too much from the exact solution. A second difficulty
is related to the choice of the value $p_\nu$ at which the transition between
the middle and the large asymptotic forms is to be made. Because the 
``order 0'' term in \eqref{eq:dirac_middle_as} contains the energy $E_p$, it 
is difficult to solve the equation $\bsq_l(p_\nu) = \bsq_m(p_\nu)$ even in
the first order. We propose, as in the scalar case, to have $p_\nu$ determined
numerically, such that the equation is satisfied. The transition between 
$\bsq_s$ and $\bsq_m$ will be done at $q = \omega e^{-\omega/2}$, because the
asymptotic form for small $p$ exhibits a violent increase, following the
increase in $\bsq$ (corresponding to the appearence of the first resonance at
$q = \pi \omega/4$, given by \eqref{eq:dirac_maxlow_p}).
\begin{gather}
	\bsq_{as} = \begin{cases}
		\bsq_s & q < \omega e^{-\omega/2}\\
		\bsq_m & \omega e^{-\omega/2} < q \text{and} p < p_\nu\\
		\bsq_l & p > p_\nu
	\end{cases}, \label{eq:dirac_bsqas}	\\
	\bsq_l(p_\nu) = \bsq_m(p_\nu). \label{eq:dirac_pnudef}
\end{gather}
Performing the integral of $\bsq_m$ brings little insight because of the terms 
containing products of $\Gamma_\pm(p,p) / E_p$, which give integrals of the form
\begin{equation}
	\int \frac{x^a}{\sqrt{x^2 + 1}} dx = \frac{x^{a+1}}{a+1}
	{}_2F_1\left(\frac{1 + a}{2}, \frac{1}{2}; 1 + \frac{1 + a}{2}; -x^2\right).
\end{equation}
For numerical values of the number of particles we rely on numerical 
computation results, which we shall present in the following subsection.
\end{subsection}
\begin{subsection}{Graphical comparison to the exact solution}\label{eq:dirac_asympt_fig}
In the following we shall present graphical illustrations of the results of
the previous subsections. We begin by representing the middle region 
(\autoref{fig:dirac_aso}) through a plot of $\bsq$ for large $\omega$. The 
plateau is described by a Fermi-Dirac distribution function 
$(e^{2\pi k} + 1)^{-1}$ \eqref{eq:dirac_middle_as_o0}, represented in dark 
colour. 
\begin{figure}[h!t]
	\centering
	\includegraphics[width=0.8\linewidth]{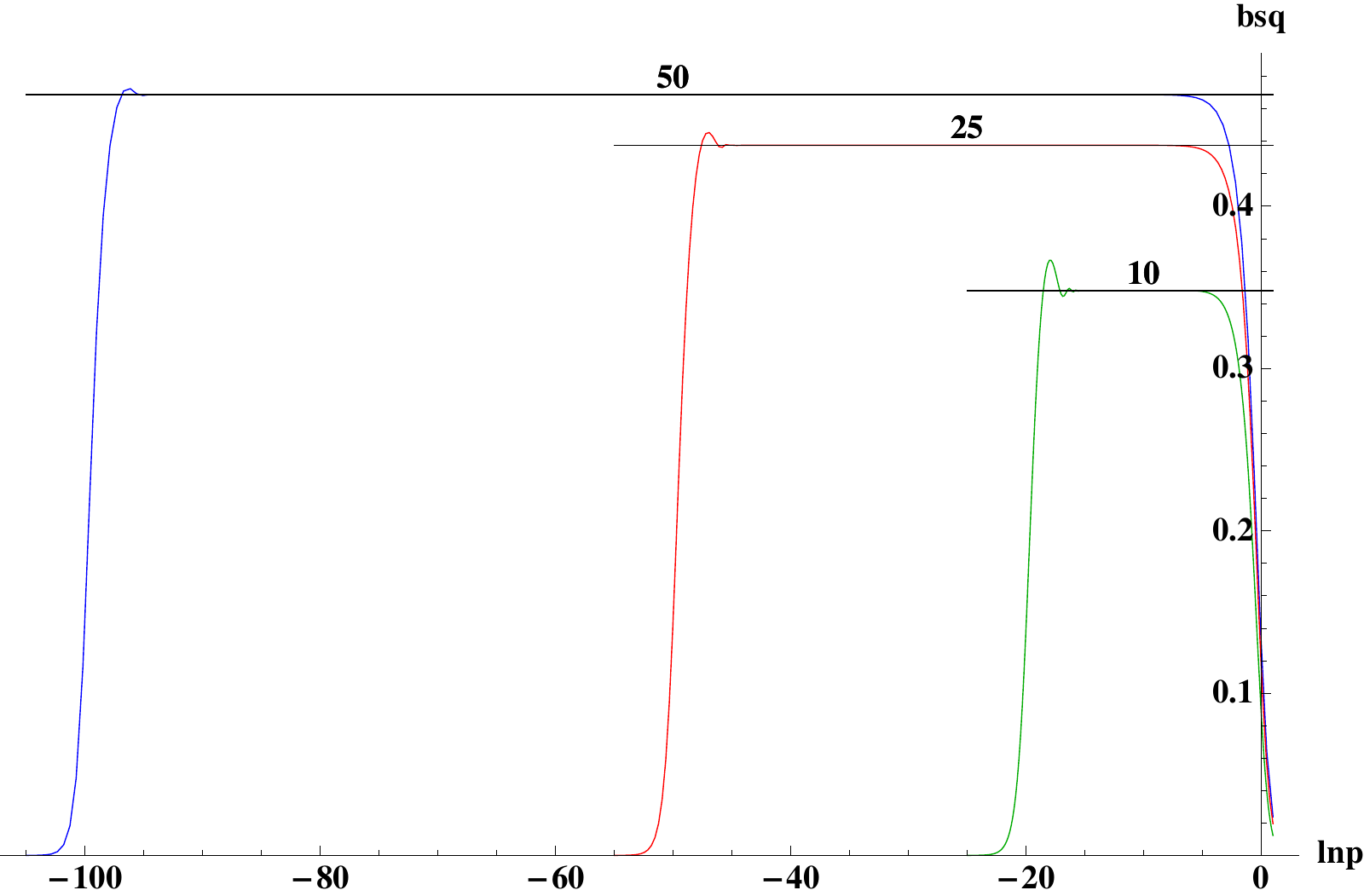}
	\caption{\small $\bsq$ represented against $\ln p$ for 
	$\omega=\colthree{50}{25}{10}$. The dark horizontal lines represent the 
	Fermi-Dirac distribution function of temperature $\omega/2\pi$ and energy
	$m$.}
	\label{fig:dirac_aso}
\end{figure}

The low-energy resonance, given by $q = \pi \omega/4$ \eqref{eq:dirac_maxlow_p}
are indicated through blue dots in \autoref{fig:dirac_t1_max}. The 
formula used is not accurate for $\omega < 1.5$, in which case the points 
have a tendency to shift left. But for high $\omega$, the approximation
becomes reliable.
\begin{figure}[h!t]
	\centering
	\includegraphics[width=0.8\linewidth]{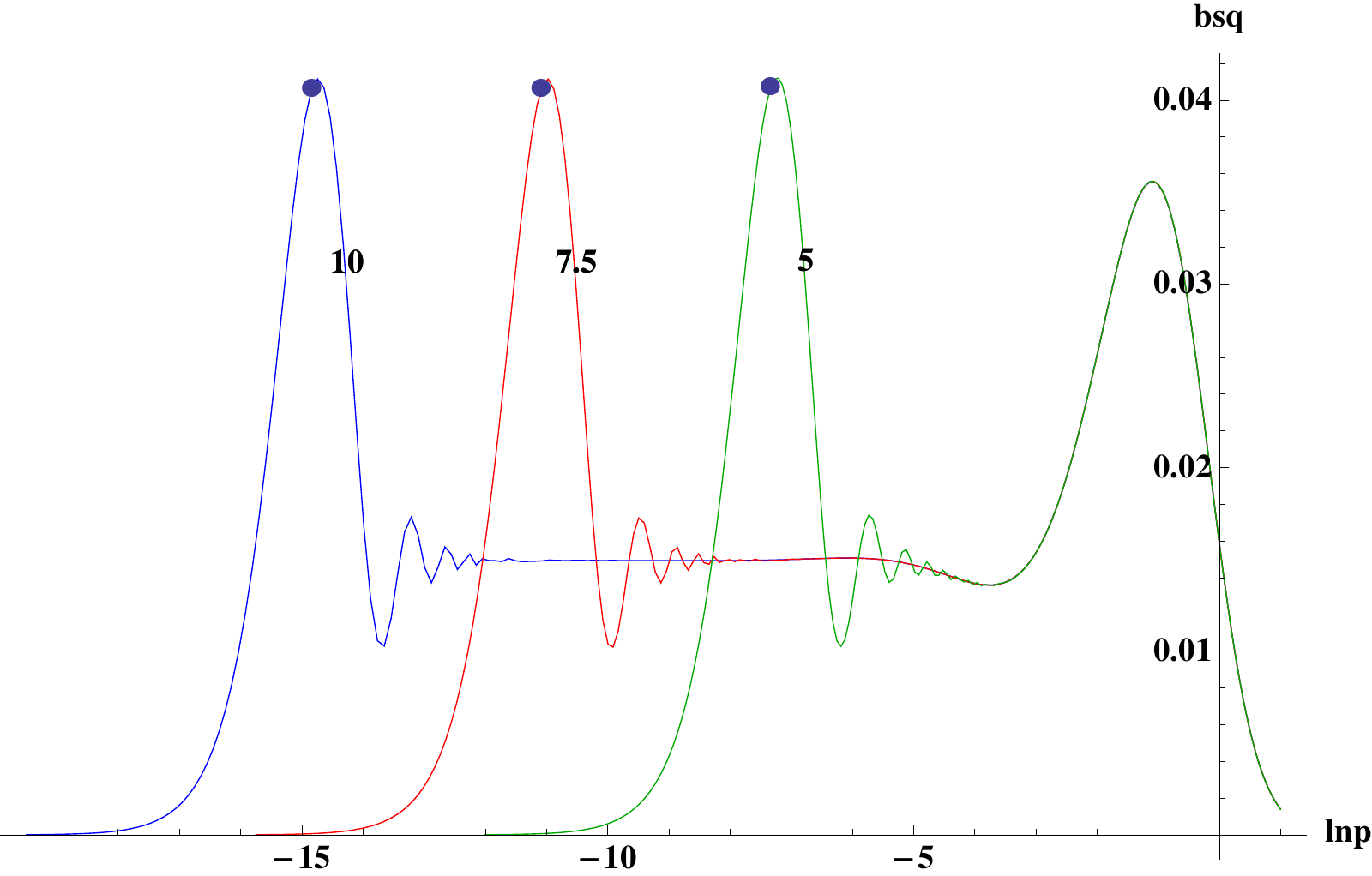}
	\caption{\small $\bsq$ plotted against $\ln p$ for $m=1$, $\tin = 0$, 
	$\omega = 1.5$ and $\tout=\colthree{10}{7.5}{5.0}$. Blue dots indicate the
	predicted points of maxima.}
	\label{fig:dirac_t1_max}
\end{figure}

The asymptotic form $\eqref{eq:dirac_bsqas}$ is compared with the exact 
solution in \autoref{fig:dirac_asallo}. The first maximum is not well captured,
and there are (important) differences near $p \sim \omega$. 
\begin{figure}[h!t]
	\centering
	\includegraphics[width=0.8\linewidth]{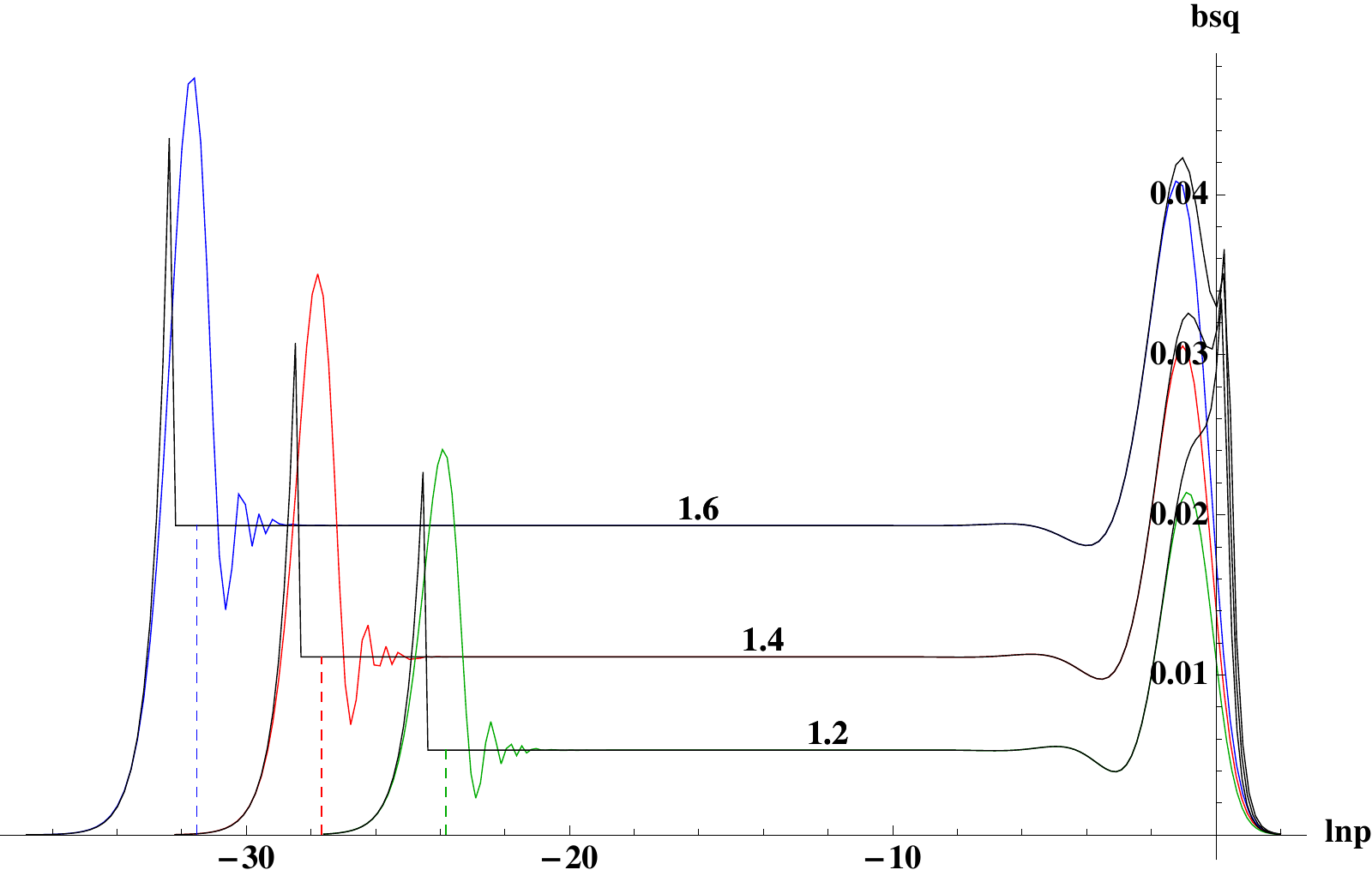}
	\caption{\small lin-log plot $\omega=\colthree{50}{25}{10}$}
	\label{fig:dirac_asallo}
\end{figure}

The value $p_\nu$
at which the transition between the middle and the large asymptotic forms 
occurs has been numerically computed using \eqref{eq:dirac_pnudef}. We give
the dependency of $p_\nu$ on $m$ (\autoref{fig:dirac_pnu_o}) and on $\omega$ 
(\autoref{fig:dirac_pnu_m}).
\begin{figure}[h!t]
  \begin{minipage}[b]{0.47\linewidth}
    \centering
    \includegraphics[width=\linewidth]{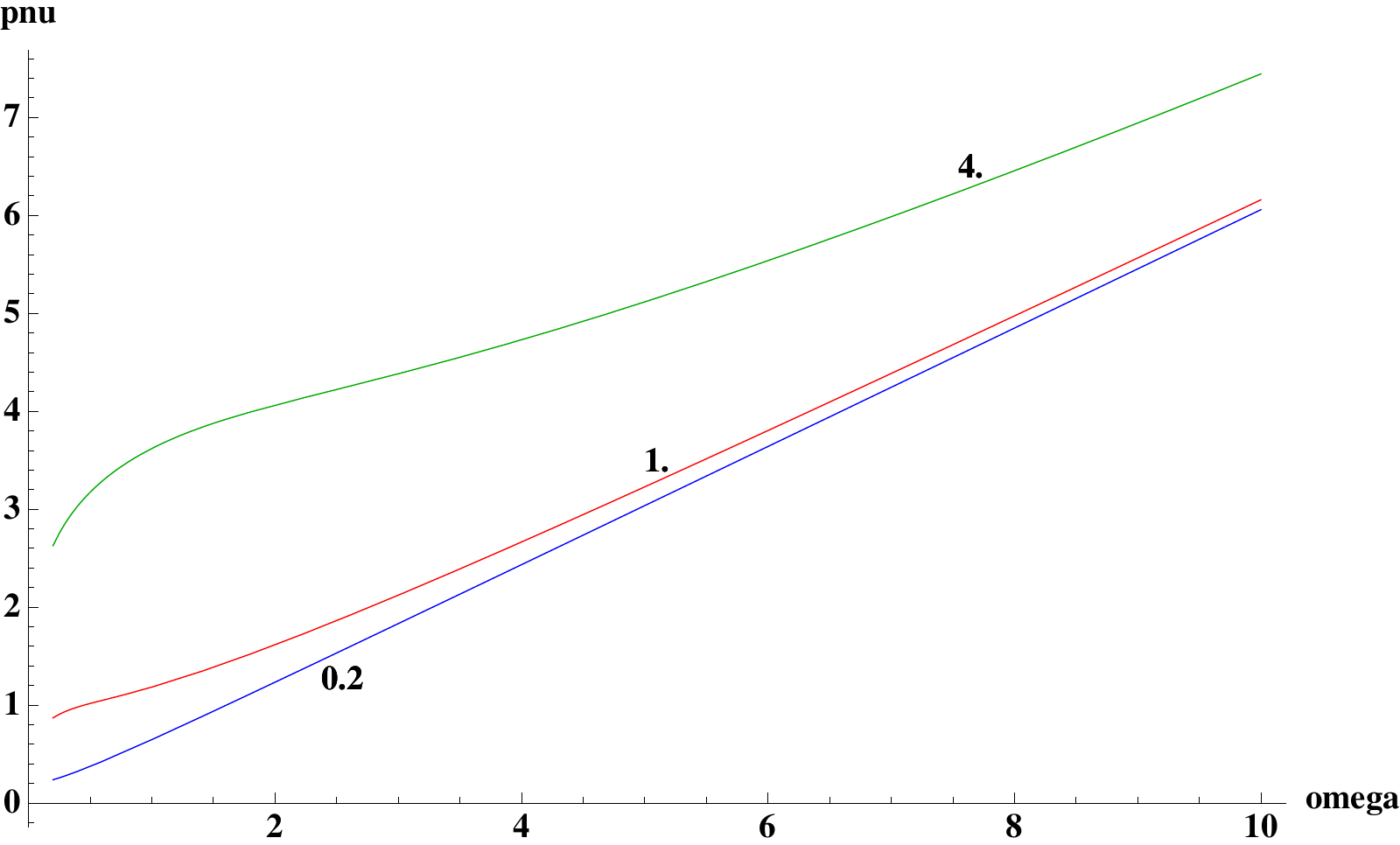}
    \caption{\small $p_\nu$ plotted against $\omega$ for 
    $m=\colthree{0.2}{1}{4}$.}
    \label{fig:dirac_pnu_o}
  \end{minipage}
  \hspace{0.03\linewidth} 
  \begin{minipage}[b]{0.47\linewidth} 
    \centering
    \includegraphics[width=\linewidth]{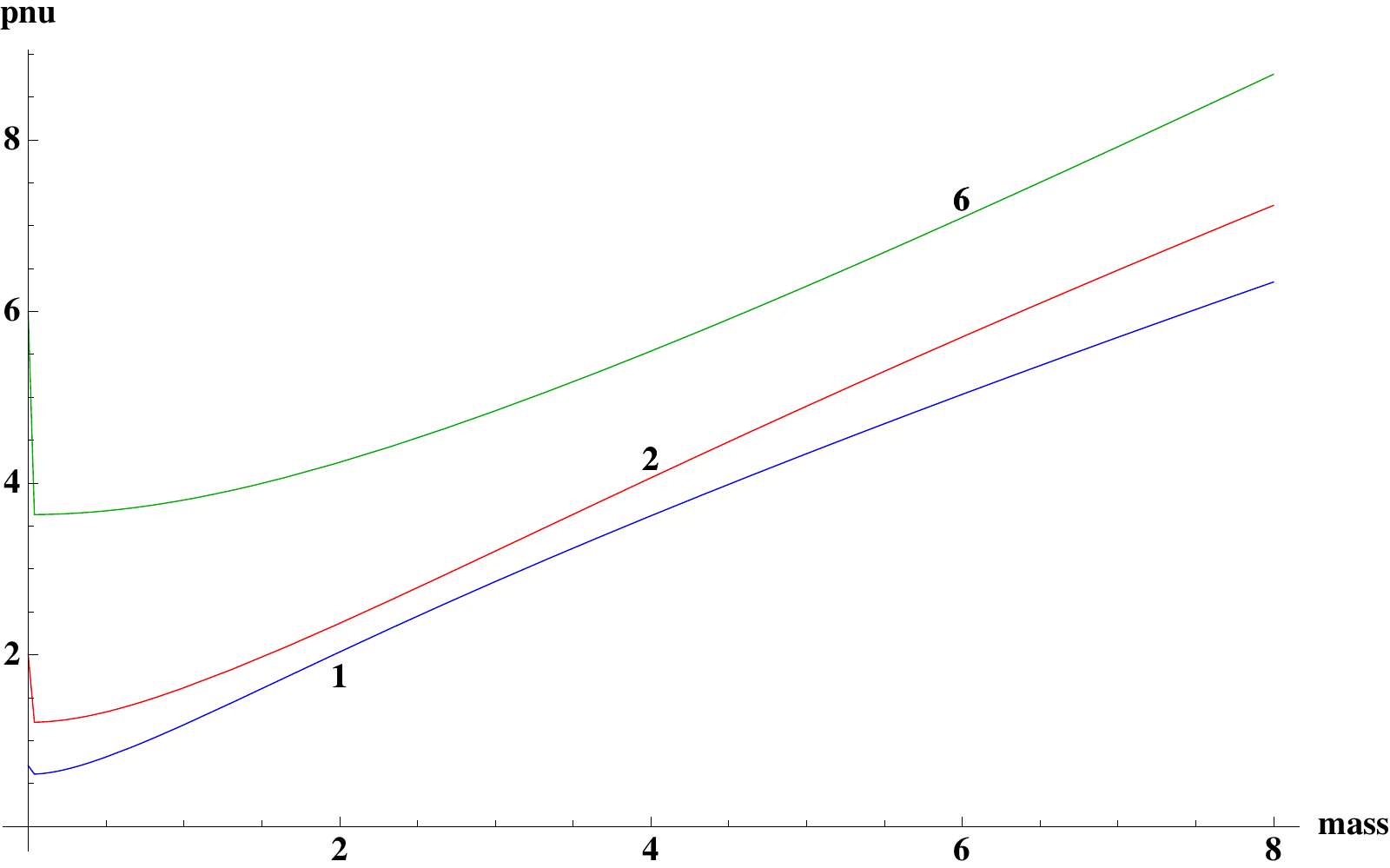}
    \caption{\small $p_\nu$ plotted against $m$ for 
    $\omega=\colthree{1}{2}{6}$.}
    \label{fig:dirac_pnu_m}
  \end{minipage}
\end{figure}

We show in more detail the amount of displacement between the asymptotic form 
and the exact solution near $p \sim \omega$ for $\bsq$ 
(\autoref{fig:dirac_aso_largep}) and for the particle number density
$2/\pi^2 p^2 \bsq$ (\autoref{fig:dirac_aso_np}).
\begin{figure}[h!t]
  \begin{minipage}[b]{0.47\linewidth}
    \centering
    \includegraphics[width=\linewidth]{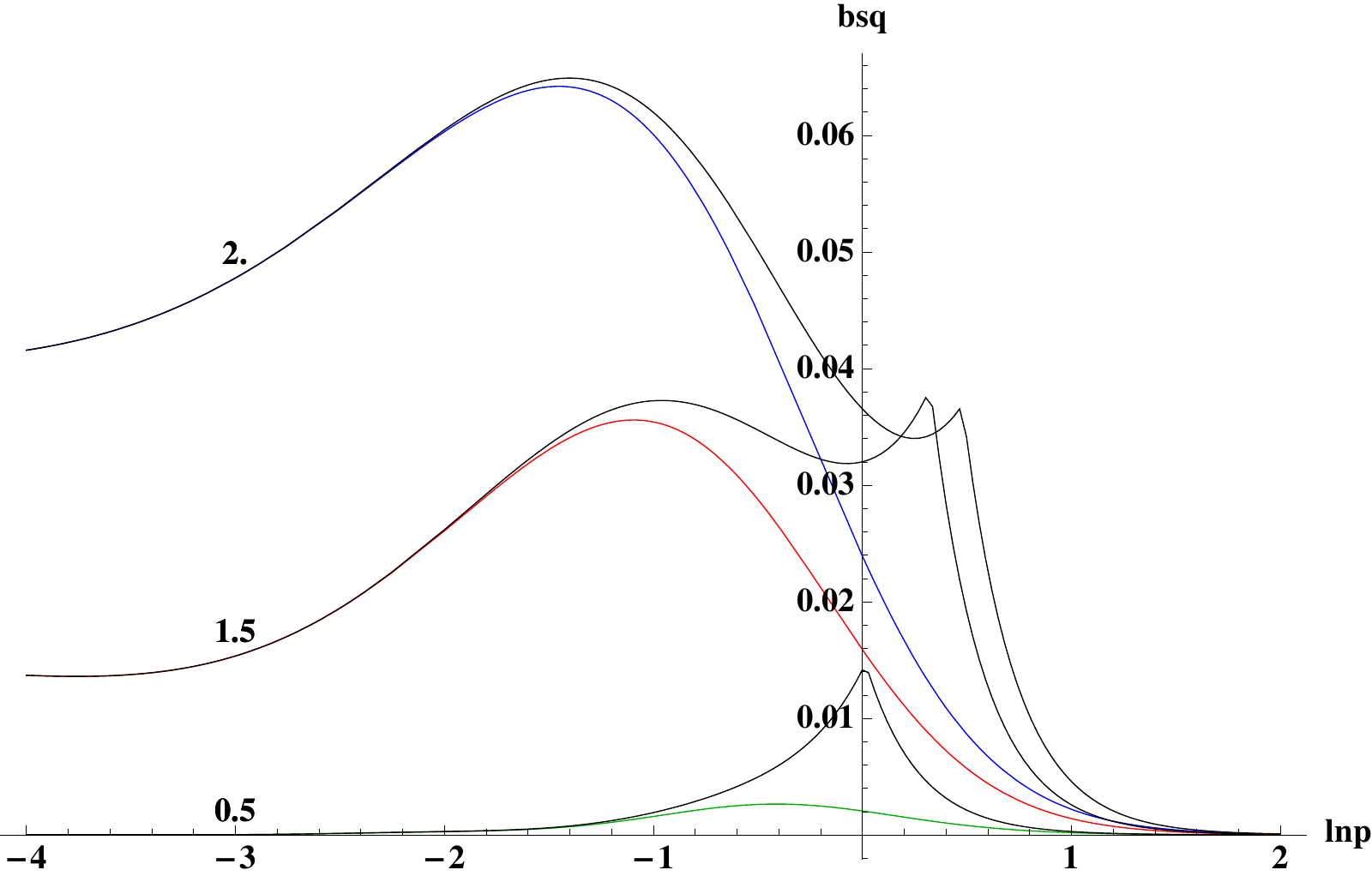}
    \caption{\small The dark lines represent $\bsq_{as}$, which we compare to 
    $\bsq$ for $\omega=\colthree{2}{1.5}{0.5}$.}
    \label{fig:dirac_aso_largep}
  \end{minipage}
  \hspace{0.03\linewidth} 
  \begin{minipage}[b]{0.47\linewidth} 
    \centering
    \includegraphics[width=\linewidth]{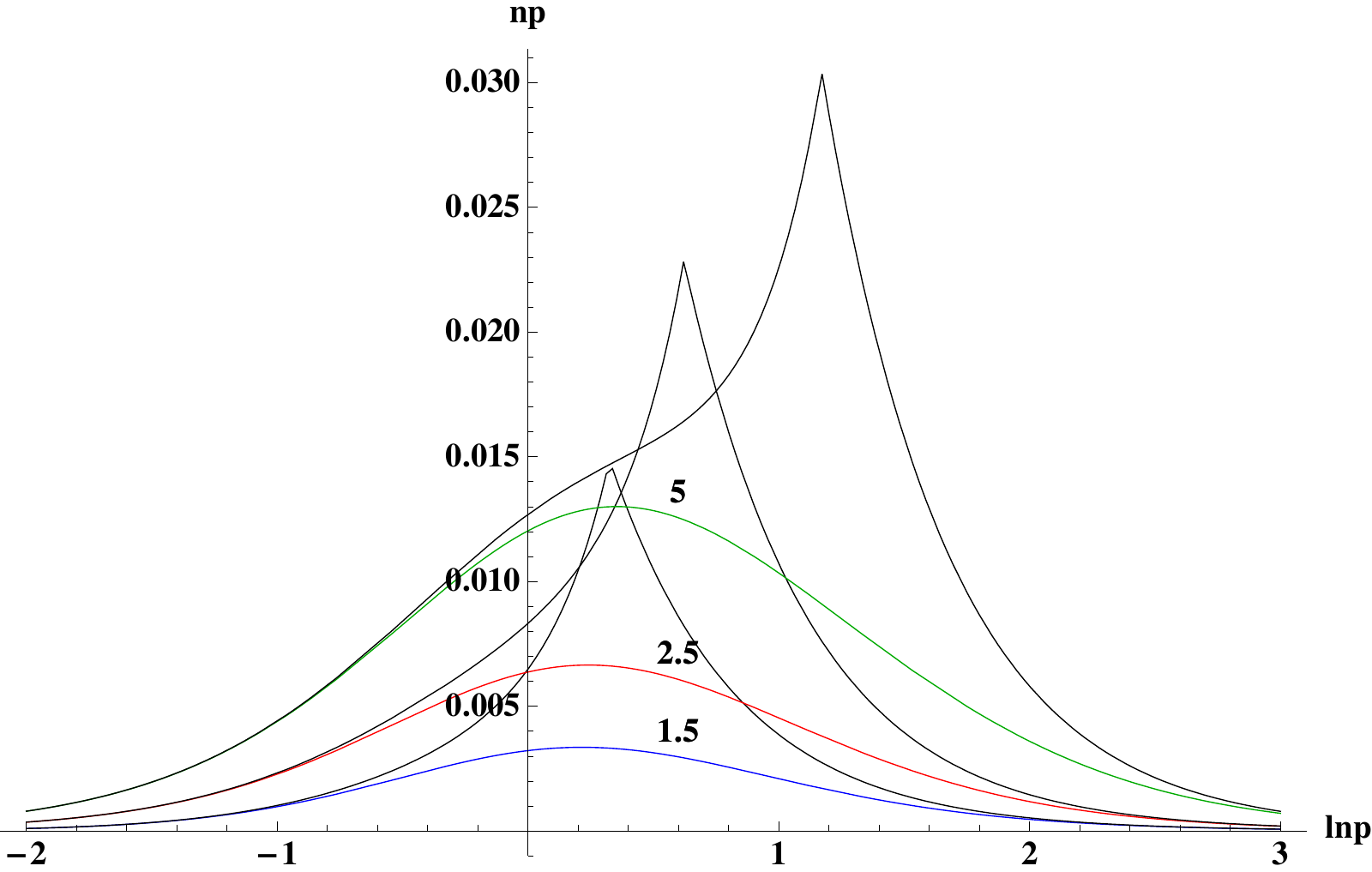}
    \caption{\small The asymptotic form of $n_p$ (plotted in black) is 
    compared to the exact $n_p$ at $\omega=\colthree{1.5}{2.5}{5}$.}
    \label{fig:dirac_aso_np}
  \end{minipage}
\end{figure}

Next we turn our attention to the particle number volumic density:
\begin{equation*}
	n(\vx) = \int_{0}^{\infty} \frac{2}{\pi^2} p^2\,\bsq.
\end{equation*}
The displacement between the asymptotic form and the exact solution causes an
increase in the asymptotic particle number density by a factor. This factor can
be read off from the logarithmic plot in \autoref{fig:dirac_lnNdiff_mass}, 
where we show the result of the numerical integration $\ln n(\vx)$ as a 
function of $m$, at $\tin = 0$, $\tout = 10$ and $\omega = 1.0$.
We find that by dividing the asymptotic value through $1.8$ the two curves
overlap on most of the domain. 
\begin{figure}[h!t]
	\centering
	\includegraphics[width=0.8\linewidth]{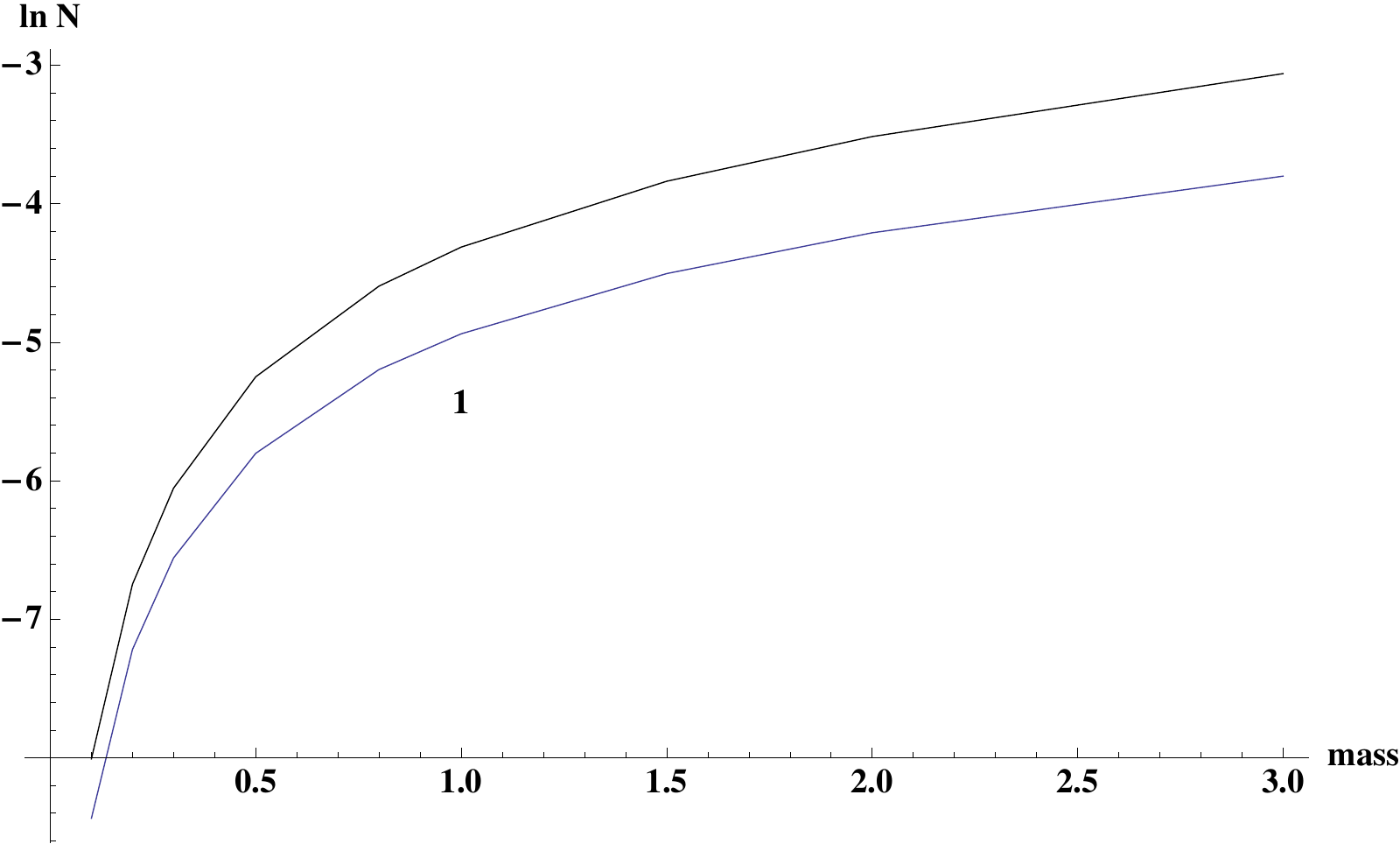}
	\caption{\small $\ln n(\vx)$ as a function of the particle mass $m$ 
	at $\omega=1$. The \cst{blue} curve represents the exact solution, while
	the asymptotic one is reprezented using dark colour.}
	\label{fig:dirac_lnNdiff_mass}
\end{figure}

\begin{figure}[h!t]
	\centering
	\includegraphics[width=0.8\linewidth]{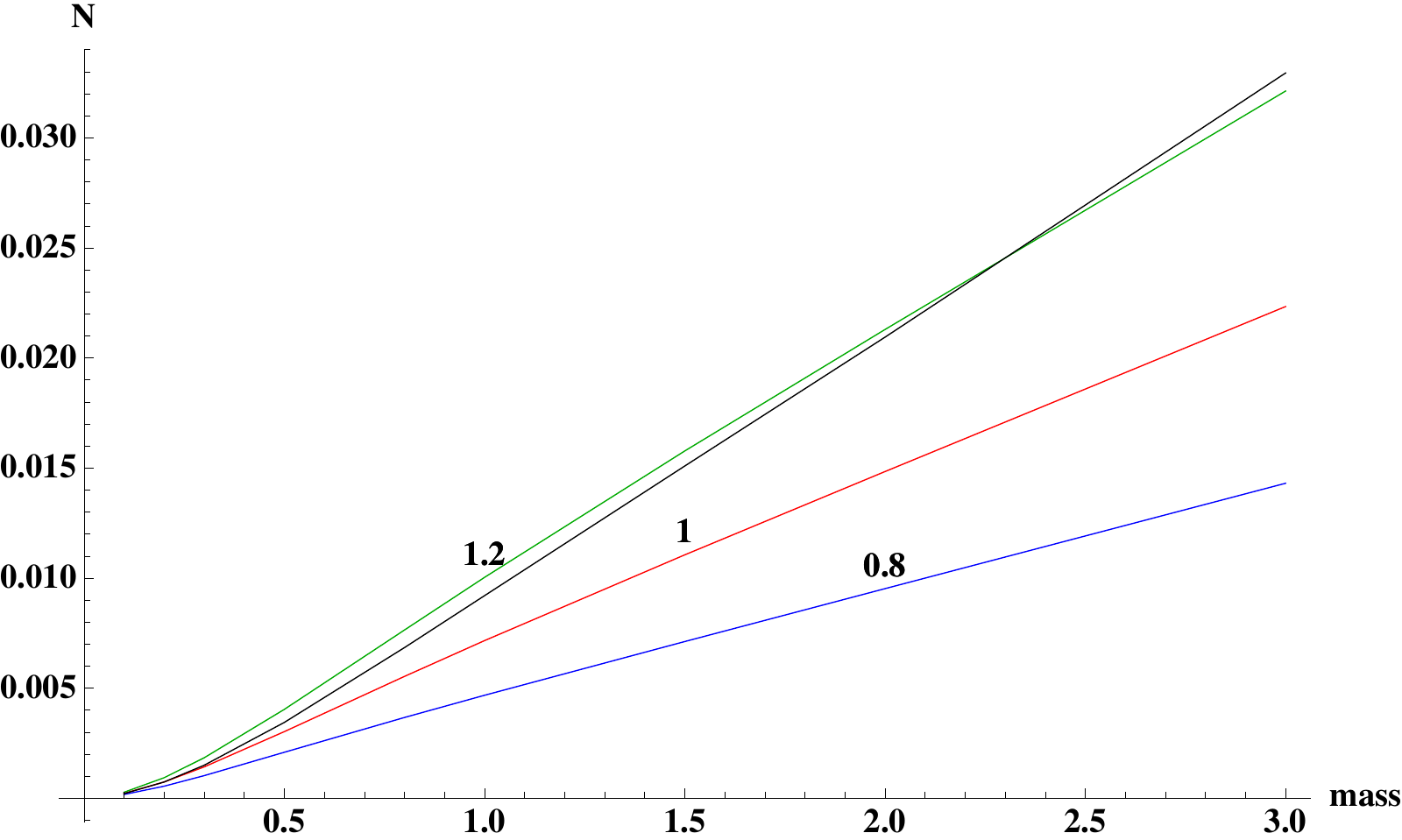}
	\caption{\small $n(\vx)$ as a function of the particle mass $m$ 
	for $\omega=\colthree{0.8}{1.0}{1.2}$ (the asymptotic solution, plotted in
	dark colour, is divided by $2.0$)}
	\label{fig:dirac_Nm}
\end{figure}
In \autoref{fig:dirac_Nm} we show the particle number density
$n(\vx)$ as a function of the particle mass for different $\omega$, and in
\autoref{fig:kg_No} we plot against $\omega$ for different $m$. The other 
parameters are $\tin = 0$ and $\tout = 10$.
\begin{figure}[!ht]
	\centering
	\includegraphics[width=0.8\linewidth]{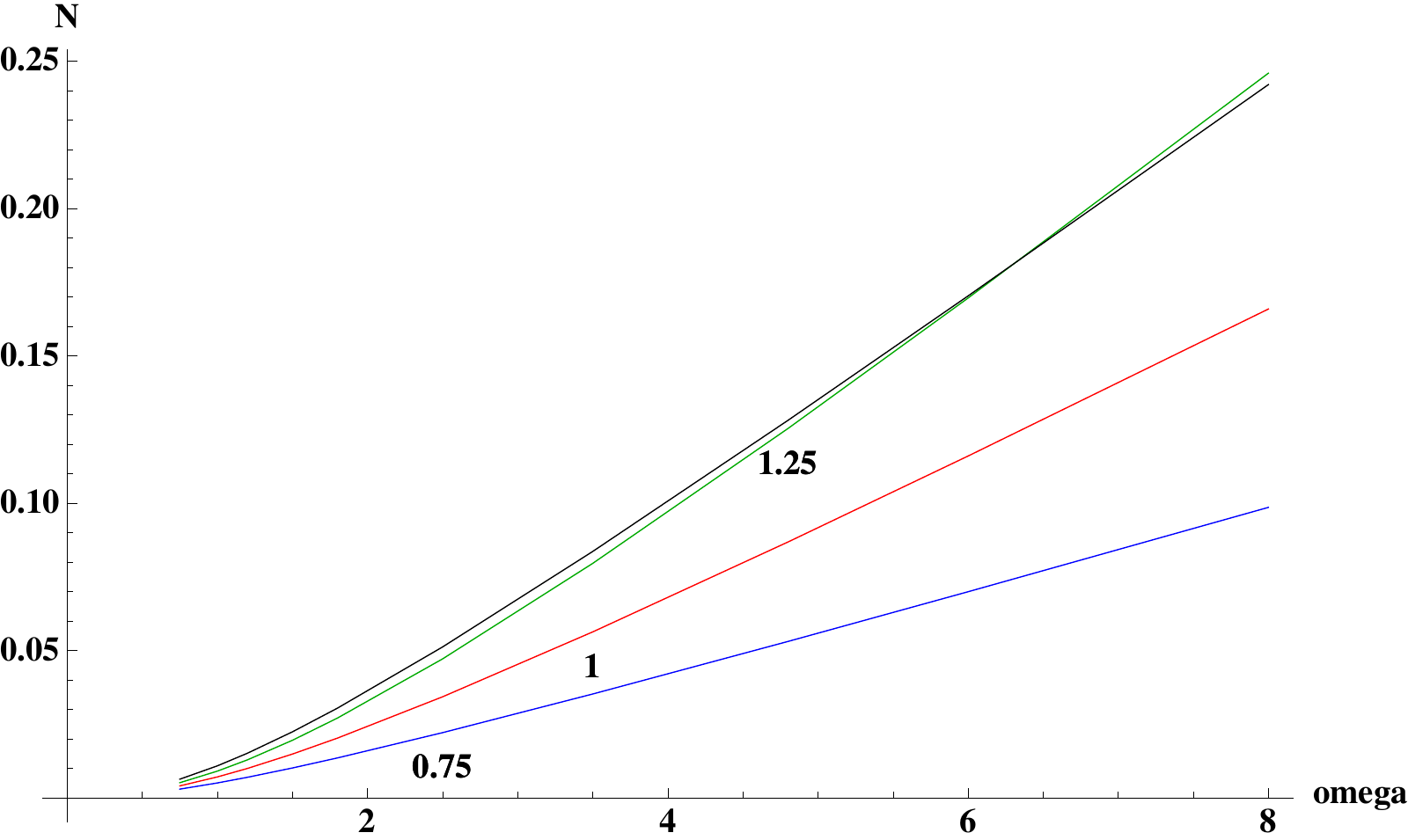}
	\caption{\small $n(\vx)$ as a function of the expansion factor $\omega$ 
	for $m=\colthree{0.75}{1.00}{1.25}$ (asymptotic solution is divided by 
	$1.5$)}
	\label{fig:dirac_No}
\end{figure}

We conclude that there is no dependency on the expansion time $\tout$, as long 
as $\omega \Delta t > 1$. There is a monotonic increase of $n(\vx)$ for 
increasing $m$ and $\omega$, and we expect it to approach $0$ as 
$m\rightarrow 0$ or $\omega \rightarrow 0$. Unfortunately, numerical 
limitations prevent us from exploring the behaviour of $\bsq$ for small
$\omega$ or for large $m$, and the numerical integration fails to converge
when using large $\omega$. Because of the logarithmic divergence of $p^3 \bsq$,
the volumic density of energy is infinite. A possible explanation for this
is that the sudden change in the metric, happening in an infinitely short
time, produces an infinite amount of energy.
\end{subsection}
\end{section}

\chapter{Conclusion}\label{chap:conc}
In this chapter we summarize our results and point out possibilities for 
further development.

We have investigated the phenomenon of particle production by a time-dependent
gravitational field of finite extension. In the following we shall present our 
results sepparatelly for the scalar and the spinorial field.

A common result, which we have confirmed in \autoref{sec:kg_massless} and 
\autoref{sec:dirac_massless}, is that there is no particle production for 
conformal fields (i.e. conformally coupled massless scalar field, and 
massless spinorial field).

In both cases we identify three different regimes. We shall refer to the 
region $q \ll \omega$ as the low momentum (or far infrared) region, 
to $p \ll \omega \ll q$ as the middle region and to the region 
$\omega \ll p$ as the large momentum (ultraviolet) region.

The quantity $\bsq$ and all subsequent quantities derived from it ($n(\vx)$, 
$\mathcal{E}(\vx)$) depend only on the time extension of the expansion phase
$\Delta t = \tout - \tin$. For $\Delta t$ of the order of the Hubble time 
$1/\omega$, quantities of the form $p^a \bsq$ (a = 2,3) approach a constant form, 
and quantities resulting from integration ($n(\vx)$, $\mathcal{E}(\vx)$) become
independent of $\Delta t$.

For the scalar field, we have identified two regimes, corresponding to 
the positiveness  of $\mu = 1/4 - m^2/\omega^2 - 12(\xi - 1/6)$. We shall refer
to the regime corresponding to $\mu > 0$ as hyperbolic, and to $\mu < 0$
as trigonometric.

In the low-momentum limit $\bsq$ approaches a constant value, which
increases exponentially with the time extension $\Delta t$ and the expansion 
factor $\omega$ in the hyperbolic regime, following the hyperbolic sine law 
\eqref{eq:kg_bsqs}. If $\mu < 0$, the argument of the hyperbolic sine is 
imaginary, and we uncover an oscillatory behaviour.

There is a transition occuring in the middle region. There is a rapid descent
from the high value attained in the hyperbolic case, following a power law 
$(2\omega/p)^{2\nu}$ \eqref{eq:kg_middle_sq_pmu_final}. In the trigonometric
regime, $\bsq$ oscillates around a constant term with a roughly constant amplitude.
The constant term resembles a Bose-Einstein distribution function, but we find
a polynomial correction of order $\omega^2/m^2$, which eliminates the strong 
suppresion characteristic to thermal distributions \eqref{eq:kg_middle_sq_nmu},
and the exact form does not diverge near $\nu = 0$, where the approximation
is not valid.

In the large-momentum limit we find that $\bsq$ goes to $0$ as 
$1/p^n$, with $n=6$ for the conformal case and $n = 4$ for any other case.
This assures a finite particle density $n(\vx)$, but the energy exhibits a 
logarithmic divergence when using non-conformal coupling. In the following 
we shall refer only to the conformally coupled case.

The particle number density $n(\vx)$ grows linearly with the particle mass
$m$ \eqref{eq:kg_N_largem} and quadratically with the expansion parameter
$\omega$ \eqref{eq:kg_N_largeo}. We have derived an asymptotic form for 
$n(\vx)$, but cannot confirm its validity in the high $\omega$ or high $m$ 
limits because of insuficient accuracy of the numerical methods used in 
evaluating the exact solution. The energy has a stronger increase with $m$.

In the low momentum limit of the spinorial case, $\bsq$ goes to $0$ as $p^2$
\eqref{eq:dirac_beta_small}.

The middle region is a plateau which follows a 
Fermi-Dirac distribution law in the approximation $p \ll \omega$ and 
$m/\omega \ll 1$. The temperature is equal to $\omega/2\pi$ and the 
particle energy is $m$ \eqref{eq:dirac_middle_as_o0}. 

For large momentum, $\bsq$ goes to $0$ as $1/p^4$, which ensures a finite 
particle number density $n(\vx)$, but gives a logarithmic divergence for 
the energy $\mathcal{E}(\vx)$.

The particle number density increases roughly linear with the mass of the 
particles (see \autoref{fig:dirac_Nm}) and with the expansion factor 
(see \autoref{fig:dirac_No}). Asymptotic analysis revealed a quadratic 
increase of the number density for large enough values of the expansion 
factors, but this prediction could not be verified because of numerical
instability.

The logarithmic divergence of the energy for the non-conformally coupled 
scalar and spinorial fields may be attributed to the unphysical instantaneous
transition between the Minkowski flat regions and the de Sitter expansion 
phase.

Further analysis can be made on the Dirac field asymptotic forms, especially
to the integrals for the particle number density and the energy density. 
Some discussion is needed to clarify the nature of the divergence of the 
energy. The constant value it approaches in the conformally coupled scalar 
case suggests a comparison between our result and the Friedmann equations of
the de Sitter space.
Our work has laid the basis for the investigation of particle 
production in external gauge fields (e.g. a Coulomb field).

\appendix
\counterwithin{equation}{section}
\chapter{Properties of Hankel functions}\label{app:hankel}
This appendix is intended to provide a reference for the properties of the 
Hankel functions required for the development of this paper.
\begin{section}{Differential equation}
In this section we present the construction of the Hankel functions as well
as some important relations between them. The Hankel functions are solutions 
to the Bessel equation \cite{bookpart:smirnov_bess}:
\begin{equation}
  \left(z^2 \frac{d^2}{dz^2} + z \frac{d}{dz} + (z^2 - \nu^2)\right) 
  Z_p(z) = 0. \label{eq:bessel_equation}
\end{equation}
The series solutions to the above equation are the Bessel functions of order
$\pm \nu$:
\begin{equation}
  J_{\pm \nu}(z) = \left(\frac{z}{2}\right)^{\pm\nu}
  \sum_{k = 0}^{\infty} \frac{(-1)^k}{k! \Gamma{k \pm \nu + 1}}
  \left(\frac{z}{2}\right)^{2k}. \label{eq:bessel_function}
\end{equation}	
The complex conjugate passes onto the order and the argument:
\begin{equation}
	J_{\nu}(z)^* = J_{\nu^*}(z^*). \label{eq:bessel_hc}
\end{equation}
The wronskian \cite{bookpart:watson_wronsk} of the Bessel functions of opposite 
order is:
\begin{equation}
  W(J_{-\nu}, J_{\nu}) = \frac{\mathcal{C}}{z} = \frac{2 \sin{\nu\pi}}{\pi z}.
  \label{eq:bessel_wronskian}
\end{equation} 
The two solutions of order $\nu$ and $-\nu$ are linearly independent when 
$\nu$ is not an integer (because the wronskian is non-zero). When 
$\nu = n$ is an integer, the two functions are linearly dependent, since
they are related by
\begin{equation}
	J_{-n}(z) = (-1)^n J_{n}(z). \label{eq:bessel_minusn}
\end{equation}
Bessel functions of the second kind, or Neumann functions, can be 
constructed using $J_{\pm \nu}$:
\begin{equation}
	N_{\nu}(z) = \frac{J_\nu(z) \cos{\nu\pi} - J_{-\nu}(z)}{\sin{\nu\pi}}.
	\label{eq:neumann_function}
\end{equation}
Using \eqref{eq:bessel_hc}, we see that the complex conjugate passes onto the 
order and the argument:
\begin{equation}
	N_{\nu}(z)^* = N_{\nu^*}(z^*). \label{eq:neumann_hc}
\end{equation}
The wronskian of Neumann functions of opposite order follows:
\begin{equation}
  W(N_{-\nu}, N_{\nu}) = \frac{-2 \cos^2{\pi\nu} + 2}{\pi z \sin{\pi\nu}}
  = \frac{2 \sin{\nu\pi}}{\pi z}. \label{eq:neumann_wronskian}
\end{equation}
The wronskian with the bessel function of the same order is
\begin{equation}	
	W(J_{\nu}, N_{\nu}) = W(J_{-\nu}, J_{\nu})/\sin{\nu\pi} = \frac{2}{\pi z}.
  \label{eq:neumann_bessel_wronskian}
\end{equation}
Therefore, the two functions are always linearly independent.

The wronskian with the bessel function of opposite order is
\begin{equation}
	W(J_{-\nu}, N_{\nu}) = W(J_{-\nu}, J_{\nu}) \cos{\nu\pi}/\sin{\nu\pi} = 
	\frac{2 \cos{\nu\pi}}{\pi z}. \label{eq:neumann_besseln_wronskian}
\end{equation}	
Bessel functions of the third kind \cite{book:jackson}, or Hankel functions 
are constructed with these two linear independent solutions:
\begin{subequations} \label{eq:hankel_functions}
\begin{align}
  \Hank{1}{\nu}(z) &= J_\nu(z) + i N_\nu(z) & 
  &\textrm{Hankel function of the first kind}\label{eq:hankel1_def}\\
  \Hank{2}{\nu}(z) &= J_\nu(z) - i N_\nu(z) &
  &\textrm{Hankel function of the second kind}\label{eq:hankel2_def}
\end{align}
\end{subequations}
	The wronskian of the two kinds of Hankel functions is
\begin{equation}
	W(\Hank{1}{\nu}, \Hank{2}{\nu}) = -2i W(J_{\nu}, N_{\nu}) 
	= -\frac{4i}{\pi z}.	\label{eq:hankel12_wronskian}
\end{equation} 
Complex conjugation passes to the order and argument and switches between
the two kinds of Hankel functions:
\begin{equation}
  \Hank{1}{\nu}(z)^{*} = \Hank{2}{\nu^*}(z^*),\qquad
  \Hank{2}{\nu}(z)^{*} = \Hank{1}{\nu^*}(z^*).
  \label{eq:hankel_hc}
\end{equation}
Replacing the Neumann function \eqref{eq:neumann_function} in 
\eqref{eq:hankel_functions} we obtain an alternative form for
the Hankel functions:
\begin{equation*}
  \Hank{1}{\nu}(z) = \frac{i}{\sin{\nu\pi}}\left(J_\nu(z) e^{-i \nu\pi} - 
  J_{-\nu}(z)\right) ,\qquad
  \Hank{2}{\nu}(z) = \frac{-i}{\sin{\nu\pi}}\left(J_\nu(z) e^{i \nu\pi} - 
  J_{-\nu}(z)\right).
\end{equation*}
From this form we can read the connection between $\Hank{1,2}{-\nu}$ 
and $\Hank{1,2}{\nu}$:	
\begin{equation}
  \Hank{1}{-\nu}(z) = e^{i\nu\pi} \Hank{1}{\nu}(z),\qquad
  \Hank{2}{-\nu}(z) = e^{-i\nu\pi} \Hank{2}{\nu}(z).
  \label{eq:hank_minusnu}
\end{equation}	
Denoting by $\Omega_{\nu}$ any of the Bessel functions introduced so far,
the following relations stand:
\begin{subequations}
\begin{gather}
	\Omega_{\nu - 1}(z) + \Omega_{\nu + 1}(z) = \frac{2\nu}{z} \Omega_{\nu}(z)
	\label{eq:bessel_rec_nu},\\
	\Omega_{\nu - 1}(z) - \Omega_{\nu + 1}(z) = 2 \Omega^\prime_{\nu}(z)		
	\label{eq:bessel_rec_nuprime},\\
	\Omega^\prime_{\nu}(z) = \Omega_{\nu - 1}(z) - \frac{\nu}{z} \Omega_{\nu}(z)
	\label{eq:bessel_rec_nuprime_minus},\\
	\Omega^\prime_{\nu}(z) = -\Omega_{\nu + 1}(z) + \frac{\nu}{z} \Omega_{\nu}(z)		
	\label{eq:bessel_rec_nuprime_plus}.
\end{gather}
\end{subequations}
A common example of Hankel functions is that of order $\nu = 1/2$:
\begin{subequations}\label{eq:hankel_1_2}
\begin{align}
	\Hank{1}{1/2} &= -i \sqrt{\frac{2}{\pi z}} e^{i z}\label{eq:hankel1_1_2},\\
	\Hank{2}{1/2} &= \phantom{-}i \sqrt{\frac{2}{\pi z}} e^{-i z}.
	\label{eq:hankel2_1_2}
\end{align}
\end{subequations}
\end{section}
\begin{section}{Asymptotic forms}
In this section we give the asymptotic forms for small and large values of the 
argument, corresponding to the Hankel functions and their derivatives.

For small values of the argument we have \cite{bookpart:zwillinger_bessel}:
\begin{subequations}\label{eq:bessel_all_as_smallz}
\begin{align}
	\Hank{1}{\nu}(z) &=
	\phantom{-}\left(\frac{2}{z}\right)^{\nu} \frac{\Gamma(\nu)}{i\pi} + 		
	\left(\frac{z}{2}\right)^{\nu} \frac{1+i\cot(\pi\nu)}{\Gamma(1+\nu)}+
	\mathcal{O}(z^{2 \pm \nu}), 
	\label{eq:hankel1_as_smallz}\\		
	\Hank{2}{\nu}(z) &=
	-\left(\frac{2}{z}\right)^{\nu} \frac{\Gamma(\nu)}{i\pi} + 
	\left(\frac{z}{2}\right)^{\nu} \frac{1-i\cot(\pi\nu)}{\Gamma(1+\nu)} +
	\mathcal{O}(z^{2\pm \nu}).
	\label{eq:hankel2_as_smallz}
\end{align}
\end{subequations}
The derivatives of the Hankel functions can be approximated by:
\begin{subequations}\label{eq:hankelp_as_smallz}
\begin{align}
	\Hank{1}{\nu}(z)^{\prime} &=
	- \frac{\nu}{z} \left(\phantom{-}
	\left(\frac{2}{z}\right)^{\nu} \frac{\Gamma(\nu)}{i\pi} -
	\left(\frac{z}{2}\right)^{\nu} \frac{1+i\cot(\pi\nu)}{\Gamma(1+\nu)}
	\right) + \mathcal{O}(z^{2\pm \nu}), \label{eq:hankel1prime_as_smallz}\\		
	\Hank{2}{\nu}(z)^{\prime} &=
	- \frac{\nu}{z} \left(
	-\left(\frac{2}{z}\right)^{\nu} \frac{\Gamma(\nu)}{i\pi} - 
	\left(\frac{z}{2}\right)^{\nu} \frac{1-i\cot(\pi\nu)}{\Gamma(1+\nu)}
	\right) + \mathcal{O}(z^{2\pm \nu}). \label{eq:hankel2prime_as_smallz}	
\end{align}
\end{subequations}
For large values of the argument we can use \cite{book:abramowitzstegun}:
\begin{subequations}\label{eq:bessel_all_as_largez}
\begin{align}
	\Hank{1}{\nu}(z) &=
	\sqrt{\frac{2}{\pi z}} e^{i(z - \nu\pi/2 - \pi/4)}
	\left(1 + \mathcal{O}(z^{-1})\right),
	\label{eq:hankel1_as_largez}\\			
	\Hank{2}{\nu}(z) &=
	\sqrt{\frac{2}{\pi z}} e^{-i(z - \nu\pi/2 - \pi/4)}
	\left(1 + \mathcal{O}(z^{-1})\right).
	\label{eq:hankel2_as_largez}
\end{align}	
\end{subequations}
The derivatives of the Hankel functions are simply
\begin{subequations}\label{eq:hankelprime_as_largez}
\begin{align}
	\Hank{1}{\nu}(z)^{\prime} &=
	i \sqrt{\frac{2}{\pi z}} e^{i(z - \nu\pi/2 - \pi/4)}
	\left(1 + \mathcal{O}(z^{-1})\right),
	\label{eq:hankel1prime_as_largez}\\		
	\Hank{2}{\nu}(z)^{\prime} &=
	-i \sqrt{\frac{2}{\pi z}} e^{i(z - \nu\pi/2 - \pi/4)}
	\left(1 + \mathcal{O}(z^{-1})\right).
	\label{eq:hankel2prime_as_largez}
\end{align}
\end{subequations}
Hankel's expansion is given by:
\begin{subequations}\label{eq:bessel_all_as_hankel_largez}
\begin{align}
	\Hank{1}{\nu}(z) &=
	\sqrt{\frac{2}{\pi z}}(P(\nu, z) + i Q(\nu, z)) e^{i\chi},
	\label{eq:hankel1_as_hankel}\\
	\Hank{2}{\nu}(z) &=
	\sqrt{\frac{2}{\pi z}}(P(\nu, z) - i Q(\nu, z)) e^{-i \chi},
	\label{eq:hankel2_as_hankel}
\end{align}		
with
\begin{align}
	P(\nu, z) &\xrightarrow[z\rightarrow \infty]{} 
	\sum_{k = 0}^{\infty} (-1)^k \frac{(\nu, 2k)}{(2z)^{2k}}\nonumber\\
	&= 1 - \frac{(\mu - 1)(\mu - 9)}{2!(8z)^2} + 
	\frac{(\mu - 1)(\mu - 9)(\mu - 25)(\mu - 49)}{4!(8z)^4} 
	+ \mathcal{O}(z^{-6}),\label{eq:P_hankel}\\
	Q(\nu, z) &\xrightarrow[z\rightarrow \infty]{} 
	\sum_{k = 0}^{\infty} (-1)^k \frac{(\nu, 2k + 1)}{(2z)^{2k + 1}}\nonumber\\
	&= \frac{\mu - 1}{8z} - 
	\frac{(\mu - 1)(\mu - 9)(\mu - 25)}{3!(8z)^3} +
	+ \mathcal{O}(z^{-5}),\label{eq:Q_hankel}\\
	(\alpha, n) &=
	- \frac{(n - 1/2)^2 - \alpha^2}{n} (\alpha, n - 1) ,\qquad
	(\alpha, 0) = 1,\nonumber\\
	\chi &= z - \frac{\pi\nu}{2} - \frac{\pi}{4},\nonumber\\
	\mu &= 4\nu^2. \nonumber
\end{align}
\end{subequations}	
Hankel's expansion for the derivatives is:
\begin{subequations}\label{eq:besselp_all_as_hankel_largez}
\begin{align}
	\Hank{1}{\nu}^\prime(z) &=
	i \sqrt{\frac{2}{\pi z}}(R(\nu, z) + i S(\nu, z)) e^{i\chi},
	\label{eq:hankel1p_as_hankel}\\
	\Hank{2}{\nu}^\prime(z) &=
	-i \sqrt{\frac{2}{\pi z}}(R(\nu, z) - i S(\nu, z)) e^{-i \chi},
	\label{eq:hankel2p_as_hankel}
\end{align}		
with
\begin{align}
	R(\nu, z) &\xrightarrow[z\rightarrow \infty]{} 
	\sum_{k = 0}^{\infty} (-1)^k 
	\frac{4\nu^2+16k^2-1}{4\nu^2-(4k-1)^2} \frac{(\nu, 2k)}{(2z)^{2k}}\nonumber\\
	&= 1 - \frac{(\mu - 1)(\mu + 15)}{2!(8z)^2} + \mathcal{O}(z^{-4}),
	\label{eq:R_hankel}\\
	S(\nu, z) &\xrightarrow[z\rightarrow \infty]{} 
	\sum_{k = 0}^{\infty} (-1)^k 
	\frac{4\nu^2+4(2k+1)^2-1}{4\nu^2-(4k+1)^2} \frac{(\nu, 2k + 1)}{(2z)^{2k + 1}}\nonumber\\
	&= \frac{\mu + 3}{8z} - 
	\frac{(\mu - 1)(\mu - 9)(\mu + 35)}{3!(8z)^3} + \mathcal{O}(z^{-5}).
	\label{eq:S_hankel}
\end{align}
\end{subequations}
\end{section}
\begin{section}{Hankel functions for the de Sitter scalar field}\label{sec:hank_kg}
Here we give some useful insight on the Hankel functions employed in the 
development of the scalar theory on de Sitter space.

The solution to the scalar field equation \eqref{eq:kg_Z_def} is chosen
such that in the conformally coupled massless case ($\mu = 1/4$) it 
approaches the Minkowski plane-wave solution. Indeed, the Hankel function
of order $\nu = 1/2$, given by \eqref{eq:hankel1_1_2}, has the exponential
form we are looking for:
\begin{equation*}
	\varphi_p(\eta) \sim_{\substack{m = 0\\\xi = 1/6}} e^{-i p \eta}.
\end{equation*}
The complex conjugate of $Z_\nu(z)$ is
\begin{equation*}
	Z^{*}_\nu(z) = \begin{cases}
	\Hank{2}{\nu}(z) & \mu > 0,\\
	e^{-\frac{\pi\nu}{2}} \Hank{2}{-i \nu}(z) & \mu < 0.
\end{cases}
\end{equation*}
The imaginary argument flipped sign. We can use \eqref{eq:hank_minusnu} to 
get the sign back:
\begin{equation*}
	Z^{*}_\nu(z) = \begin{cases}
		\Hank{2}{\nu}(z) & \mu > 0,\\
		e^{\frac{\pi\nu}{2}} \Hank{2}{i \nu}(z) & \mu < 0.
	\end{cases}
\end{equation*}
which is just what \eqref{eq:kg_Z_def} states. It is important to note
that the Wronskian of $Z_{\nu}$ and $Z^*_\nu$ is the same in both branches, 
irrespective of the value of $M^2$, and is given by 
\eqref{eq:hankel12_wronskian}:
\begin{equation}
	W(Z_\nu(z), Z^*_\nu(z)) = - \frac{4 i}{\pi z}.
	\label{eq:kg_wronski}
\end{equation}
Although the normalization factor is dependent on the value of $\mu$,
it cancels in products of the form $Z_\nu Z^*_\nu$.
\end{section}
\begin{section}{Hankel functions for the de Sitter spinorial field}\label{sec:hank_dirac}
In this section we briefly discuss the Hankel functions used for the 
construction of the solutions to the Dirac equation on de Sitter space 
\eqref{eq:dirac_fieldeq}, which are written in terms of Hankel functions
$\Hank{1/2}{\nu_\pm}(z)$, with $\nu_{\pm} = \nicefrac{1}{2} \pm k$ and 
$z = -p\eta$. In the massless limit, the Dirac equation field is conformal 
to the Minkowski case and the solutions \eqref{eq:dirac_Usol_frw} reduce to
\begin{equation}
	U_{\vp, \lambda}(\eta,\vx) \xrightarrow[m\rightarrow 0]{} \frac{1}{(2\pi)^{3/2} \sqrt{2}} 
	(-\omega\eta)^{3/2} 
	\begin{pmatrix}
		\xi_{\lambda}(\vp)\\2\lambda \xi_{\lambda}(\vp)
	\end{pmatrix} 
	e^{-i p \eta + i \vp \vx}.
	\label{eq:dirac_Usol_m0}
\end{equation}
Up to a factor $(-\omega \eta)^{3/2}$, this is the polarized plane wave 
solution on the Minkowski space.

Under complex conjugation, the Hankel functions change order and kind:
\begin{equation}
	\Hank{1}{\nu_\pm}^*(z) = \Hank{2}{\nu_\mp}(z).
	\label{eq:hank_nupms}
\end{equation}
Using \eqref{eq:bessel_rec_nuprime_minus}, \eqref{eq:hank_minusnu} and 
$-\nu_\pm = 1 - \nu_\mp$ we find:
\begin{subequations}\label{hank_nupm_prime}
\begin{align}
	\Hank{1}{\nu_\pm}^\prime(z) &= \phantom{-}
	i e^{\pm\pi k} \Hank{1}{\nu_\mp}(z)-\frac{\nu_{\pm}}{z}\Hank{1}{\nu_\pm}(z),
	\label{eq:hank1_nupm_prime}\\
	\Hank{2}{\nu_\pm}^\prime(z) &= 
	-i e^{\mp\pi k}\Hank{2}{\nu_\mp}(z)-\frac{\nu_{\pm}}{z}\Hank{2}{\nu_\pm}(z).
	\label{eq:hank2_nupm_prime}
\end{align}
\end{subequations}
Replacing this in the wronskian \eqref{eq:hankel12_wronskian} we find
\begin{equation*}
	iW(\Hank{1}{\nu_-}, \Hank{2}{\nu_-}) = 
	e^{\pi k} \Hank{2}{\nu_+}(z) \Hank{1}{\nu_-}(z) + 
	e^{- \pi k} \Hank{2}{\nu_-}(z) \Hank{1}{\nu_+}(z),
\end{equation*}
and we arrive at the identity \cite{art:cota_polarized_fermions}:
\begin{equation}
	e^{\pi k} \Hank{2}{\nu_+}(z) \Hank{1}{\nu_-}(z) + 
	e^{- \pi k} \Hank{2}{\nu_-}(z) \Hank{1}{\nu_+}(z) = \frac{4}{\pi z}.
	\label{eq:hank_idcota}
\end{equation}
Because the order is $\nu_{\pm} = 1/2 \pm ik$, the small argument 
approximation \eqref{eq:bessel_all_as_smallz} is
\begin{subequations}\label{eq:hankelnu_as_smallz}
\begin{align}
	\Hank{1}{\nu_\pm}\pparant &\xrightarrow[z\rightarrow 0]{} \phantom{-}
	\sqrt{\frac{2\omega}{p}} \frac{1}{i\pi} \Gamma_\pm(p,p) + 
	\sqrt{\frac{p}{2\omega}}\frac{e^{\pm \pi k}}{\pi\nu_\pm}
	\Gamma_\mp(p,p),
	\label{eq:hankelnu1_as_smallz}\\		
	\Hank{2}{\nu_\pm}\pparant &\xrightarrow[z\rightarrow 0]{}
	-\sqrt{\frac{2\omega}{p}} \frac{1}{i\pi} \Gamma_\pm(p,p) + 
	\sqrt{\frac{p}{2\omega}}\frac{e^{\mp \pi k}}{\pi\nu_\pm}
	\Gamma_\mp(p,p).
	\label{eq:hankelnu2_as_smallz}
\end{align}
\end{subequations}
The shorthand notation $\Gamma_{\pm}(p, q)$ stands for
\begin{equation}
	\Gamma_{\pm}(p, q) = \left(\frac{2\omega}{\sqrt{pq}}\right)^{\pm ik}
	\Gamma(\nu_{\pm}),\qquad
	\Gamma_{+}^*(p, q) = \Gamma_{-}(p, q) \quad,\quad
	\abs{\Gamma_{\pm}(p, q)}^2 = \frac{\pi}{\cosh{\pi k}}.
	\label{eq:def_gammapm}
\end{equation}
We have used the identity
\begin{equation}
	\Gamma(x) \Gamma(1-x) = \frac{\pi}{\sin{\pi x}}.
	\label{eq:gamma_id}
\end{equation}
\end{section}

\chapter{Pauli spinors} \label{chap:pauli_spinors}
\begin{section}{Spinor construction}
	The Pauli spinors are two-component eigenvectors of the Pauli matrices:
	\begin{align}
		\sigma_1 = \begin{pmatrix} 
		0 & 1\\
		1 & 0 \end{pmatrix}
		,\qquad
		\sigma_2 = \begin{pmatrix} 
		0 & -i\\
		i & 0 \end{pmatrix}
		,\qquad
		\sigma_3 = \begin{pmatrix} 
		1 & 0\\
		0 & -1 \end{pmatrix}.
		\label{eq:pauli_matrices}
	\end{align}
	These matrices have the following (anti)-commutation rules:
	\begin{equation}
		\acomm{\sigma_i}{\sigma_j} = 2 \delta_{ij},\qquad
		\comm{\sigma_i}{\sigma_j} = 2i \varepsilon_{ijk} \sigma_k.
		\label{eq:sigma_comm_anticomm}
	\end{equation}	
	The goal of this section is to construct the Pauli spinors
	$\xi_\lambda(\vp)$ and $\eta_\lambda(\vp)$ defined by \eqref{eq:dirac_xietadef}. 
	Let $\vn = \vp/p$ be the 
	unit vector along $\vp$. The eigenvalue equations for $\xi$ and $\eta$ are
	\begin{equation}
		\vn \cdot \vsigma \xi_{\lambda}(\vn) = 2 \lambda \xi_{\lambda}(\vn),\qquad
		\vn \cdot \vsigma \eta_{\lambda}(\vn) = -2 \lambda \eta_{\lambda}(\vn).
		\label{eq:xieta_eigeneq}
	\end{equation}
	First, we solve the equation for $\vn=(0,0,1)$. If we denote 
	$\xi_\lambda(\ve3) = (\xi^1_\lambda, \xi^2_\lambda)^{\rm T}$, 
	we arrive at the equation
	\begin{equation*}
		(1 - 2\lambda) \xi^1_\lambda = 0,\qquad 
		(1 + 2\lambda) \xi^2_\lambda = 0.
	\end{equation*}
	From this we conclude that $\lambda = \pm 1/2$, so
	$1 - 4\lambda^2 = (1 - 2\lambda)(1 + 2\lambda) = 0$. This gives the 
	natural solution for $\xi$ (and $\eta$ through (\ref{eq:dirac_etaxicc})):
	\begin{equation}
		\xi_{\lambda}(\ve3) = \begin{pmatrix}
			\frac{1}{2} + \lambda \\
			\frac{1}{2} - \lambda
		\end{pmatrix},\qquad
		\eta_{\lambda}(\ve3) = \begin{pmatrix}
			\frac{1}{2} - \lambda \\
			-\frac{1}{2} - \lambda
		\end{pmatrix}.
		\label{eq:xieta_e3}
	\end{equation}
	Explicitly, the spinors have the following form:
	\begin{equation}
	\begin{array}{c|cc}
		& \xi_{\lambda}(\ve3) & \eta_{\lambda}(\ve3)\\\hline
		\lambda=\frac{1}{2} & \begin{pmatrix}	1\\ 0	\end{pmatrix} & 
		\begin{pmatrix}	0\\ -1	\end{pmatrix}\smallskip\\
		\lambda=-\frac{1}{2} & \begin{pmatrix}	0\\ 1	\end{pmatrix} & 
		\begin{pmatrix}	1\\ 0	\end{pmatrix}		
	\end{array}
	\end{equation}		
	These spinors are already orthonormal in the sense that
	\begin{equation}
		\xi_{\lambda}^{\dagger}(\ve3)\xi_{\lambdap}(\ve3) = \delta_{\lambda\lambdap},\qquad
		\eta_{\lambda}^{\dagger}(\ve3)\eta_{\lambdap}(\ve3) = \delta_{\lambda\lambdap}.
		\label{eq:xieta_ortho_e3}
	\end{equation}
	The spinors corresponding to arbitrary orientations $\vn$ can be 
	constructed from the above by using the spinorial representation
	of the $SU(2)$ group, $D_{\vn}(\theta) = \exp(-i \vn \vsigma/2)$.
	We shall use the {\em Euler angles} parametrization of the rotation
	group:
	\begin{equation*}
		R(\alpha, \beta, \gamma) = R_3(\alpha)R_2(\beta)R_3(\gamma).
	\end{equation*}
	$R$ stands for the standard $3$-dimensional representation of the rotation
	group, and $D(R)$ is the spinorial representation of the rotation $R$.
	The subscripts of the form $i$ denote rotations about the $i$ coordinate
	axis.	The rotation which brings a vector on the third axis given by
	spherical coordinates $\theta=0, \varphi=0$ to an arbitrary position
	$\vn(\theta, \varphi)$ is $R(\varphi, \theta, 0)$, and thus we expect 
	that the eigenvector of the operator $\vn \cdot \vsigma$ is given by
	\begin{equation*}
		\vn \vsigma \xi_{\lambda}(\vn) = 2\lambda \xi_{\lambda}(\vn),\qquad
		\xi_{\lambda}(\vn) = D_3(\varphi) D_2(\theta) \xi_{\lambda}(\ve3).
	\end{equation*}
	We can prove the validity of this construction by considering the action of
	the rotation $D(\varphi,\theta,0)$ over the eigenvalue equation 
	\eqref{eq:xieta_eigeneq} for $\vn = \ve3$. The following identity 
	($A$ and $B$ are matri\-ces) 
	is useful:
	\begin{equation}
	e^{A} B e^{-A} = B + \comm{A}{B} + \frac{1}{2!} \comm{A}{\comm{A}{B}} + \dots
	= B + \sum_{n} \frac{1}{n!} \ad{A}^n(B).\label{eq:id_eA_B_eA}
	\end{equation}
	With \ad{X} given by
	\begin{equation}
		\ad{X}(Y) = \comm{X}{Y},\qquad \ad{X}^{n+1}(Y) 
		= \comm{X}{\ad{X}^n(Y)}. \label{eq:ad_def}
	\end{equation}
	By using the commutation relations \eqref{eq:sigma_comm_anticomm} between 
	the $\sigma$ matrices, we can evaluate
	\begin{equation}
		D_j(\theta) \sigma_i D_j^{\dagger}(\theta) 
		= \sigma_j \cos{\theta} + \varepsilon_{ijk} \sigma_k \sin{\theta},
		\qquad i\neq j.
		\label{eq:sigmai_rotj}
	\end{equation}
	Succesively applying relation \eqref{eq:sigmai_rotj} for the two rotations
	$D(\vn) = D_3(\varphi) D_2(\theta)$ yields
	\begin{equation*}
		D(\varphi,\theta,0) \sigma_3 \xi_{\lambda}(\ve3) = 
		(\sigma_1 \sin{\theta} \cos{\varphi} + \sigma_2 \sin{\theta} \sin{\varphi}
		+ \sigma_3 \cos{\theta}) D_3(\varphi) D_2(\theta) \xi_{\lambda}(\ve3),
	\end{equation*}
	which is equivalent to the eigenvalue equation \eqref{eq:xieta_eigeneq}.
	This enables us to define the spinors of arbitrary orientation $\vn$ as
	\begin{subequations}\label{eq:xieta_vn_def}
	\begin{gather}
		\vn \cdot \vsigma \xi_{\lambda}(\vn) = 2\lambda \xi_{\lambda}(\vn),\qquad
		\xi_{\lambda}(\vn) = D(\vn) \xi_{\lambda}(\ve3), \label{eq:xi_vn_def}\\
		\vn \cdot \vsigma \eta_{\lambda}(\vn) = -2\lambda \eta_{\lambda}(\vn),\qquad
		\eta_{\lambda}(\vn) = D(\vn) \xi_{\lambda}(\ve3). \label{eq:eta_vn_def}
	\end{gather}
	\end{subequations}
	The explicit form of these spinors can be obtained by evaluating the rotation
	matrices $D_3(\varphi)$ and $D_2(\theta)$ and applying them following the 
	prescription \eqref{eq:xieta_vn_def}:
	\begin{subequations} \label{eq:xi_eta_vn}
	\begin{align}
		\xi_{\lambda}(\theta,\varphi) &= \begin{pmatrix}
			e^{-i\varphi/2} \left( 
			\cos{\frac{\theta}{2}} \left(\frac{1}{2} + \lambda\right) -
			\sin{\frac{\theta}{2}} \left(\frac{1}{2} - \lambda\right) \right)\\
			e^{i\varphi/2} \left( 
			\cos{\frac{\theta}{2}} \left(\frac{1}{2} - \lambda\right) +
			\sin{\frac{\theta}{2}} \left(\frac{1}{2} + \lambda\right) \right)
		\end{pmatrix}, \label{eq:xi_vn}\\
		\eta_{\lambda}(\theta,\varphi) &= \begin{pmatrix}
			e^{-i\varphi/2} \left( 
			\cos{\frac{\theta}{2}} \left(\frac{1}{2} - \lambda\right) +
			\sin{\frac{\theta}{2}} \left(\frac{1}{2} + \lambda\right) \right)\\
			e^{i\varphi/2} \left( 
			-\cos{\frac{\theta}{2}} \left(\frac{1}{2} + \lambda\right) +
			\sin{\frac{\theta}{2}} \left(\frac{1}{2} - \lambda\right) \right)
		\end{pmatrix}.\label{eq:eta_vn}
	\end{align}
	\end{subequations}
	Explicitly, these spinors are
	\begin{equation}
	\begin{array}{c|cc}
		& \xi_{\lambda}(\vn) & \eta_{\lambda}(\vn)\\\hline
		\lambda=\frac{1}{2} & 
		\begin{pmatrix}
			e^{-i\varphi/2} \cos{\frac{\theta}{2}}\\
			e^{\phantom{-}i\varphi/2} \sin{\frac{\theta}{2}}
		\end{pmatrix}
		& 
		\begin{pmatrix}
			\phantom{-}e^{-i \varphi/2} \sin{\frac{\theta}{2}}\\
			-e^{\phantom{-}i \varphi/2} \cos{\frac{\theta}{2}} 
		\end{pmatrix}
		\smallskip\\
		\lambda=-\frac{1}{2} & 
		\begin{pmatrix}
			-e^{-i \varphi/2} \sin{\frac{\theta}{2}}\\
			\phantom{-}e^{\phantom{-}i \varphi/2} \cos{\frac{\theta}{2}}
		\end{pmatrix}
		&
		\begin{pmatrix}
			e^{-i\varphi/2} \cos{\frac{\theta}{2}}\\
			e^{\phantom{-}i\varphi/2} \sin{\frac{\theta}{2}} 
		\end{pmatrix}
	\end{array}
	\end{equation}				
	The appearance of half angles should not be a surprise, since we are working
	in a spinorial representation of $\mathcal{S}\mathcal{U}(2)$ (spin $\frac{1}{2}$). The
	interesting feature of this representation is that 
	$\xi_\lambda(\theta, \phi+2\pi) = -\xi_{\lambda(\theta,\phi)}$. This property
	will be very important in understanding the Bogoliubov coefficients defined
	in \autoref{chap:dirac_bogo}.
	
	In order to be convinced that the above definition for $\eta$ still obeys 
	the conjugation relation \eqref{eq:dirac_etaxicc}, we use the anticommutation 
	relations \eqref{eq:sigma_comm_anticomm} and the explicit forms of the 
	$\sigma$ matrices \eqref{eq:pauli_matrices} to obtain
	\begin{equation}
		\sigma_2 \sigma_i = - \sigma_i^* \sigma_2. \label{eq:sigma_cc}
	\end{equation}
	We apply the conjugation relation \eqref{eq:dirac_etaxicc} 
	to the equation defining $\xi_\lambda(\vn)$ \eqref{eq:xieta_vn_def} 
	and we obtain
	\begin{equation}
		i \sigma_2 D^{*}(\vn)\xi^*_{\lambda}(\ve3) = 
		D(\vn) (i \sigma_2 \xi^*_{\lambda}(\ve3)),
	\end{equation}
	since $D^{*}_k(\varphi) = \exp(i \varphi \sigma_k^* /2)$.
	
	Finaly, these spinors obey the same orthonormalization relations as their 
	$\ve3$ counterparts (\eqref{eq:xieta_ortho_e3}): 
	\begin{equation}
		\xi^\dagger_{\lambdap}(\vn) \xi_{\lambda}(\vn) = \delta_{\lambda\lambdap}
		,\qquad
		\eta^\dagger_{\lambdap}(\vn) \eta_{\lambda}(\vn) = \delta_{\lambda\lambdap}
		\label{eq:xieta_ortho_vn}
	\end{equation}
\end{section}
\begin{section}{Behaviour under parity transformations}
	It is instructive to consider the parity-transformed spinor corresponding 
	to $-\vn$. In spherical coordinates, the easiest way to define this 
	reflection about the origin of the coordinate axes is 
	\begin{equation}
		\vn(\theta, \varphi) \mapsto -\vn(\pi - \theta, \varphi + \pi)
		\mapsto -(-\vn)(\theta, \varphi + 2\pi).
		\label{eq:sph_rev}
	\end{equation}
	We can see that two consecutive reversals (as defined above)
	are equivalent to a shift of $2\pi$ in the angle $\varphi$, which has the
	effect of introducing a $-$ sign in the corresponding spinor. To prove this,
	we must investigate the transformed spinor. Let's analyze first the spinor
	$\xi_{\lambda}(-\ve3)$ (which we read from \eqref{eq:xi_vn}):
	\begin{equation}
		\xi_{\lambda}(-\ve3) = i \xi_{-\lambda}(\ve3).
		\label{eq:xi_me3}
	\end{equation}
	The $i$ factor is the hallmark of the spinorial representations
	of the rotation group. Next, in order to evaluate de general case, we
	must use
	\begin{equation}
		D_3(\pi) D_2(\theta) D_3^{\dagger}(\pi) = D_2(-\theta).
	\end{equation}	
	The above follows from the unitarity of the representation 
	matrices and relation \eqref{eq:sigmai_rotj}. We can now evaluate the $-\vn$
	spinor:
	\begin{subequations}\label{xi_eta_mvn}
	\begin{multline}
		\xi_{\lambda}(-\vn) = D_3(\pi + \varphi) D_2(\pi - \theta) \xi_{\lambda}(\ve3) = 
		D_3(\varphi) D_2(\theta) D_3(\pi) D_2(\pi) \xi_{\lambda}(\ve3)\\
		= i \xi_{-\lambda}(\vn).
		\label{eq:xi_mvn}
	\end{multline}
	The $\eta$ spinor has a similar behaviour, which we investigate by applying
	the charge conjugation \eqref{eq:dirac_etaxicc} to the above expression:
	\begin{equation}
		\eta_{\lambda}(-\vn) = -i \eta_{-\lambda}(\vn).
		\label{eq:eta_mvn}
	\end{equation}
	\end{subequations}
\end{section}

\begin{section}{Relation between \texorpdfstring{$\xi$ and $\eta$}{xi and eta}}
	If we take a closer look at the forms of the spinors $\xi$ and $\eta$ 
	\eqref{eq:xieta_e3}, we see that there is an easy relation between them:
	\begin{equation}
		\eta_{\lambda}(\ve3) = -2 \lambda \xi_{-\lambda}(\ve3).
	\end{equation}
	Rotating the above relation to an arbitrary vector $\vn$, we obtain 
	\begin{equation}
		\eta_\lambda(\vn) = - 2\lambda \xi_{-\lambda}(\vn).
	\end{equation}
	If we let $\vn \mapsto -\vn$ as defined in \eqref{eq:sph_rev}, we find the
	relevant relation
	\begin{equation}
		\eta_{\lambda}(-\vn) = -2 i \lambda \xi_{\lambda}(\vn).
		\label{eq:eta_mvn_xi}
	\end{equation} 
	With this we can evaluate the inner products
	\begin{equation}
		\eta^\dagger_{\lambda}(-\vn) \xi_{\lambdap}(\vn) = 
		-2 i \lambda \delta_{\lambda\lambdap},\qquad
		\xi^\dagger_{\lambda}(-\vn) \eta_{\lambdap}(\vn) = 
		2 i \lambda \delta_{\lambda\lambdap}\label{eq:eta_mvn_xi_vn_prod}
	\end{equation}
	These relations will be used in the derivation of the matching coefficients
	between de Sitter and Minkowski modes in \autoref{chap:dirac_bogo}.
\end{section}

\bibliographystyle{agsm}
\bibliography{thesisbib}
\end{document}